\documentclass[traditabstract]{aa}
\usepackage{txfonts}
\usepackage{graphicx}
\usepackage{natbib}
\usepackage{rotating}

\usepackage{natbib,twoopt}
 \usepackage[breaklinks=true]{hyperref} 
 \bibpunct{(}{)}{;}{a}{}{,}    
 \newcommandtwoopt{\citeads}[3][][]{\href{http://adsabs.harvard.edu/abs/#3}%
                                        {\citealp[#1][#2]{#3}}}
 \newcommandtwoopt{\citepads}[3][][]{\href{http://adsabs.harvard.edu/abs/#3}%
                                        {\citep[#1][#2]{#3}}}
 \newcommandtwoopt{\citetads}[3][][]{\href{http://adsabs.harvard.edu/abs/#3}%
                                        {\citet[#1][#2]{#3}}}
\newcommandtwoopt{\citeyearads}[3][][]%
   {\href{http://adsabs.harvard.edu/abs/#3}{\citeyear[#1][#2]{#3}}}

\bibpunct{(}{)}{;}{a}{}{,}

\begin{document}

\title{Understanding synthesis imaging dynamic range}

\author{Robert Braun\inst{1}}

\offprints{Robert Braun, \email{robert.braun@csiro.au}}

\institute{CSIRO -- Astronomy and Space Science, PO Box 76, Epping, NSW 1710, Australia}

\date{Received $nnn$ / Accepted $nnn$}

\abstract{We develop a general framework for quantifying the many
  different contributions to the noise budget of an image made with an
  array of dishes or aperture array stations. Each noise contribution
  to the visibility data is associated with a relevant
  correlation timescale and frequency bandwidth so that the net impact
  on a complete observation can be assessed when a particular
    effect is not captured in the instrumental calibration. All
  quantities are parameterised as function of observing frequency and
  the visibility baseline length. We apply the resulting noise budget
  analysis to a wide range of existing and planned telescope systems
  that will operate between about 100 MHz and 5 GHz to ascertain the
  magnitude of the calibration challenges that they must overcome to
    achieve thermal noise limited performance. We conclude that
    calibration challenges are increased in several respects by small
  dimensions of the dishes or aperture array stations. It will be more
  challenging to achieve thermal noise limited performance using 15~m
  class dishes rather than the 25~m dishes of current arrays. Some of
  the performance risks are mitigated by the deployment of phased
  array feeds and more with the choice of an $(alt,az,pol)$ mount,
  although a larger dish diameter offers the best prospects for risk
  mitigation. Many improvements to imaging performance can be
  anticipated at the expense of greater complexity in calibration
  algorithms. However, a fundamental limitation is ultimately imposed
  by an insufficient number of data constraints relative to
  calibration variables. The upcoming aperture array systems will be
  operating in a regime that has never previously been addressed,
  where a wide range of effects are expected to exceed the thermal
  noise by two to three orders of magnitude. Achieving routine thermal
  noise limited imaging performance with these systems presents an
  extreme challenge. The magnitude of that challenge is inversely
  related to the aperture array station diameter. }

\maketitle
\section{Introduction\label{sec:intro}}
Dynamic range limitations in synthesis imaging arise from three
distinct categories of circumstance: 1. Instrumental artefacts.
2. Imaging artefacts. 3. Incomplete calibration of the instrumental
response. The first category can be addressed by insuring a linear
system response to signal levels together with other design measures
within the receiver and correlator systems that minimise spurious
responses. While challenging to achieve, the engineering requirements
in this realm are moderately well defined and this class of circumstance
will not be considered further in the current discussion. Some of the
specific relevant effects, such as ``closure errors'' and quantisation
corrections are discussed in \citetads{1999ASPC..180..275P}. The
second category includes a range of effects, from the well understood
smearing effects that result from a finite time and frequency sampling
to the imaging challenges associated with non-coplanar baselines as
discussed in \citetads{2008ISTSP...2..647C}. The third category is one
that is less well understood and documented. Several important effects
in this area are also discussed by \citetads{1999ASPC..180..275P};
including imperfect polarisation calibration, inadequate (u,v)
coverage, and numerical modelling errors. In this study we will attempt
to identify and quantify the phenomena that influence synthesis image
dynamic range more generally.

The approach we adopt is the consideration of a wide range of
  calibration issues that influence interferometeric imaging. For each
  of these effects we develop a parameterised model that quantifies
  the fluctuation level due to that effect on the measured
  visibilities, together with its correlation timescale and frequency
  bandwidth. We then assess the equivalent image noise due to each
  effect. While some effects contribute directly to the image noise,
  many contribute to image noise in an indirect manner via the
  self-calibration process. We demonstrate that both direct (in-field)
  and indirect (out-of-field) noise contributions typically have a
  comparable image magnitude if it has proven necessary to employ
  self-calibration. By tracking each calibration effect individually,
  it becomes possible to quantify how each contributes to a final
  image noise level. The purpose of calibration and imaging strategies
  is to accurately model all of the effects that might limit imaging
  performance. We do not attempt to evaluate how well any specific
  algorithm or strategy performs in this regard. We only provide an
  indication of the precision with which each of these calibration
  effects must be addressed for an assumed telescope performance
  specification, so as not to form an ultimate limitation on
  performance. 

\section{Calibration errors\label{sec:cal}}
A key factor that determines image dynamic range is calibration of the
instrumental gain. Instrumental gain, in this context, should be
considered to be the (spatial, polarisation, frequency) response in
(amplitude, phase) as function of time of each of the antennas (be
they aperture array stations, or the phased-array fed- or single pixel
fed- dish beams) of a synthesis array. Traditional array calibration
methods, employing semi-continuous (minute scale) noise source
injection and occasional (hour scale) observations of
well-characterised calibration sources routinely achieve a calibration
precision at GHz frequencies of a few percent in amplitude and about
10 degrees of phase. This gain calibration precision, $\phi_C =
0.2$, is only sufficient to achieve an image dynamic range of less
than about 1000:1 in full track observations with existing arrays. A
rough estimate of the relation between these quantities is derived by
\citetads{1999ASPC..180..275P}. The dynamic range, D, is about,
\begin{equation}
D \approx M_T^{0.5} M_F^{0.5} N / \phi, 
\label{eqn:dr}
\end{equation}
for $M_T$ independent time and $M_F$ frequency intervals and $N$
antennas that each have independent gain errors of magnitude
$\phi$. What constitutes an independent time sample depends critically
on the nature of the gain error together with both baseline length and
field of view; varying from hours to fractions of a second. Similar
considerations also apply to what constitutes an independent frequency
interval. Improvements of the image dynamic range from this base level
rely on self-calibration. Self-calibration
\citepads[e.g.][]{1999ASPC..180..187C} consists of the iterative
modelling of the sky brightness distribution together with the
instrumental calibration. Self-calibration only works when two
conditions are met:
\begin{enumerate}
\item{There are a sufficiently small number of degrees of freedom in
  this model relative to the number of independent data constraints. }
\item{The signal-to-noise ratio in the data for each antenna is
  sufficiently high within each solution interval.}
\end{enumerate}

The number of data constraints (for a given frequency channel and
polarisation state) is $N_D \approx N^2/2$ over the solution timescale
over which all model parameters are constant. Significant
instrumental/ atmospheric/ ionospheric gain fluctuations typically
occur on (sub-) minute timescales. If the sky brightness is dominated
by a single compact source, then the number of degrees of freedom,
$N_F$, is $N_F \approx 1 + N$, and its clear that the system is highly
over-determined ($N_D >> N_F$) for the solution of the $N$ antenna
gains in a single source direction within each solution interval. If,
on the other hand, the sky brightness is characterised by a large
number of widely separated components, $N_C$, each of which is
represented by multiple variables (eg. a peak brightness, two
positions, two sizes and an orientation) that approaches $N^2$ in
number, then its clear that difficulties can arise. In that case, it
has become customary to assume that the spatial instrumental response
to the sky brightness remains time invariant, so that the total number
of degrees of freedom, $N_F \approx 6 N_C + M_T N$ still remains small
relative to the complete observation that comprises $N_T = M_T N^2/2$
total independent data constraints.  We will explore below the
circumstances under which this assumption breaks down.  In practise,
the number of degrees of freedom will be closely tied to the specific
calibration algorithm that is employed. There are clearly great
benefits to minimising $N_F$ with an algorithm that is optimally
matched to the problem.

The requirement of sufficient signal-to-noise
in the data from each antenna has a direct bearing on the precision of
the self-calibration solution that can be
achieved. \citetads{1999ASPC..180..187C} give an estimate for the
residual calibration error for a given solution timescale, $\tau_S$, as
$\phi(\tau_S) = \sigma(\tau_S)/[(N-3)^{0.5}S]$ in terms of the fluctuation
level of a single visibility $\sigma(\tau_S)$ and a single dominant,
unresolved source of flux density, $S$. In the case of a more
complicated sky brightness distribution composed of $N_C$ widely
separated discrete
components of total flux density $S_{Tot}$, this becomes:
\begin{equation}
\phi(\tau_S) = \sigma(\tau_S)N_C^{0.5}/[(N-3)^{0.5}S_{Tot}] , 
\label{eqn:phi}
\end{equation}
since the $N_C$ components will not contribute coherently to the net
visibility vector. A practical solution timescale in the single,
on-axis source case would be tied to the intrinsic fluctuation
timescale of about a minute, while for the multiple component case, it
will be determined by the off-axis distance on the sky of these
components. As we will see below, this will be comparable to a single
integration time; typically one to ten seconds, for sources within the
main beam of the antenna.  When both these conditions are satisfied to
a sufficient degree, it has proven possible to achieve image dynamic
range, $D = S_{Max}/\sigma_{Map}$, defined to be ratio of peak
brightness, $S_{Max}$, to the image root mean square (RMS) fluctuation
level, $\sigma_{Map}$, of more than one million to one with current
arrays after considerable effort by ``black belt'' level
interferometrists. Its vital to note that the highest dynamic range
imaging that has been achieved to date
\citep[eg.][]{2011A&A...527A.108S} is associated with fields dominated
by particularly bright, yet simple sources. Fields that are dominated
by more typical complex sources have not achieved anything comparable.

Rather than specifying some arbitrarily high numerical value of the dynamic
range as a goal for a particular instrument or design, a more useful
requirement might be that the instrument routinely achieves thermal noise
limited performance even after long integrations. Since the noise in a
naturally weighted synthesis image is simply related to the visibility noise
as,
\begin{equation}
\sigma_{Map} = \sigma(\tau_S)/[M_T M_F N (N-1)/2]^{0.5} , 
\label{eqn:sens}
\end{equation}
we can assess individual contributions, denoted by subscript $i$, to
the final image noise by considering their magnitude on a self-cal
solution timescale, $\sigma_i(\tau_S)$, together with the values of
$M_{Ti}$ and $M_{Fi}$ which apply to each particular contribution. It
is important to stress that the equivalent image noise calculated in
eqn.~\ref{eqn:sens} will only represent an actual image noise when it
relates to objects within the image field of view. For effects that
pertain to sources beyond the field of view the influence is more
indirect as will be discussed in detail below. However, when
self-calibration is employed, even the out-of-field error
contributions will result in this magnitude of image fluctuation due
to their adverse impact on the self-calibration error. 
Equation~\ref{eqn:sens} will be used to track the magnitude of
the different contributions to self-calibration error while taking
into account the very different correlation timescales and bandwidths
that apply to each.

As noted above, a major simplifying assumption that is often
implicitly, if not explicitly, invoked is that of a time-invariant
observed ``sky'' during the period of data acquisition. The underlying
assumption is that neither the radio sky nor the (spatial,
polarisation, frequency) components of the instrumental response vary
with time and that only a single (amplitude, phase) term is sufficient
to capture the time variable nature of the instrumental response. What
if this is not the case?

\subsection{Far sidelobes}\label{sec:far}

The assumption of a stationary sky brightness will break down in
practise for several reasons. Firstly, there is the residual all-sky
response due to the far sidelobe pattern of the antenna. The far
sidelobe attenuation of an antenna relative to the on-axis response,
$\epsilon_F$, is given approximately by the ratio of the effective area of a
single dipole to that of the entire antenna, 
\begin{equation}
\epsilon_F = \eta_F (\lambda/d)^2,
\label{eqn:epf}
\end{equation}
with a proprotionality constant $\eta_F$ that is tied to the quality
of aperture illumination and varies between about $0.1 < \eta_F < 1$
in practise. The $(\lambda/d)^2$ dependence is a manifestation of
physical optics and applies to both blocked and unblocked
apertures. Far sidelobe levels have been accurately determined for the
25 m diameter Very Long Baseline Array (VLBA) antennas at 1717~MHz by
\citet{2002.Dhawan}. Peak sidelobe levels of about -53dB were measured
at 30--110 degrees off-axis, yielding $\eta_F \approx
0.1$. \citet{2002.Dhawan} note that while these VLBA levels are in
good agreement with theoretical expectations, the values for Very
Large Array (VLA) antennas were about 6 dB higher at that time
($\eta_F \approx 0.4$), likely due to nonoptimum illumination by the
20cm feed lens system that has since been replaced in the Jansky Very
Large Array (JVLA) upgrade. The prime focus fed 25~m diameter dishes
of the Westerbork Synthesis Radio Telescope (WSRT) have measured far
sidelobe levels at $\lambda =$49 and 92~cm that are consistent with
$\eta_F \approx 0.1$ \citep{1993.Braun}. All radio sources above the
horizon are attenuated by only this factor when they appear in a far
sidelobe peak of the antenna. As the mainbeam of the antenna is used
to track a source on the sky, the apparent brightness of the sky will
vary due to the rising and setting of specific sources relative to the
antenna horizon, but also due to changes in the far sidelobe response
including, but not limited to, rotation of that pattern relative to
the sky for an antenna mount that does not compensate for parallactic
angle. The far sidelobe voltage response pattern of an antenna
consists of local maxima and minima that are comparable to the main
beam in size and separated by nulls. The basic pattern will have a
radial frequency scaling, but may also contain additional frequency
dependence, for example when multi-path propagation conditions are
present.

The properties and density of radio sources are summarised in
Table \ref{tab:nvss}. Values are based on the NRAO VLA Sky Survey (NVSS)
\citepads{1998AJ....115.1693C} for source densities and median flux
densities in each bin at GHz frequencies, and
\citetads{2003NewAR..47..357W} for the median angular size and
spectral index. Some care needs to be exercised with the
interpretation of the angular size, since in many cases this refers to
the separation of more compact components, rather than to the
effective diameter of a diffuse source. This is particularly true for
the brightest GHz source counts that are dominated by luminous
classical double sources down to about 30 mJy. Below about 30 mJy, the
source counts begin to be dominated by lower luminosity, edge-darkened
radio galaxies and below about 0.1 mJy by star forming galaxies
\citepads[e.g.][]{2008MNRAS.388.1335W}.

\begin{table}
\caption{Statistical source properties of the extragalactic sky at 1.4
  GHz in bins of 0.5 dex.}
\label{tab:nvss}
\centering
\begin{tabular}{c c c c c}
\hline\hline
S$_{1.4GHz}$ & $\mu_{1/2}$ & N($>$S) & FWHM & Spectral Index \\
(Jy) & (Jy) & (sr$^{-1}$) & (arcsec) &  \\
\hline
30	&272	&2.4	&160	&-0.80 \\
10	&14	&5.7	&100	&-0.85 \\
3	&4.2	&33	&25	&-0.87 \\
1	&1.4	&203	&15	&-0.90 \\
0.3	&0.42	&1360	&10	&-0.87 \\
0.1	&0.14	&5920	&8	&-0.83 \\
0.03	&0.042	&21700	&5	&-0.80 \\
0.01	&0.014	&56100	&4	&-0.75 \\
0.003	&0.0042	&136000	&3	&-0.70 \\
\hline
\end{tabular}
\end{table}

\begin{figure*}
\resizebox{\hsize}{!}{\includegraphics{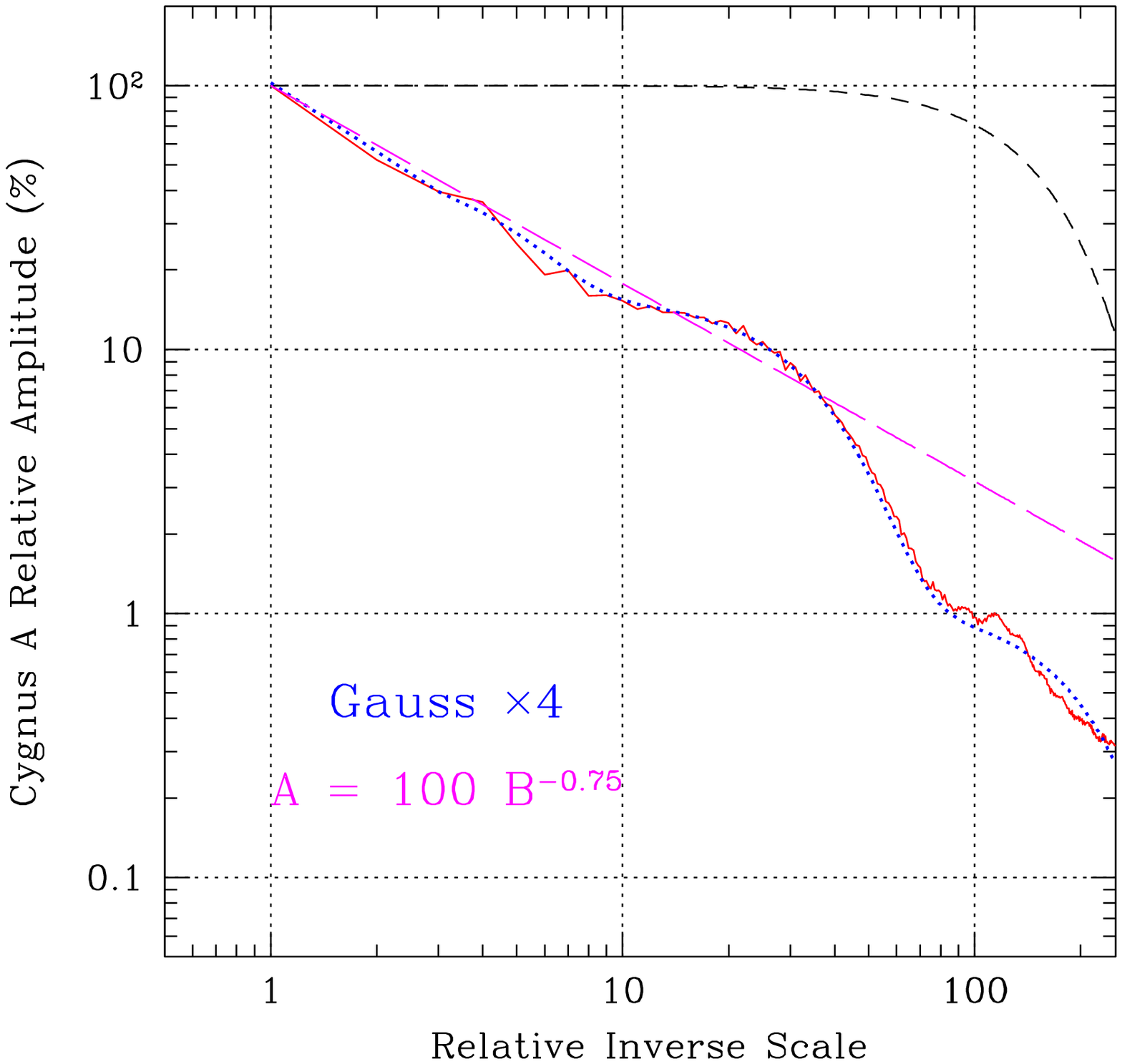},\includegraphics{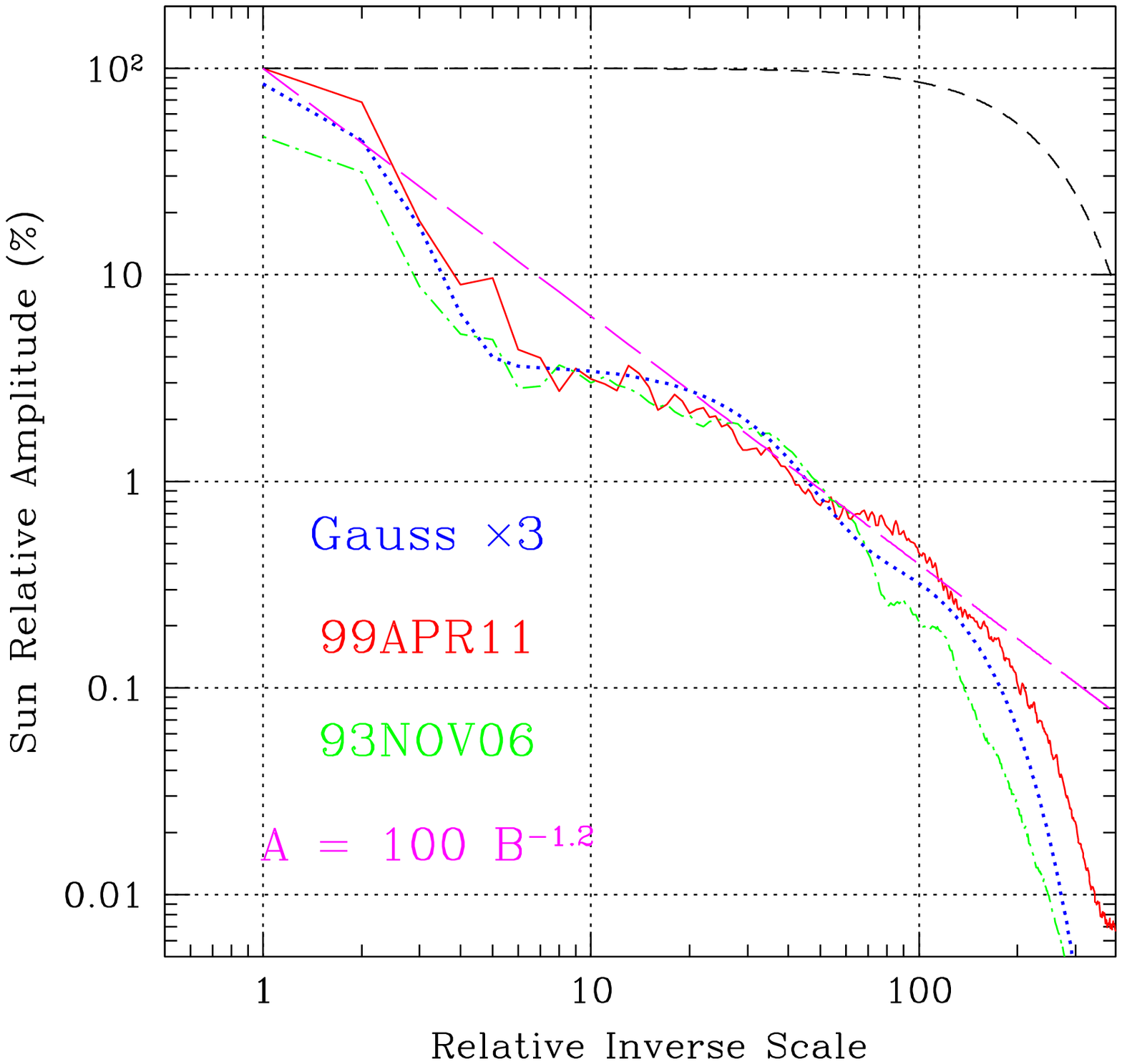}}
\caption{Relative visibility amplitude of Cygnus A (left) and the
  Sun (right) are shown as the solid curves. The short-dashed curve
  shows the point spread function of the observations, while the
  long-dashed curve a power-law approximation. The dotted curve is a
  sum of four (left) and three (right) Gaussians that approximates the
  measured spectra. Both the active (1999APR11) and quiet (1993NOV06)
  Sun are shown.}
\label{fig:powspec}
\end{figure*}

\begin{figure*}
\resizebox{\hsize}{!}{\includegraphics{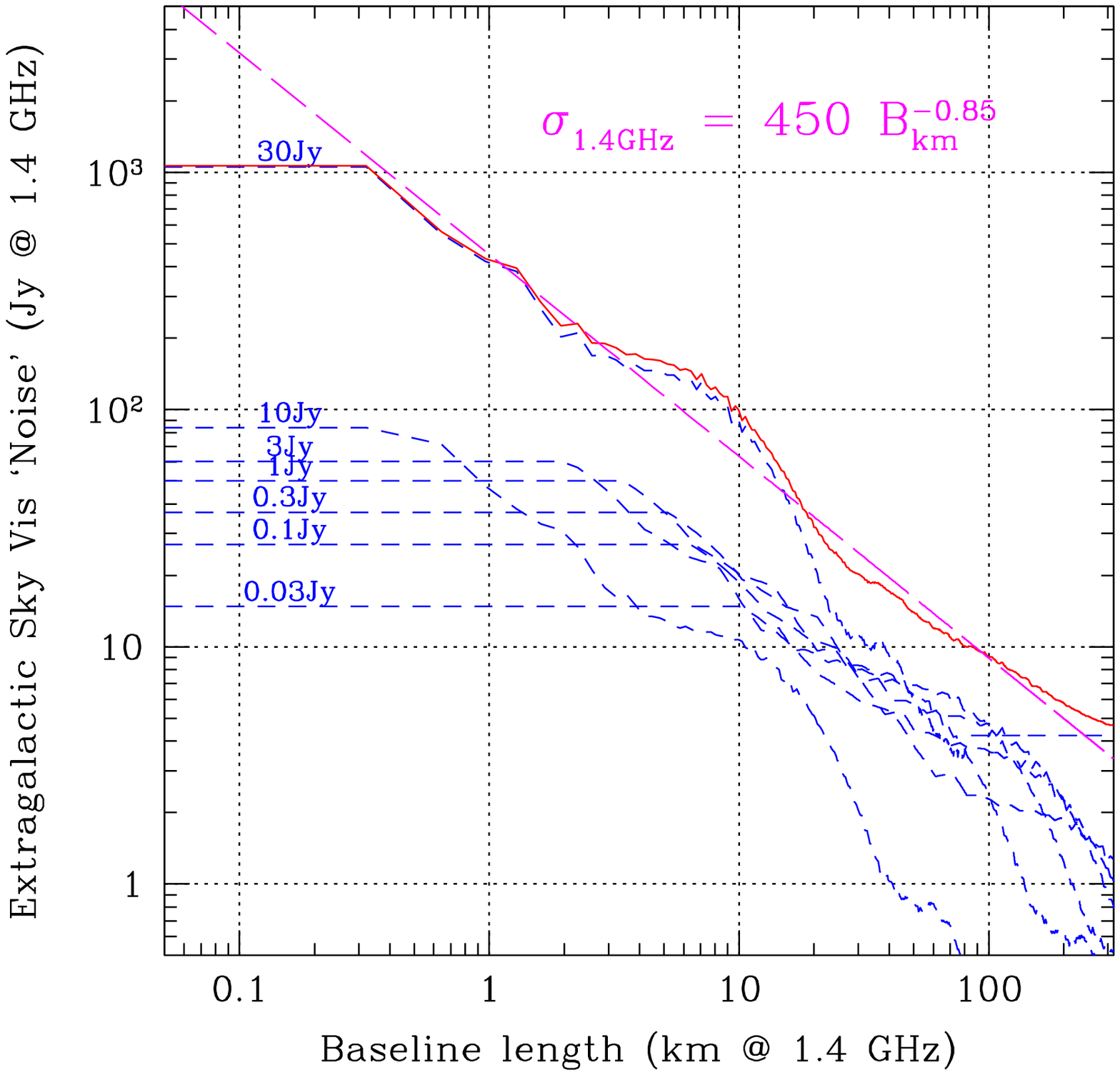},\includegraphics{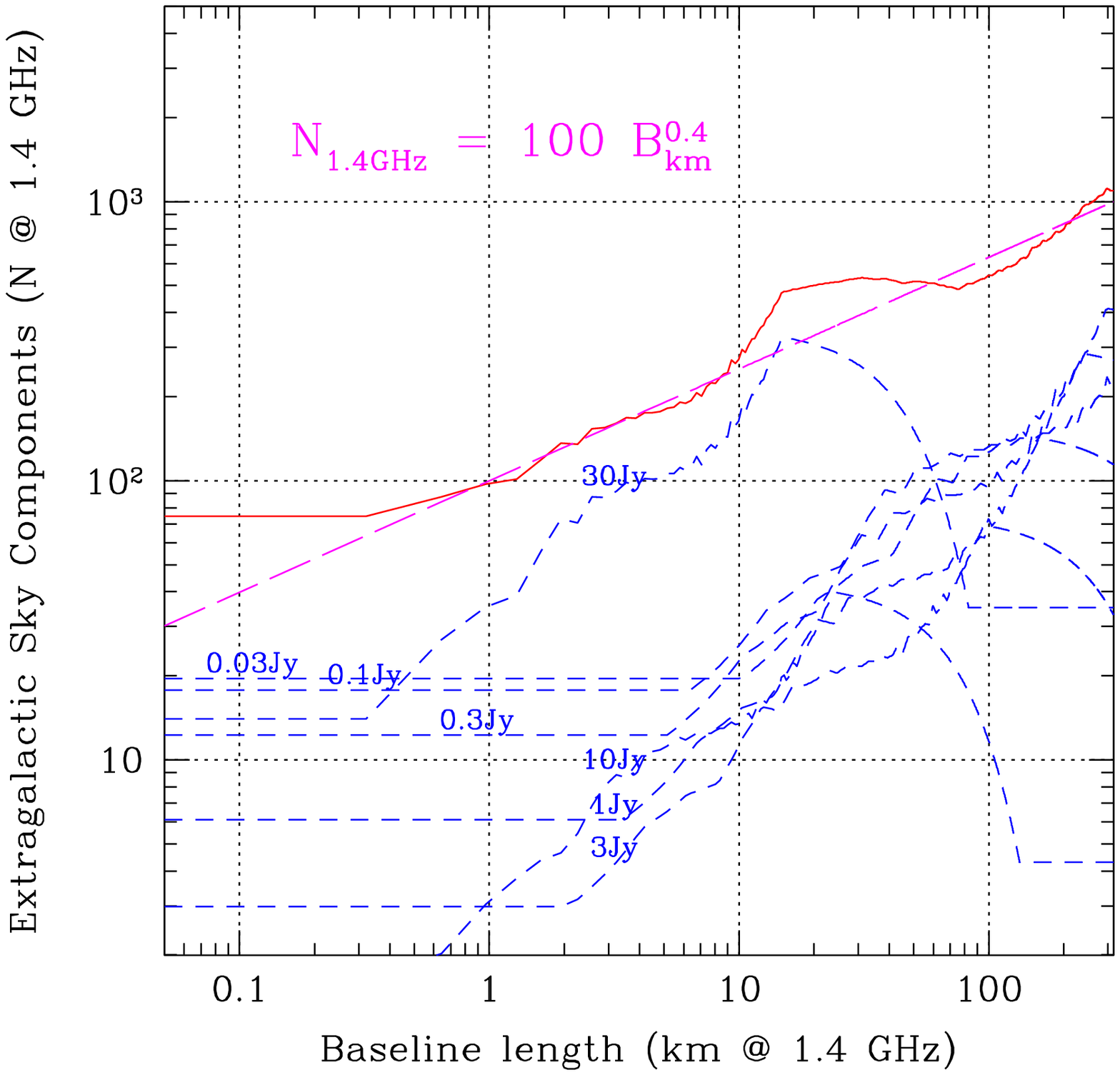}}
\caption{Incoherent visibility ``noise'' due to 2$\pi$ steradians
  of the extragalactic sky (left) and the corresponding number of
  flux-weighted components (right) as function of baseline length
  (angular scale). Visibility ``noise'' is completely dominated by the
  brightest source bin out to baselines of about 30 km that would
  require several hundred components to adequately model.}
\label{fig:sign}
\end{figure*}

A good indication for how the flux density of high luminosity radio
galaxies varies with scale is given in Figure~\ref{fig:powspec}
showing the power spectrum of Cygnus A based on a VLA A+B+C+D
configuration 6 cm image of \citetads{1984ApJ...285L..35P}. The solid
line is the average normalised power as function of inverse scale,
deconvolved for PSF tapering in the observation based on the
short-dashed curve.  The long-dashed curve is a power law of slope
-0.75 which provides a reasonable representation of the trend for
scales up to 30 times smaller than the source size. For statistical
purposes, the Cygnus A power spectrum can be represented by the sum of
four Gaussian components as shown by the dotted curve that has
relative angular scales of (1: 0.29: 0.039: 0.0067) and amplitudes of
(0.648: 0.237: 0.107: 0.008). Such a representation is clearly
inadequate for detailed modelling of individual sources.  From the
combination of number densities and source sizes in Table 1 together
with the power spectrum in Figure~\ref{fig:powspec}, one can construct
a composite fluctuation spectrum due to the high luminosity
extragalactic background sources within $2\pi$ steradians. This is
shown in Figure~\ref{fig:sign} as the solid line, while the
short-dashed curves are the contributions from the individual flux
bins down to 30 mJy from Table 1.  Within each flux bin, we form the
incoherent sum of the relevant sources by taking the square root of
the source number multiplied by the median flux of that
bin. Similarly, the composite fluctuation level is taken to be the
square root of the sum of the squares of the bin contributions. This
is intended to describe the net composite fluctuation level in an
interferometric visibility for a given baseline length. The net
composite fluctuation level is well fit by a power law of slope -0.85
in baseline length from a base level of 450 Jy at 1 km as shown by the
long-dashed curve.

The other important parameter to quantify is the effective number of
model components that would be needed to represent the sky brightness
distribution. Within each flux bin of Table~\ref{tab:nvss} we have
taken the number of sources within 2$\pi$ steradians and scale this
with the inverse of the normalised power spectrum shown in
Figure~\ref{fig:powspec}. In this way, as each source becomes resolved
it is distributed into a proportionally larger number of fainter
components. We then form the weighted sum of components over flux bins
using a weighting factor equal to the median flux of that bin. The
total effective component number is shown as the solid line in
Figure~\ref{fig:sign}, while the individual bin contributions are
plotted as the short-dashed lines. As expected, only a few hundred
components dominate the sky on all baselines between about 0.5 and 50
km. On longer baselines the total number increases approximately as a
power law of baseline with slope $\sim$0.4 from a base level of 100 on
1 km baselines as shown with the long-dashed curve.

With the -50dB peak far sidelobe level of a 25 m dish observing at 25 cm,
one hemisphere of the extragalactic radio sky would contribute some 5
mJy of noise-like fluctuations to visibility measurements on 1 km
baselines. Adequate modelling of these sources would require about 100
components on this same scale. On 10 km scales such fluctuations will
have declined to the 0.7 mJy level and would require some 100's of
components to model.

\begin{figure*}
\resizebox{\hsize}{!}{\includegraphics{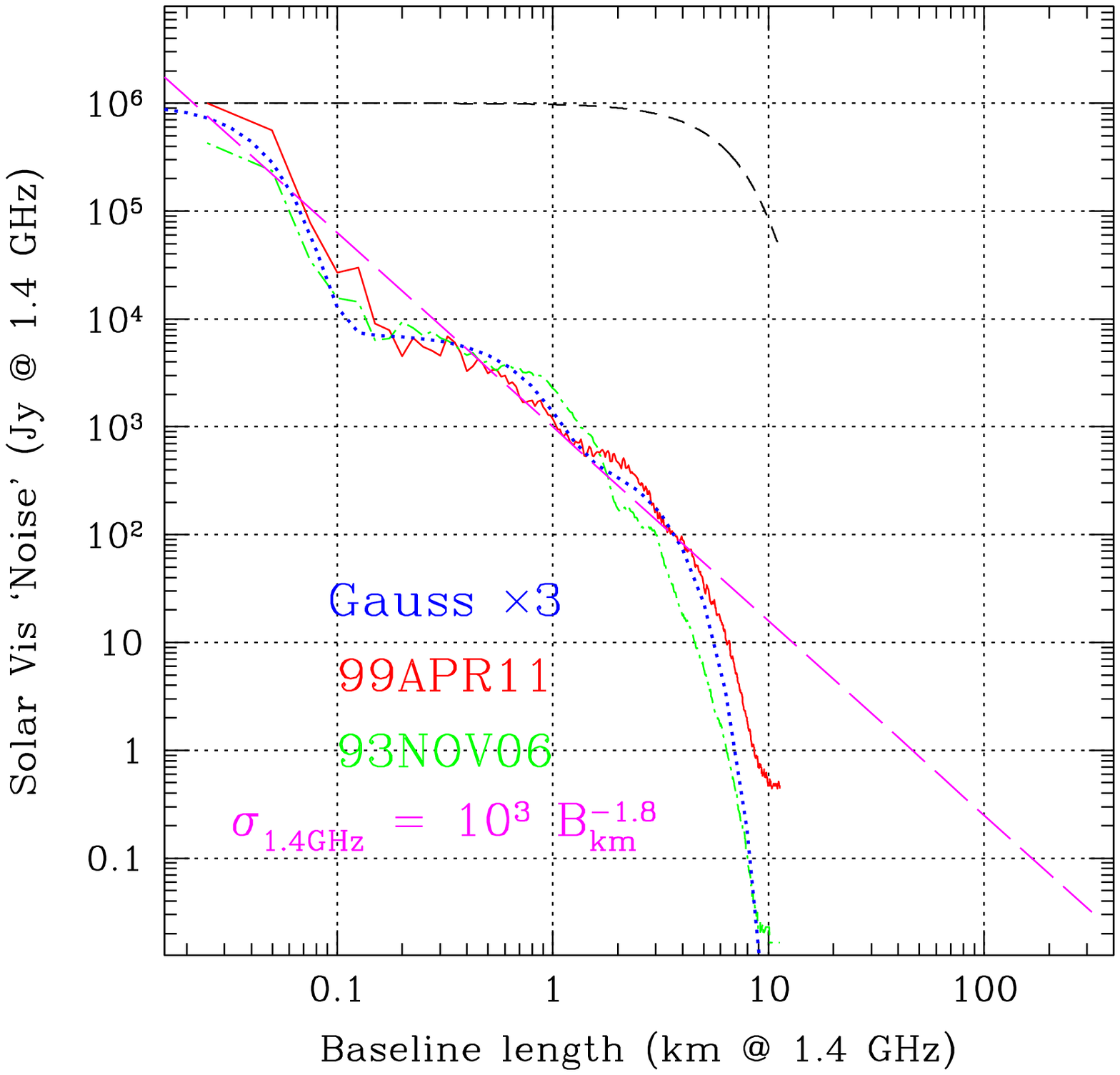},\includegraphics{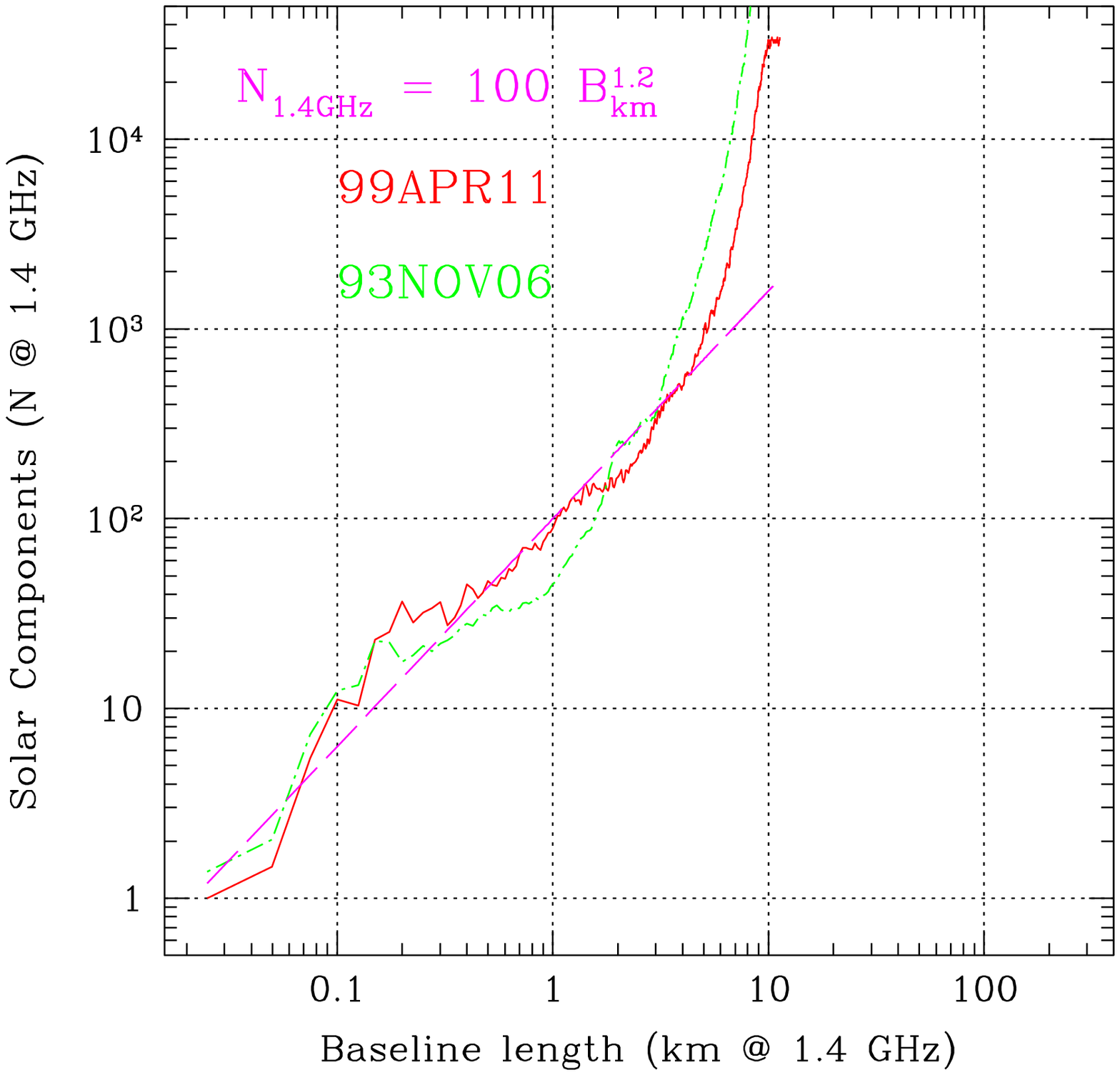}}
\caption{Visibility ``noise'' due to the active and quiet Sun
  (left) and the corresponding number of components (right) as
  function of baseline length (angular scale).}
\label{fig:sigd}
\end{figure*}

The Sun is the other major contributor to residual visibility
fluctuations when not adequately modelled. We plot the relative
visibility amplitude of the Sun in Figure~\ref{fig:powspec}, based on
6cm VLA mosaic images kindly provided by
\citetads{2002ASPC..277..299W} and 2012, private communication). The
two observing dates correspond to relatively active and quiet portions
of the solar cycle. The solar power spectrum can be well modelled by
the sum of three Gaussian components, as shown by the dotted curve,
that have relative angular scales of (1: 0.0625: 0.0156) and
amplitudes of (0.9645: 0.0302: 0.0053). The corresponding visibility
``noise'' and component number are plotted in
Figure~\ref{fig:sigd} as function of baseline length. The
visibility modulation level was taken to be the power spectrum of
Figure~\ref{fig:powspec} divided by the square root of component
number to account for the fact that the individual solar flux
components will not sum coherently. The 1.4 GHz integral flux on these
dates has been scaled to an assumed total of 100 sfu on 99APR11 and 50
sfu on 93NOV06 in line with the daily solar monitoring data from those
epochs. With 50dB of attenuation, there is still a residual
fluctuation level of about 10 mJy on 1 km scales that would require
about 100 components for adequate modelling.

\subsubsection{Radio frequency interference} 

Another contribution to the far sidelobe response from the telescope
environment is that of radio frequency interference (RFI) from 
ground-based, aerial and space-based sources. Clearly, the greater the
degree of attenuation outside of the main beam, the more
straightforward it becomes to deal with the power levels of
non-astronomical transmissions. In view of the great diversity of RFI
attributes, we will not attempt to characterise them here. We simply
note that adequate RFI recognition, excision or mitigation will be
essential to achieving thermal noise limited performance. There are
many potential telescope locations and frequency bands in which RFI
will preclude high performance imaging.

\subsubsection{The ionosphere and atmosphere} 

A significant complicating factor in wide-field imaging is the
direction dependent nature of propagation effects relative to a
ground-based telescope array. The diffractive and refractive effects
of the ionosphere and troposphere (\citetads{2009A&A...501.1185I},
\citetads{2011A&A...527A.107S}) make it necessary to determine
distinct calibration solutions for sources separated by more than
about 1~degree and timescales of about 1 minute. The situation is
further complicated for an array that is spatially extended relative
to the physical size of density fluctuations, since the array is
  in the near-field of these fluctuations so that different
telescopes have distinct propagation paths through the medium and
a distant thin screen approximation is no longer valid. The relevant
size scale for traveling ionospheric disturbances (TIDs) is a few
km. While we will not attempt to quantify the impact of residual
solution errors, we note the necessity for distinct, time and position
dependent calibration solutions due to these effects.

\subsubsection{The role of a global sky model} 

One might consider that a major aid to the self-calibration problem
would be the provision of a high quality model of the entire radio sky
at each frequency of an observation. To the extent that intrinsic
source variability or propagation effects (in the interstellar,
interplanetary, ionospheric or tropospheric media) were negligible
this might eliminate many degrees of freedom in the sky model. While
undoubtedly of great benefit, a global sky model will only partially
reduce the number of degrees of freedom in the model (source
positions), since a remaining unknown is generally the instrumental
response in the direction of each source at each solution time. The
accuracy with which far sidelobe responses of an antenna system can be
predicted at any specific location is, conservatively, no better
than a factor of two, implying that this prediction has little value
in constraining a solution for the apparent brightness of each source
within an observation. Although very significant effort is being
expended to improve the prediction of far sidelobe responses in the
aperture array domain \citepads{5355494} it is not yet clear how
effective these will be in practise. Far sidelobe responses are
influenced by the surface irregularities of dishes, or equivalently,
the gain and radiation pattern variations of individual aperture array
elements. As such they will be different for each dish or station of
an array and are likely to be time variable due to many environmental
effects, including gravitational distortions and temperature
gradients.

In the event that the global sky model also incorporates an extremely
high fidelity, high resolution representation of every relatively
bright source in the sky, it would provide an invaluable resource for
accurate self-calibration. As will become apparent subsequently, the
accurate modelling of the random sources that happen to occur within
the field of view is often the primary limitation to the dynamic range
that can be achieved. While this has been put forward as the preferred
calibration method for survey projects such as planned with the
Low-Frequency Array for Radio astronomy (LOFAR)
\citepads[eg.][]{2010iska.meetE..57H} in practise it is likely to be
the ultimate outcome of a multi-year data acquisition and source
modelling effort, rather than a convenient starting point for a random
field of interest.

\subsubsection{Time and bandwidth smearing/correlation}
\label{sec:tbsmear}
The visibility response to distant off-axis sources will be modified
by several effects, most notably time- and bandwidth- smearing for the
finite integration times and frequency resolution used in an
observation. This can be viewed both positively and negatively. The
peak visibility response will be diminished by these smearing effects
and this will be beneficial in reducing noise-like
modulations. However, if the source is still bright enough that it
must be modelled, then the smearing will adversely impact the quality
of that modelling and contribute to residual gain errors in the
calibration. The angular extent of time smearing of the response due to
Earth rotation is given approximately by, 
\begin{equation}
\delta_\tau = \omega \tau_S \theta, 
\label{eqn:tsm}
\end{equation}
for an angular rotation rate $\omega$, (15~$\deg$/hr) an
integration time, $\tau_S$, and an offset angle from the phase centre,
$\theta$. For a source on the celestial pole viewed by an east-west
baseline, this smearing is along a circular track, while for other
circumstances the smearing direction is more complex. The bandwidth
smearing has an extent, 
\begin{equation}
\delta_\nu = \theta \Delta \nu_S / \nu, 
\label{eqn:bsm}
\end{equation}
for a fractional bandwidth $\Delta \nu_S / \nu$ and is in a radial
direction. For example, an unresolved source that is offset from the
pointing direction by 1 radian that is sampled with an integration
time of 1 second with a fractional bandwidth of $10^{-4}$ will suffer
about 15 arcsec of both time smearing and bandwidth smearing. The
specific choice of integration times and averaging bandwidths is
determined by the need to keep these effects at the level of a small
fraction, $\eta_S$, of the synthesised beamwidth within the antenna main
beam. In terms of a dish/station diameter, $d$, and array of baseline,
$B_{Max}$, this implies an integration time no longer than,
\begin{equation}
\tau^\star \leq \eta_S d / (\omega B_{Max}) = 1.38\times10^4 \eta_S d / B_{Max}
\quad \rm{(sec)},  
\label{eqn:timesm}
\end{equation}
and a fractional bandwidth no larger than,
\begin{equation}
\Delta \nu^\star / \nu \leq \eta_S d / B_{Max}. 
\label{eqn:bandsm}
\end{equation}
For example, for $\eta_S = 0.1, d = 25$ m and $B_{Max} = 25$ km, the
requirements are $\Delta \nu^\star / \nu \leq 10^{-4}$ and $\tau^\star \leq 1$
sec. Exactly the same considerations can be used to evaluate the
likely time and frequency intervals over which particular types of
calibration errors might be correlated. If the calibration error
relates to a source that is near the half power point of the main
beam, this corresponds to $\delta_\tau = \delta_\nu = \lambda/B$ and $\theta
= \lambda /(2d)$ in eqns.~\ref{eqn:tsm} and ~\ref{eqn:bsm}, which will
be correlated over time intervals of,
\begin{equation}
\tau \approx 2 d / (\omega B) = 1.38\times10^4 2 d / B
\quad \rm{(sec)},  
\label{eqn:timecn}
\end{equation}
and bandwidths,
\begin{equation}
\Delta \nu / \nu \approx  2 d / B. 
\label{eqn:bandcn}
\end{equation}
For example, for $d = 25$ m and $B_{Max}$ = 25 km, the calibration
errors associated with sources near the half power point of the main
will be correlated over $\Delta \nu / \nu \approx 2\times10^{-3}$ and
$\tau \approx 28$ sec. In the far sidelobe regime, with $\theta = 1$
radian, the correlation time reduces to,
\begin{equation}
\tau \approx \lambda / (\omega B) = 1.38\times10^4 \lambda /
B \quad \rm{(sec)},  
\label{eqn:timecf}
\end{equation}
and bandwidth to,
\begin{equation}
\Delta \nu / \nu \approx  \lambda / B.
\label{eqn:bandcf}
\end{equation}

\subsubsection{Visibility consequence} 

The consequence of unmodelled off-axis sources in the visibility data
is a modulation of the visibilities with a magnitude equal to that of
the unmodelled source. Unresolved and unsmeared sources will have the
same modulation amplitude on all observed baselines, while resolved
and/or smeared sources will have a modulation amplitude that decreases
with increasing baseline length. Since these are periodic modulations,
with a modulation wavelength, $m = \lambda / \theta$, that is very
short for large pointing offsets, they will partially cancel when
averaged over finite time intervals or within gridded visibility
cells. However, the presence of such residual modulations will
negatively influence the quality of the self-calibration that can be
achieved, particularly on relatively short timescales over which the
cancellation will be increasingly imperfect. In the case where
successive integration times sample visibilities that are offset from
one another by a significant fraction of $\lambda/\theta$,
i.e. baselines of length greater than, $B > \lambda B_{Max} / (\eta_S
\theta d)$, or about a km for GHz observations with $\theta$ = 1
radian, such unmodelled sources will add a noise-like contribution in
quadrature to the visibility noise $\sigma(\tau)$ and directly impact
the calibration precision $\phi$, and ultimately the final image noise
level. 

\begin{figure*}
\resizebox{\hsize}{!}{\includegraphics{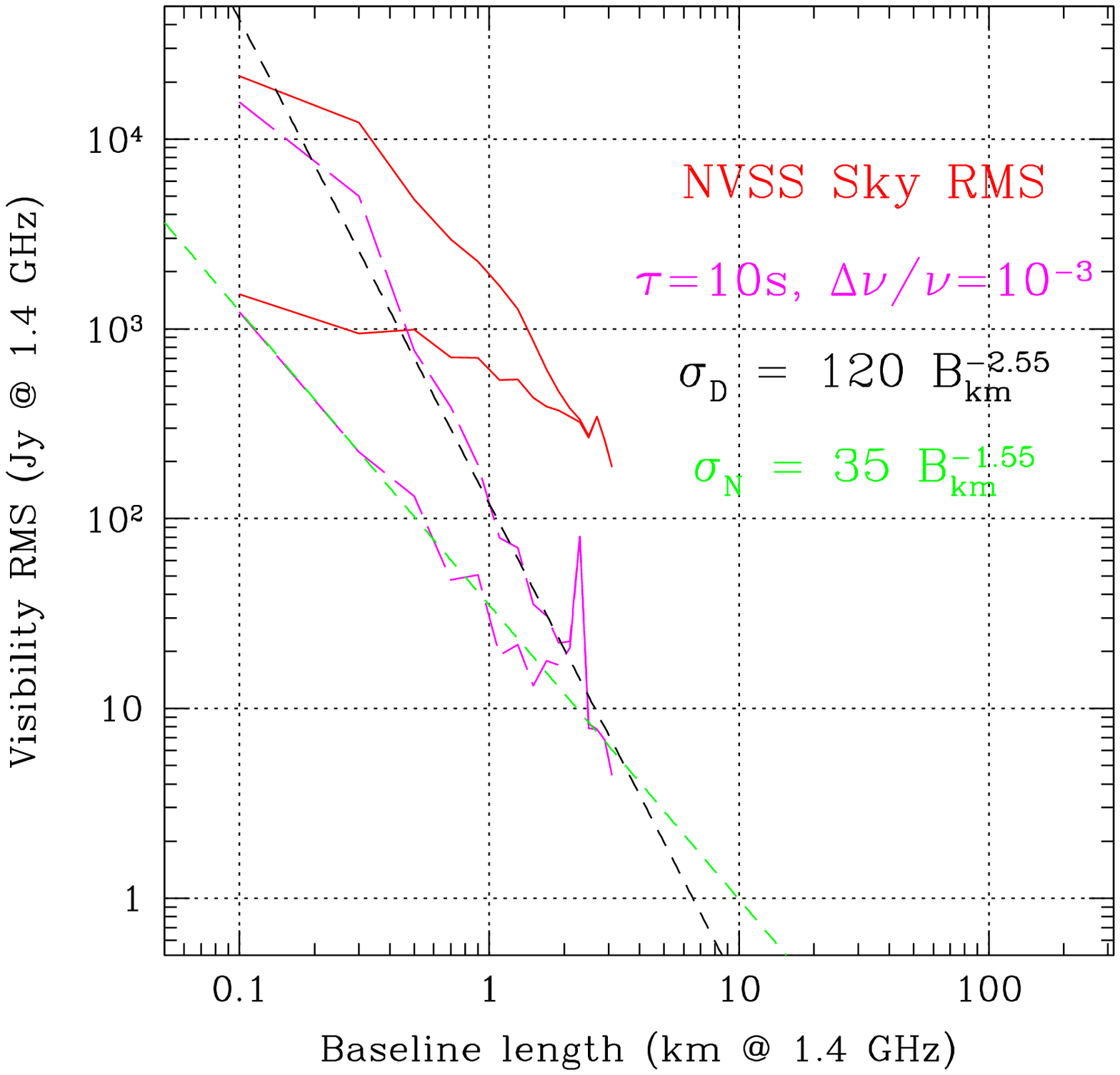},\includegraphics{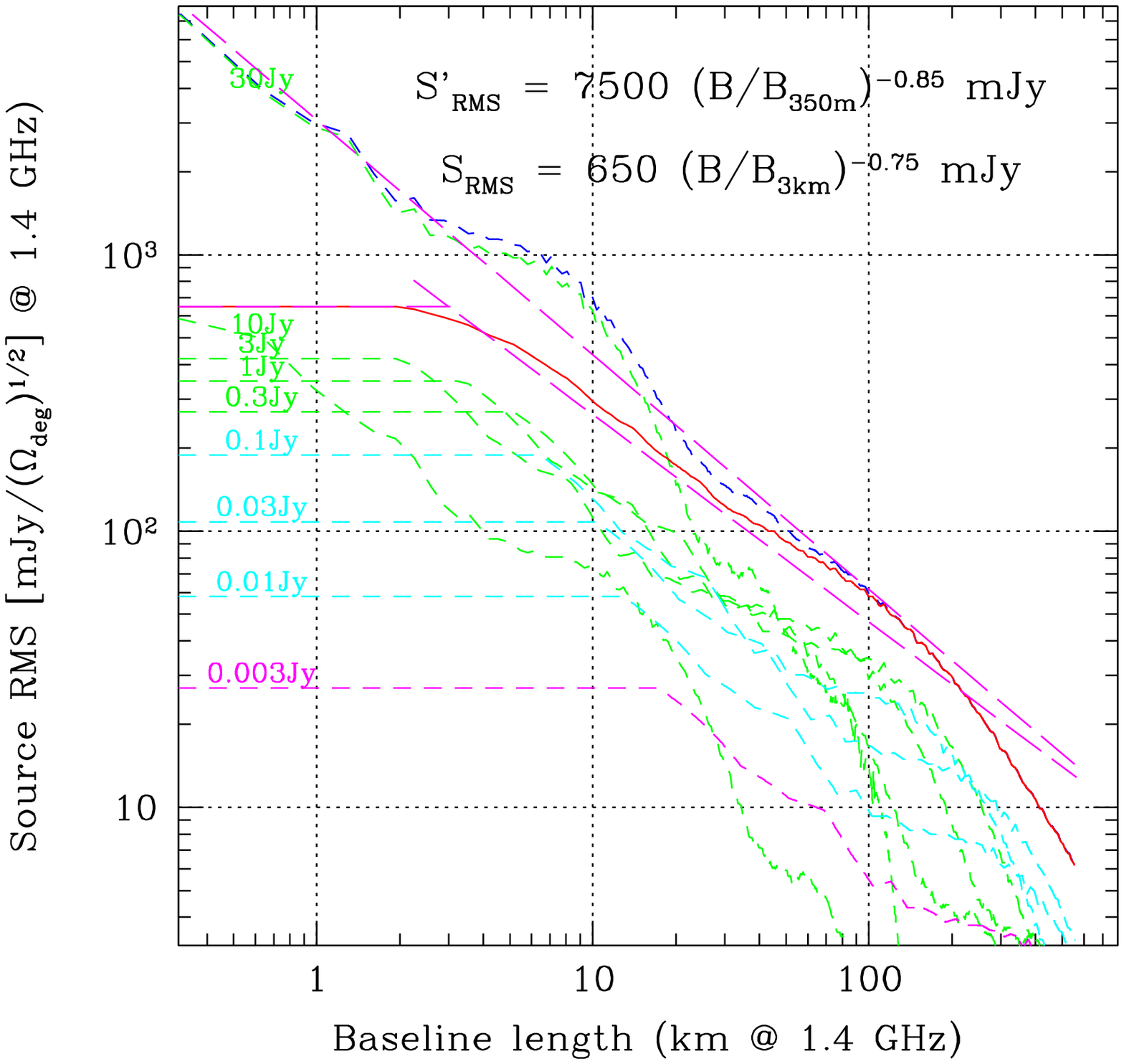}}
\caption{Measured visibility ``noise'' due to the NVSS $>$ 10 Jy sky
  as seen by the VLA C configuration without primary beam attenuation
  (left). The two solid curves show the fluctuation level with and
  without the inclusion of the Sun. The two dashed curves include the
  effects of time and frequency smearing and are overlaid with
  representative power-law approximations. Visibility ``noise'' per
  solid angle due to the NVSS sky (right). The RMS contribution
  of NVSS sources, normalised to a one square degree field of view, is
  plotted as function of baseline length. Flux bins from
  Table~\ref{tab:nvss} are shown individually with the labelled
  curves. The root squared sum of contributions of the bins $<$ 10 Jy
  is shown as the solid line, together with an overlaid powerlaw fit.}
\label{fig:simsig}
\end{figure*}

We have attempted to verify the analytic estimates of far sidelobe
effects by simulating model visibilities as observed by the VLA C
configuration at 1.4 GHz. All radio sources within the NVSS database
brighter than 10 Jy and above the local VLA horizon at LST 12:00:00
were added to an otherwise empty visibility file of 12 hour duration
using the $miriad$ \citepads{1995ASPC...77..433S} task $uvgen$. Each
catalogued source was assumed to be composed of the sum of four
Gaussian contributions with angular sizes of (1: 0.29: 0.039: 0.0067)
relative to the tabulated Full Width at Half Maximum (FWHM) and
amplitudes of (0.648: 0.237: 0.107: 0.008) relative to the integrated
flux density. The response was calculated relative to a pointing
centre at (RA, Dec)=$(12^h, 34\deg)$ both with and without a three
Gaussian component model of the Sun (assumed to be at (RA, Dec) =
$(9^h, -11\deg)$ when present) as illustrated in
Figure~\ref{fig:sigd}. The intrinsic fluctuation levels are indicated
by the solid curves in Figure~\ref{fig:simsig}. The residual
fluctuations including the smearing effects of a 10 sec integration
time and $\Delta \nu / \nu = 10^{-3}$ frequency sampling are indicated
by the dashed curves. (We note that since the $uvgen$ task does not
provide explicit support for time and bandwidth smearing, these
effects were introduced by generating a highly oversampled simulation
that was vector averaged appropriately after calculation.) The
intrinsic fluctation spectra are flatter by unit slope compared to
those which include both smearing effects, as expected. No primary
beam attenuation has been applied. Representative power law
approximations of the post-smearing fluctuation levels are,
\begin{equation}
\label{eqn:sign}
 \sigma_N = \epsilon_F 35 \bigg({\nu \over {\rm 1.4\ GHz}}\bigg)^{-0.8}
  \bigg({B \over B_k}\bigg)^{-1.55} \bigg({\tau \over {\rm 10\ sec}}
  \bigg)^{-0.5} \bigg({\Delta\nu/\nu \over 10^{-3}}\bigg)^{-0.5} {\rm Jy} 
\end{equation}
for nighttime observing and, 
\begin{equation}
\label{eqn:sigd}
\sigma_D= \epsilon_F 120 
  \bigg({B \over B_k}\bigg)^{-2.55} \bigg({\tau \over {\rm 10\ sec}}
  \bigg)^{-0.5} \bigg({\Delta\nu/\nu \over 10^{-3}}\bigg)^{-0.5} {\rm Jy} 
\end{equation}
with the Sun present. The $\epsilon_F$ term represents the far side-lobe
response from equation \ref{eqn:epf}. The extragalactic contribution
scales roughly as $\nu^{-0.8}$, given the steep spectra of the
brightest sources, while the solar spectrum at GHz frequencies is
relatively flat. The baseline length is normalised with $B_k =
(\nu/{\rm 1.4\ GHz})^{-1}$ km. These measured fluctuation spectra are
quite similar to those obtained by the analytic estimate shown
previously in Figure~\ref{fig:sign} and \ref{fig:sigd} although
with a slightly flatter slope when the individual source attributes
and positions are employed rather than the tabulated binned source
properties. Clearly, the daytime power-law approximation is only valid
for baselines less than about $B/B_k = 3$, where the solar
contribution dominates over that of the extragalactic sky.

An important aspect of the far sidelobe contribution to visibility
``noise'' is that it is essentially uncorrelated between adjacent
frequency channels and time stamps. Eqns.~\ref{eqn:bandcf} and
\ref{eqn:timecf} yield $\Delta \nu / \nu \approx 10^{-4}$ and $\tau \approx
1.5$ sec at a typical off-axis distance of 1 radian. While this noise
contribution may adversely influence individual solutions, at least it
will average down substantially during long tracks and over significant
bandwidths. The relevant values of $M_{Ti}$ and $M_{Fi}$ are given
simply by the total number of solution intervals in time and
frequency, as,
\begin{equation}
\label{eqn:mtf}
 M_{TF} = \tau_T/\tau_S 
\end{equation}
and
\begin{equation}
\label{eqn:mff}
 M_{FF} = \Delta\nu_T/\Delta\nu_S 
\end{equation}
for a total observing time $\tau_T$ and bandwidth $\Delta\nu_T$.

\subsubsection{Image consequence} 

The image plane consequence of unmodelled sources is influenced by some
additional factors. Sources that lie beyond the edges of a synthesis
image will be aliased back into that image. The degree of suppression
of such aliased responses is determined by the convolving function
that is used to grid the visibility data. This constitutes a degree of
averaging of the highly modulated visibilities (as discussed above) to
provide partial cancellation of off-axis responses. Gridding functions
that are currently used in synthesis imaging provide about 40 dB
suppression of the aliased response in the far sidelobe regime
\citepads[e.g.][]{1999ASPC..180..127B}. What the gridding function
will not suppress is any synthesised beam sidelobes of distant sources
that would naturally occur within the image. Since the apparent
brightness of a far sidelobe source will often be highly time
variable, it is the instantaneous synthesised beam quality which
becomes important in this regard. In the most extreme case of a one
dimensional array configuration, the instantaneous synthesised beam
far sidelobes can have amplitudes of 100\% so that unmodelled sources
can add spurious structure to the image with very little suppression
beyond that provided by the main beam. Even in the case of a regular
three arm configuration such as used in the VLA, the instantaneous
synthesised beam has peak sidelobe levels of 10's of percent. What
prevents this factor from becoming dominant in most broad-band
continuum applications is the radial frequency scaling of the
synthesised beam pattern, including its sidelobes. The sidelobe
pattern will partially cancel in the frequency averaged response. 

In addition to the ``direct'' image consequence of unmodelled
  far side-lobe sources discussed above, there is the ``indirect'' consequence of
  such sources. Within each self-cal solution interval, the noise-like
  visibility modulation of unmodelled sources will limit the precision
  of the self-cal solution as shown in eqn.~\ref{eqn:phi}. The net
  image plane noise due to degradation of the self-cal solution can be
  estimated from the visibility fluctuation level $\sigma(\tau_S)$, by
  combining eqns.~\ref{eqn:dr} and \ref{eqn:phi}
  giving, 
\begin{equation}
\sigma^\prime_{Map} = {\sigma(\tau_S) S_{Max} N_C^{0.5} \over S_{Tot} N
  (N-3)^{0.5} M_T^{0.5} M_F^{0.5} }. 
\label{eqn:farsens}
\end{equation}
The ratio of ``indirect'' to ``direct'' contributions
(eqns. \ref{eqn:farsens} and \ref{eqn:sens}) is then,
\begin{equation}
{\sigma^\prime_{Map} \over \sigma_{Map}} = { S_{Max}  \over S_{Tot}} \bigg({
  N_C \over 2 (N-3) }\bigg)^{0.5},
\label{eqn:far2near}
\end{equation}
which is approximately unity for arrays with $N\approx30$.

\subsection{Near-in sidelobes}\label{sec:near}

In addition to the issue of far-sidelobes, is that of the near-in
sidelobe pattern of the antenna response. Any aperture blockage and
non-zero illumination taper at the edges of the antenna will give rise
to near-in sidelobes in the antenna response. For a uniform aperture
illumination, such as might be employed for an aperture array station,
these will have a peak value, $\epsilon_S = 0.1$, while the more
tapered, but ``blocked'' illumination patterns of prime focus and
symmetric Cassegrain-fed dishes yield peak near-in sidelobes of about
$\epsilon_S = 0.02$. The unblocked 100 m Green Bank Telescope (GBT)
aperture has near-in peak sidelobe levels of less than 1\% at 1.4 GHz
\citepads{2009PASP..121..272R}, although these have significant
azimuthal modulation.

Near-in sidelobes complicate the calibration and imaging problem in a
number of ways. Significant sidelobe responses will increase the
region of sky that must be modelled with high precision to enable
self-calibration. The increase in the number of degrees of freedom
that are absorbed by sky components may then limit the precision of
the self-cal solution as outlined above. Since these components are
significantly further off-axis than main-beam components, they will
also require proportionally shorter solution timescales to be
employed, which impacts the solution precision via the visibility
signal-to-noise constraint. This is particularly problematic if the
sidelobe pattern is not fixed on the sky during source tracking, since
one then encounters the under-determined self-cal solution condition
more rapidly. Some deconvolution aspects of this problem have been
addressed by \citetads{2008A&A...487..419B} with extensive modelling
of both the VLA main beam and its near-in sidelobe and polarisation
response, including its changing orientation relative to the
sky. A self-cal implementation has been described in
\citet{2004.Bhat}. When it proves too challenging to model sources
occuring within the near-in sidelobes they will yield a residual
visibility modulation of,
\begin{equation}
\label{eqn:sigs}
 \sigma_S = \epsilon_S S_{RMS}(\Omega_S) 
\end{equation}
in terms of the net RMS source brightness, $S_{RMS}$, within the area
of the sidelobe pattern, $\Omega_S$. The net visibility fluctation
level due to the NVSS source population per unit area is shown in the
right hand panel of Fig.~\ref{fig:simsig}. This plot is similar to
Fig.~\ref{fig:sign} but is calculated for a reference solid angle of 1
deg$^2$ rather than $2\pi$ sr, and the summation excludes the two
brightest flux bins so that it better represents a random, but
``quiet'', piece of the sky that does not include the 100 or so
brightest sources per hemisphere. A power-law representation of the
net visibility fluctuation level is given by,
\begin{eqnarray}
\label{eqn:srms}
 S_{RMS} & = & 650\ \Omega_{deg}^{0.5} \bigg({\nu \over {\rm
     1.4\ GHz}}\bigg)^{-0.8} \qquad (B < B_S) \\
\nonumber & = & 650\ \Omega_{deg}^{0.5} \bigg({\nu \over {\rm
     1.4\ GHz}}\bigg)^{-0.8} 
  \bigg({B \over B_S}\bigg)^{-0.75} \hfil (B > B_S)  \quad {\rm
    mJy},
\end{eqnarray}
where the RMS brightness scales roughly as $\nu^{-0.8}$, and the
baseline length is normalised to the break at about $B_S = 3 (\nu/{\rm
  1.4\ GHz})^{-1}$ km, below which the RMS flux saturates. Of note in
equation~\ref{eqn:srms} is the square root dependence on solid angle,
which accounts for the incoherent vector summation of sources in the
field of view. In cases where the main beam field of view exceeds
about 200 deg$^2$, the ``quiet'' sky approximation is clearly no
longer relevant, and the complete source population must be considered
yielding, 
\begin{eqnarray}
\label{eqn:srmsp}
 S_{RMS}^\prime & = & 7500\ \Omega_{deg}^{0.5} \bigg({\nu \over {\rm
     1.4\ GHz}}\bigg)^{-0.8} \qquad (B < B_S^\prime) \\
\nonumber & = & 7500\ \Omega_{deg}^{0.5} \bigg({\nu \over {\rm
     1.4\ GHz}}\bigg)^{-0.8} 
  \bigg({B \over B_S^\prime}\bigg)^{-0.85} \hfil (B > B_S^\prime)  \quad {\rm
    mJy},
\end{eqnarray}
with $B_S^\prime = 0.35 (\nu/{\rm 1.4\ GHz})^{-1}$ km. As will be seen below,
the values of $S_{RMS}$ and $S_{RMS}^\prime$ are intermediate between
the peak brightness, $S_{Max}$, and the integrated brightness,
$S_{Tot}$, within a field of size $\Omega_{deg}$.

The telescope main beam has its first diffraction null at a radius of
$1.22 \lambda / d$ radians which is also approximately the main beam
FWHM. This corresponds to a main beam solid angle,
\begin{equation}
\label{eqn:om}
\Omega_M = \pi (1.22 \lambda /
d)^2/4. 
\end{equation}
The first diffraction sidelobe extends from radius $(1.22
\lambda / d - 2.44 \lambda / d)$ and so has solid angle of, 
\begin{equation}
\label{eqn:oms}
\Omega_S
= 3 \Omega_M = 3 \pi (1.22 \lambda /
d)^2/4.
\end{equation}

The correlation timescale and bandwidth for these modulations due to
sources offset from the pointing direction by $\theta = \lambda/d$
will be $\tau \approx d/(\omega B)$ and $\Delta\nu/\nu \approx d/B$,
yielding,
\begin{equation}
\label{eqn:mts}
 M_{TS} = {\tau_T B \over 1.38\times 10^4 d} 
\end{equation}
and,
\begin{equation}
\label{eqn:mfs}
 M_{FS} = {\Delta\nu_T B \over \nu d }.
\end{equation}
Near-in sidelobe patterns of dishes have been shown to have
substantial frequency modulation at the 10\% level
\citepads{2008A&A...479..903P} with a periodicity of about $f_S
\approx c/2l_C$ Hz, for a vertex to prime focus ``cavity'' separation,
$l_C$. For Cassegrain optics the relevant separation, $l_C$, appears
to be approximately that of the primary and subreflector, while for
offset Gregorian designs it is that of the feed and subreflector.
While such sidelobe modulation has only been demonstrated to date with
frequency resolved holography on the WSRT dishes, this effect is
intimately related to the ``standing wave'' phenomenon that afflicts
all current dish designs. The correlation bandwidth will be about half
a period, $\Delta\nu \approx c/(4l_C)$ which yields,
\begin{equation}
\label{eqn:mfsp}
 M_{FS}^\prime = 4 l_C \Delta\nu_T/c . 
\end{equation}
The larger of the two values given by eqns.~\ref{eqn:mfs} and
\ref{eqn:mfsp} provides the better estimate of the independent sample
number in the full observation.

\subsection{Main-beam issues}\label{sec:mainbm}

There are several ways in which attributes of the main beam of the
antenna will influence image dynamic range. In the case of aperture
arrays, it is conceivable that the antenna main beam may be
continuously changing as a field is tracked on the sky due to the
changing geometric foreshortening of the aperture. In this case, the
main beam sky brightness model will be highly time variable and one
will need to operate essentially in the snap-shot self-cal mode,
without the benefit of the factor of $M_T$ additional data constraints
that come with a time-invariant main beam model. In the event that
different aperture array main beams were to apply to different
antennas (as implemented for the three variants of the LOFAR high band
antennas) this is further complicated, since there are only smaller
subsets of the visibilities (e.g. $N^2/8$) for which the same
snap-shot sky brightness model would apply, greatly diminishing the
number of relevant data constraints relative to sky brightness
components. A means of compensating for different main beam types as
well as main beam variability has been termed the {\it A-Projection}
method and is decribed by \citetads{2008A&A...487..419B}. The method
employs a gridding convolution function for the visibilities that
compensates for modelled main beam properties. This provides an exact
method for the ``reverse'' calculation of visibilities given a sky
model, but only an approximate ``forward'' calculation of the dirty
image from the acquired data. In cases where there are substantial
differences between- or time variability in- the main beam properties
of the visibilities, there will be a position dependent synthezied
beam and signal-to-noise ratio which may limit the quality of
deconvolution which can be achieved.

\subsubsection{Pointing errors}\label{sec:point}

A second important class of main beam influences relates to the
antenna pointing accuracy (either mechanical for dishes or electronic
for aperture arrays). Pointing variations will introduce a strongly
time variable component to the radio sky brightness model for
components that are significantly off-axis. If the sky brightness
model is dominated by such widely distributed components, then
pointing variations may again push one into the snap-shot self-cal
mode. For the simplest form of pointing error, that preserves
main-beam shape, the model variations could in principle be captured
with only two additional degrees of freedom per antenna (the two
perpendicular pointing offsets). \citetads{2008A&A...487..419B} have
also considered this class of error in their study. Pointing errors
that are a fraction, $\beta_P$, of the main beam of size, $P = \beta_P
\lambda / d$ radians, are most serious where they introduce the
strongest absolute modulation in apparent source brightness. For a
Gaussian approximation to the main beam, $M = exp(-0.5(\theta/s)^2)$,
the maximum spatial derivative occurs at a bore-sight offset equal to
the dispersion, $s = \lambda / (2d)$, and has amplitude $exp(-0.5)/s$
rad$^{-1}$, giving an amplitude modulation, $\epsilon_P = \beta_P \lambda
\ exp(-0.5)/(sd) = 2 \beta_P exp(-0.5) \approx \beta_P$. The maximum
amplitude modulation expected on the flanks of the main beam is thus
comparable to the pointing error expressed as a fraction of the main
beam size, $\epsilon_P \approx \beta_P$. Actual main beam shapes can only be
well-fit by a Gaussian out to about 10\% of the peak response. Beyond
this radius lies a deep response minimum followed by the near-in
sidelobes. The main beam power pattern of an interferometer arises
from the product of the complex voltage response patterns of each pair
of antennas. The deep response minimum in the power pattern
corresponds to the region where the sign of the voltage pattern of
each antenna changes from positive to negative. A uniformly
illuminated circular aperture has a voltage response pattern given by
$V(x) = 2 J_1(x)/x$, where $J_1(x)$ is the Bessel function of the
first kind of order 1. Evaluation of the radius at which the product
of sky area $(\pi x^2)$ with $V(x)$ with the derivative of $V(x)$ is
maximised yields a similar result to that found above for the Gaussian
approximation. The largest and most probable absolute modulation still
occurs on the main beam flanks (at about 20\% of peak power in this
case) and has amplitude comparable to the pointing error expressed as
a fraction of the main beam size, $\phi_P \approx \beta_P$. The
resulting visibility modulation due to mechanical pointing errors is given by,
\begin{equation}
\label{eqn:sigp}
 \sigma_P = 0.7\ P\ d\ S_{RMS}(\Omega_M)/\lambda
\end{equation}
after scaling with the RMS source brightness that occurs within
the main beam, $S_{RMS}(\Omega_M)$, and a typical attenuation on the
main beam flank of 0.7. The same analysis will also apply to the case
of variations in the main beam shapes for different stations in an
array. The failure to account for such variations that are a fraction,
$\beta_P$, of the main beam size will give rise to the same magnitude of
residual visibility based errors.

An important attribute of pointing errors is that they are likely to
be intrinsically broad-band in nature and so will not diminish
significantly with frequency averaging. The correlation timescale of
pointing errors will depend on their nature. Errors due to wind
buffeting will have characteristic timescales, $\tau_P$, of seconds while errors
in the global pointing model or due to differential solar heating are
likely to be correlated over many minutes to hours. Estimates for the
independent time and frequency intervals are therefore simply,
\begin{equation}
\label{eqn:mtp}
 M_{TP} = {\tau_T / \tau_P} 
\end{equation}
and,
\begin{equation}
\label{eqn:mfp}
 M_{FP} = 1.
\end{equation}
In the event that there are also electronic pointing errors of
magnitude, $\epsilon_P^\prime$, and correlation timescale
$\tau_P^\prime$, these will yield an additional error contribution with,  
\begin{equation}
\label{eqn:sigpp}
 \sigma_P^\prime = 0.7 \epsilon_P^\prime S_{RMS}(\Omega_M),
\end{equation}
\begin{equation}
\label{eqn:mtpp}
 M_{TP}^\prime = {\tau_T / \tau_P^\prime}, 
\end{equation}
and,
\begin{equation}
\label{eqn:mfpp}
 M_{FP}^\prime = 1.
\end{equation}

\subsubsection{Polarisation beam squint and squash}\label{sec:pbss}

A third main-beam issue is that of the wide-field polarisation
response and its possible variation with time. Polarisation
calibration, in its traditional form, calibrates only the relative
gain and apparent leakage of polarisation products in the on-axis
direction. Any differences in the spatial response patterns of the two
perpendicular polarisation products that are measured by the antenna
will translate into so-called ``instrumental polarisation'' in
off-axis directions. One of the more extreme examples of this
phenomenon is termed ``beam squint'' \citepads{2008A&A...486..647U,
  2008A&A...487..419B} which occurs in cases where the feed does not
illuminate a parabola symmetrically. This condition even applies to
the Cassegrain optics of the JVLA, since the feeds are not located on
the symmetry axis of the main reflector, resulting in RR and LL beams
that are similar in size and shape, but are significantly offset from
one another, by 5.5\% of the main beam FWHM, on the sky. Since this
offset is introduced by the antenna optics, it rotates on the sky
during source tracking of the $(alt,az)$ mount. In this case, the sky
brightness model used for self-calibration will need to be similarly
time variable. Since this variability is systematic in nature it could
in principle be modelled analytically, and this has been undertaken
with some success by \citetads{2008A&A...486..647U}. It has also
proven possible to simultaneously correct for antenna pointing errors
together with the ``squinted'' polarisation response during
deconvolution \citepads{2008A&A...487..419B}.  The beam squint can be
minimised in secondary focus systems by satisfying the condition
\citepads{1147539},
\begin{equation}
\label{eqn:alphaq}
{\rm tan} \alpha = { |1-e^2| {\rm sin} \beta \over  (1+e^2) \rm{cos} \beta - 2 e},
\end{equation}
where $\alpha$ is the tilt angle between the subreflector and the
feed, $\beta$, the tilt angle between the subreflector and main
reflector, and $e$, the eccentricity of the subreflector. This
condition forms the basis for current offset Gregorian designs,
including that of the GBT, and delivers very good performance, with
cross-polarisation amplitudes of better than -40 dB when the
subreflector size exceeds about 25$\lambda$. For smaller subreflector
dimensions, the polarisation performance is diminished by diffractive
effects \citepads{1147944}, increasing to about -32 dB for a
10$\lambda$ subreflector. The corresponding measured values of beam
squint for the GBT \citepads{2009PASP..121..272R} of about 1.5 arcsec
at 1420 MHz, represents a 0.3\% FWHM offset at a frequency where the
7.5~m secondary is about 38$\lambda$.

Even for fully symmetric optics designs, there will inevitably be
significant differences in the polarisation beams from single pixel
fed dishes \citepads[e.g.][]{2008A&A...479..903P}, with the XX and YY
beams having an elliptical shape with a perpendicular relative
orientation that has been termed ``beam squash''. Ellipticities of
between 3 and 4\% are measured for the XX and YY WSRT beams at 1.5
GHz. Beam squash of the linear polarisation products for the off-axis
GBT design has been measured to be about 20~arcsec between 1200 and
1800~MHz \citepads{2003.Heiles} or about 4\% FWHM. In the case of an
equatorial (WSRT) or parallactic axis (ASKAP) mount the instrumental
polarisation pattern will at least be fixed relative to the sky, so
that additional degrees of freedom (or modelling complexity) are not
introduced by time variability. Aperture arrays and phased array feeds
offer interesting prospects for greatly suppressing instrumental
polarisation via the choice of beam-forming weights that accurately
match the spatial response patterns and relative alignment of the two
polarisation beams. When unmodelled beam squint or squash of magnitude,
$\epsilon_Q$, is affecting an observation it will contribute a
visibility modulation of about,
\begin{equation}
\label{eqn:sigq}
 \sigma_Q = 0.7 \epsilon_Q S_{RMS}(\Omega_M).
\end{equation}
The rate of change of parallactic angle with hour angle for an
$(alt,az)$ mount depends on the source declination, but in general the
rate is about twice the sideral one, so for a source on the main beam
flank the number of independent time samples will be about, 
\begin{equation}
\label{eqn:mtq}
 M_{TQ} = {\tau_T B \over d 1.38\times 10^4}, 
\end{equation}
while this effect is fully coherent with frequency,
\begin{equation}
\label{eqn:mfq}
 M_{FQ} = 1.
\end{equation}

\subsubsection{Multi-path frequency modulation}\label{sec:mpfm}

\begin{figure}
\resizebox{\hsize}{!}{\includegraphics{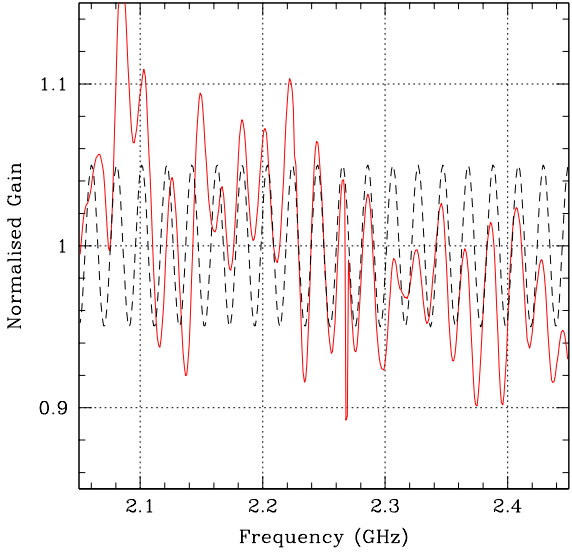}}
\caption{Normalised gain of the ATCA 22m antennas as function of
  frequency. The solid curve is the mean gain over all interferometer
  baselines while the dashed curve is a sinusoid with 20.5 MHz
  periodicity and 5\% peak modulation about the mean.}
\label{fig:atrip}
\end{figure}

Finally is the issue of frequency modulation of the main beam response
pattern. Just as noted above for the near-in sidelobe pattern, the
main beam dimensions of symmetrically fed dishes have been shown to be
modulated at about the $\epsilon_B = 0.05$ level
\citepads[e.g.][]{2008A&A...479..903P} with a periodicity of about
$c/2l_C$ Hz. What is particularly pernicious about this phenomenon is
that it represents a position dependant amplitude response which is
not removed by an on-axis calibration of the instrumental bandpass.
While the \citetads{2008A&A...479..903P} result was documented for the
prime focus WSRT dishes, a comparable phenomenon is seen in Cassegrain
systems. We illustrate the normalised gain of the 22~m diameter,
Cassegrain focus, ATCA system in the 2--3 GHz band in
Figure~\ref{fig:atrip}. What's shown is the ratio of the
cross-correlated amplitude towards a compact calibration source,
PKS~1934-638, normalised by the square root of the product of the two
relevant auto-correlations for each interferometer baseline. As shown
in \citetads{2008A&A...479..903P} this quantity is proportional to the
interferometer gain and is tied to a comparable modulation of the main
beam effective area. Even in the case of the off-axis optics of the
100~m Green Bank Telescope, there are a wide range of quasi-sinusoidal
baseline effects that have been documented by \citet{2003.Fish}. The
most significant of these is consistent with multipath effects between
the subreflector and feed horns and has an amplitude at 1.4 GHz of
$\epsilon_B = 0.005$ and frequency period $c/2l_C$ Hz, for a
subreflector-feed separation, $l_C$. The visibility modulation of this
effect is given by,
\begin{equation}
\label{eqn:sigb}
 \sigma_B = 0.7 \epsilon_B S_{RMS}(\Omega_M)
\end{equation}
for a typical attenuation of 0.7 on the beam flank, while the number
of independent time samples is 
\begin{equation}
\label{eqn:mtb}
 M_{TB} = {\tau_T B \over 2 d 1.38\times 10^4}.
\end{equation}
As discussed above for the near-in sidelobes, there are two relevant
frequency modulation intervals to consider, corresponding to,
\begin{equation}
\label{eqn:mfb}
 M_{FB} = {\Delta\nu_T B \over 2 \nu d }.
\end{equation}
and
\begin{equation}
\label{eqn:mfbp}
 M_{FB}^\prime = 4 l_C \Delta\nu_T/c . 
\end{equation}
The larger of these two values would apply to the full observation.

\subsection{Synthesised beam issues}\label{sec:synbm}

Some of the influences of the synthesised beam have already been
mentioned previously; in particular the far sidelobes of the
instantaneous synthesised beam that will be present in an image, due
to the residual response of all inadequately modelled sources on the
sky. The frequency scaling of the far sidelobe pattern provides
significant attenuation of these fluctuations once broad-band
frequency averaging has been applied. Another issue in this area is
the quality of $(u,v)$ coverage in relation to the sky brightness
distribution that requires modelling; both its main-beam and off-axis
components. If the $(u,v)$ coverage is inadequate, for example due to
a dearth of short interferometer spacings, this will introduce a
limitation on the fidelity of the self-calibration model that is
deduced, for example via spurious source components, which in turn
will limit the precision of the solution. An interesting challenge to
consider in this regard is the Sun, with its $10^6$ Jy, 30 arcmin disk
and intermittent active regions. If the far sidelobe antenna response
does not provide sufficient suppression of this source, it will
constitute a significant modelling challenge in any self-calibration
solution, as already demonstrated in Figure~\ref{fig:sigd}. Of
more general concern is the challenge of modelling sources within the
main beam of an observation that are only partially resolved by the
synthesised beam. From Table~\ref{tab:nvss} it is clear that the
typical sources encountered randomly in a self-cal context are
unlikely to be point-like in nature, but rather partially extended for
baselines exceeding a few km. Such sources provide a particular
modelling challenge, since intrinsic source structure must be
distinguished from the general image-plane smearing effects induced by
imperfect calibration of the visibilities. An estimate of the maximum
modelling error can be made from the Gaussian decomposition of the
Cygnus A power spectrum shown in Figure~\ref{fig:powspec}. The
fine-scale substructure within this powerful radio galaxy accounts for
between 10 and 30\% of the total flux. This suggests that a modelling
error, $\epsilon_M > 0.1$, might be the consequence of simply using a
point-like model for such sources.

\begin{figure*}
\resizebox{\hsize}{!}{\includegraphics{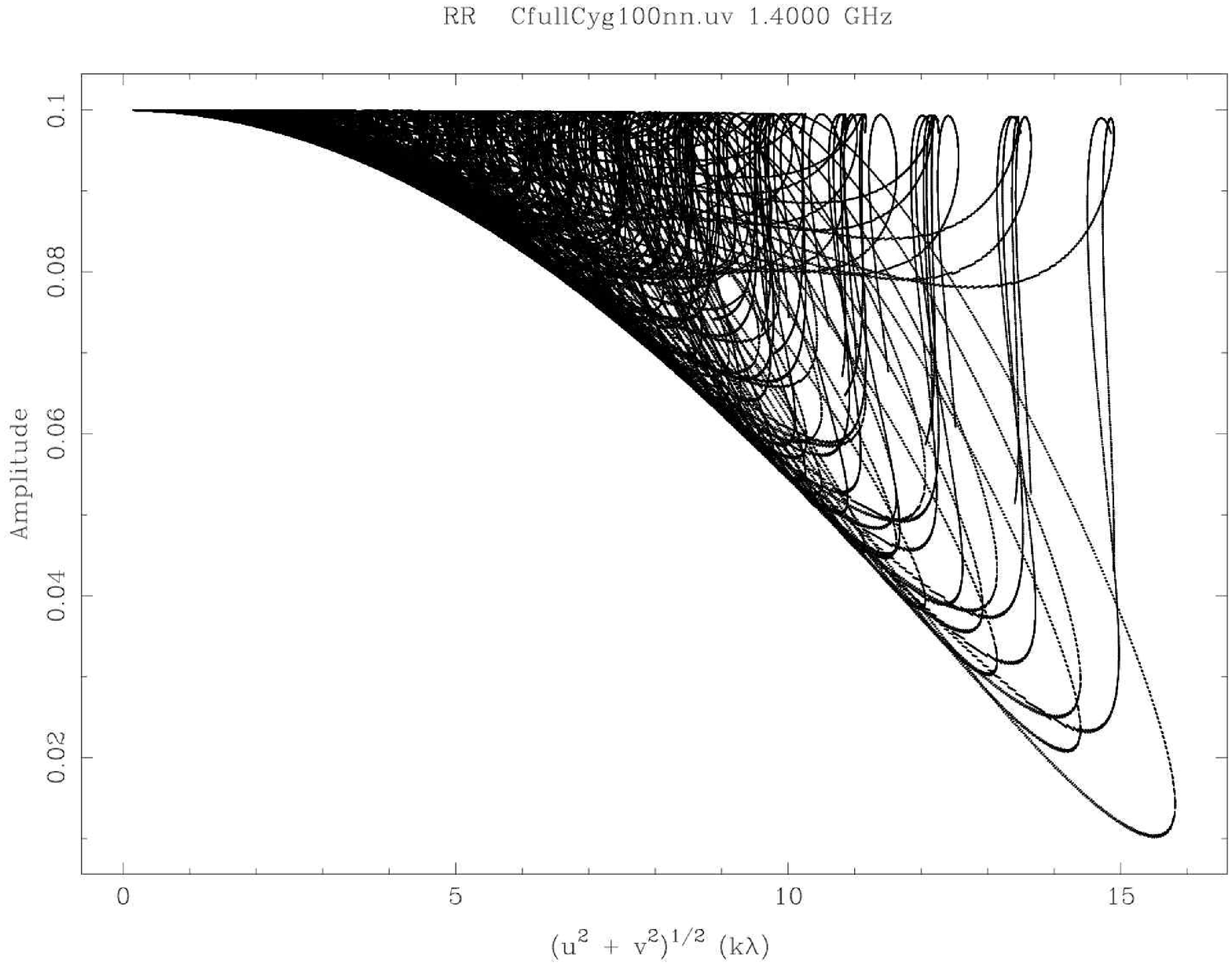},\includegraphics{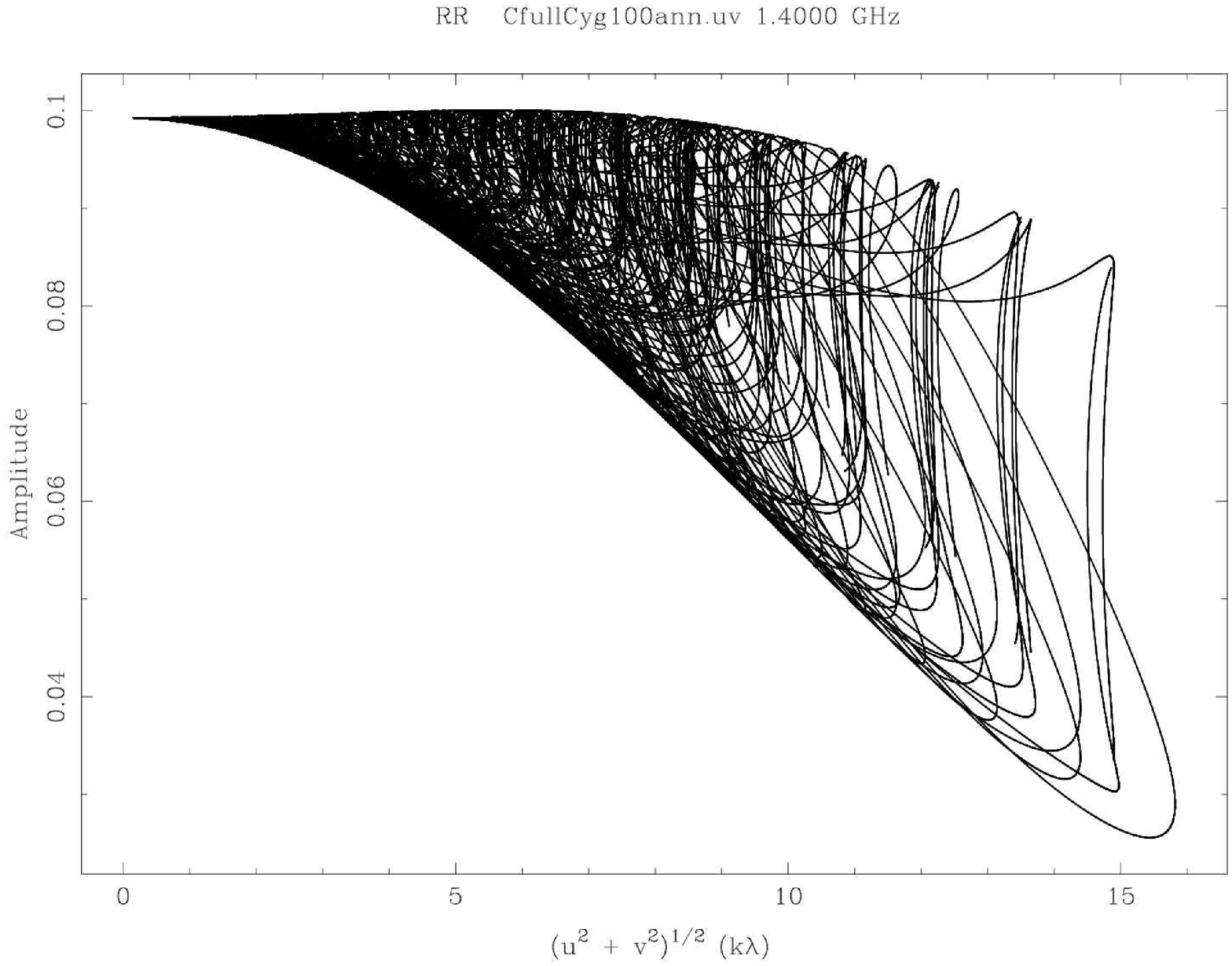},\includegraphics{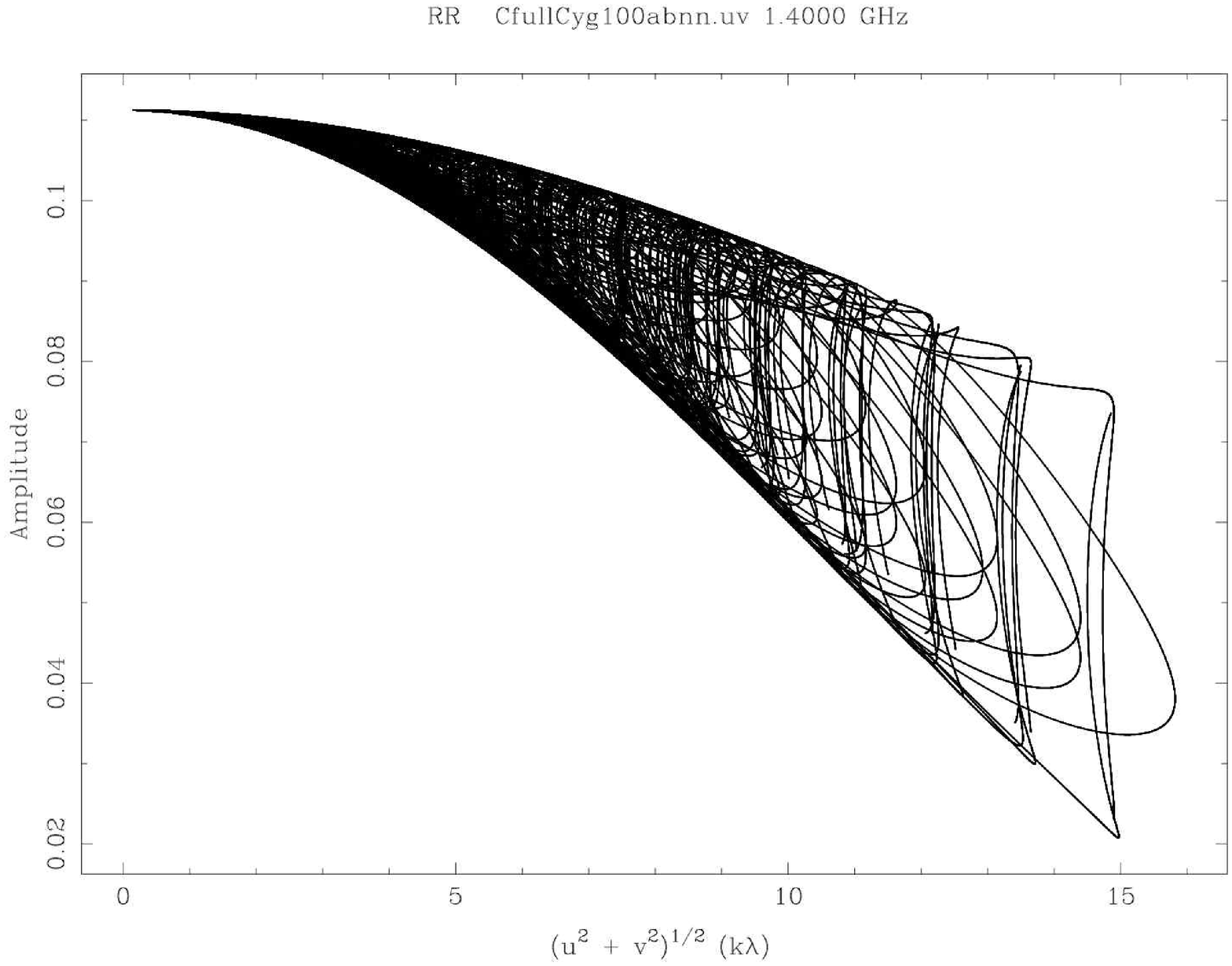}}
\caption{Comparison of input (left) and reconstructed (centre, right)
  visibilities of a 100 mJy Cygnus A, when observed on-axis in the marginally
  resolved regime (8 arcsec source size and 12 arcsec beam) with
  negligible time and bandwidth smearing. The model visibilities in
  the right hand panel were constrained to only positive source
  components. The mean visibility error is 0.7\% (centre) and 7\%
  (right) relative to the integrated flux of the source.}
\label{fig:modelon}
\end{figure*}

We have attempted to better quantify source modelling errors by
considering the quality of a CLEAN reconstruction of Cygnus A when it
has been assigned an integrated flux of 100 mJy and an angular size of
8 arcsec and is observed for 12 hours with the VLA in the C
configuration in a single spectral band at 1.4 GHz.  The noiseless
input and reconstructed visibilities for an on-axis source location
are contrasted in Figure~\ref{fig:modelon}. Realistic thermal noise
was included in the imaging and deconvolution process but no
additional calibration errors, implying that this is in many ways a
best case scenario. Two versions of the reconstructed visibility
amplitudes are shown, one in which both positive and negative apparent
brightness components found in a CLEAN deconvolution are retained and
the other in which only positive brightness components were
retained. The input and derived model visibilities have a mean
difference of $\epsilon_M = 0.007$ when all components are retained,
but this increases to $\epsilon_M = 0.07$ when only positive
components are permitted. The positive source component constraint
has historically been used as a means to discriminate against the
inclusion of instrumental calibration errors into the source model
that is used for self-calibration. The large modelling errors which
arise from this constraint are cause for substantial concern. The
unphysical nature of the negative source model is apparent in the
visibility amplitudes seen in the centre panel of
Figure~\ref{fig:modelon} which peak at a finite radius rather than at
the origin. Despite the unphysical nature of such a source model, it
provides a significantly better match for the source visibilities at
locations where they have been measured. 

\begin{figure*}
\resizebox{\hsize}{!}{\includegraphics{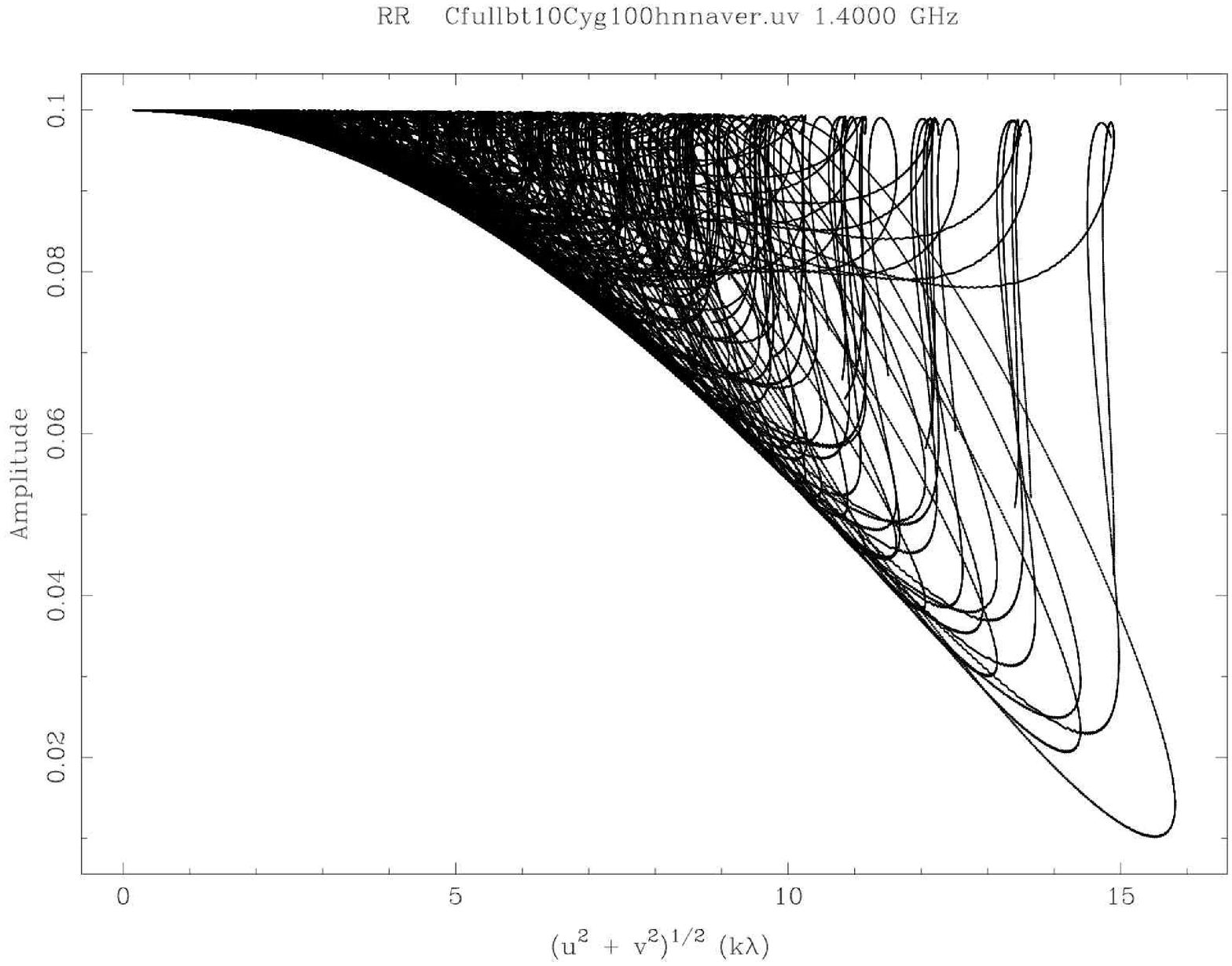},\includegraphics{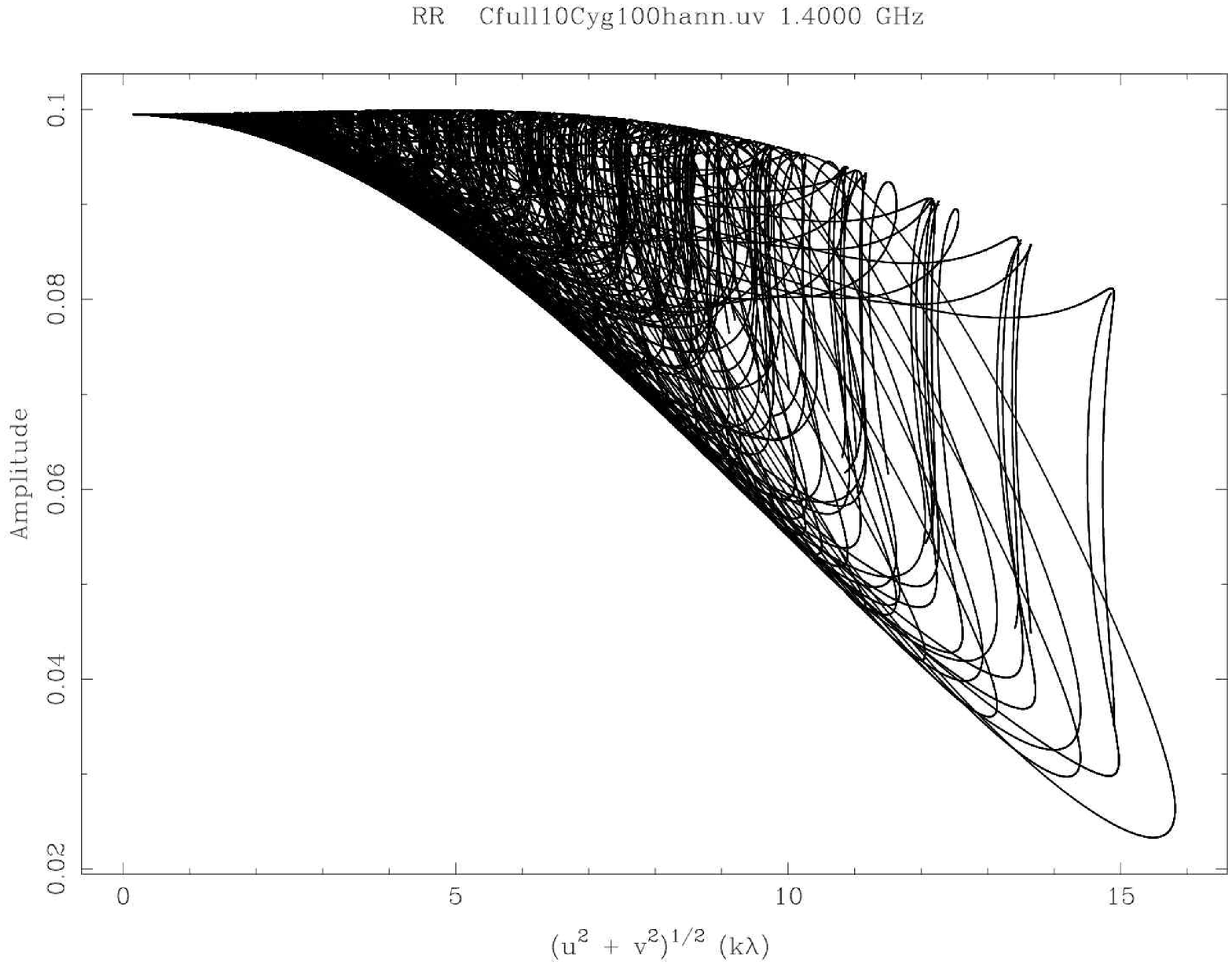},\includegraphics{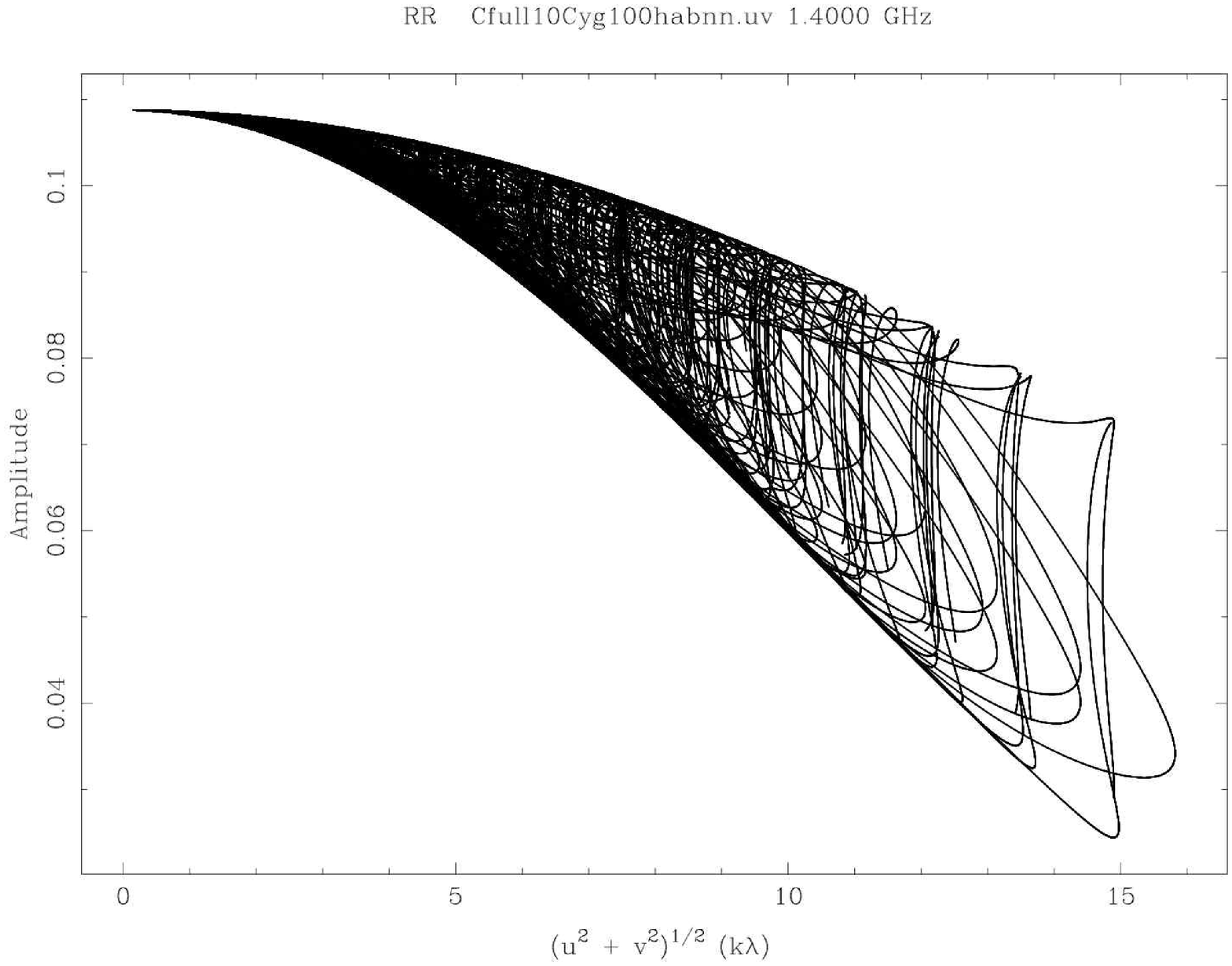}}
\caption{Comparison of input (left) and reconstructed (centre, right)
  visibilities of a 100 mJy Cygnus A, when observed significantly
  off-axis in the marginally resolved regime (8 arcsec source size and
  12 arcsec beam). The integration time and channel bandwidth were
  chosen to keep smearing below 10\% of the synthesised beam at the
  edge of the main beam. The model visibilities in the right hand
  panel were constrained to only positive source components. The mean
  visibility modelling error is 0.9\% (centre) and 10\% (right).}
\label{fig:modeloff}
\end{figure*}

\begin{figure}
\resizebox{\hsize}{!}{\includegraphics{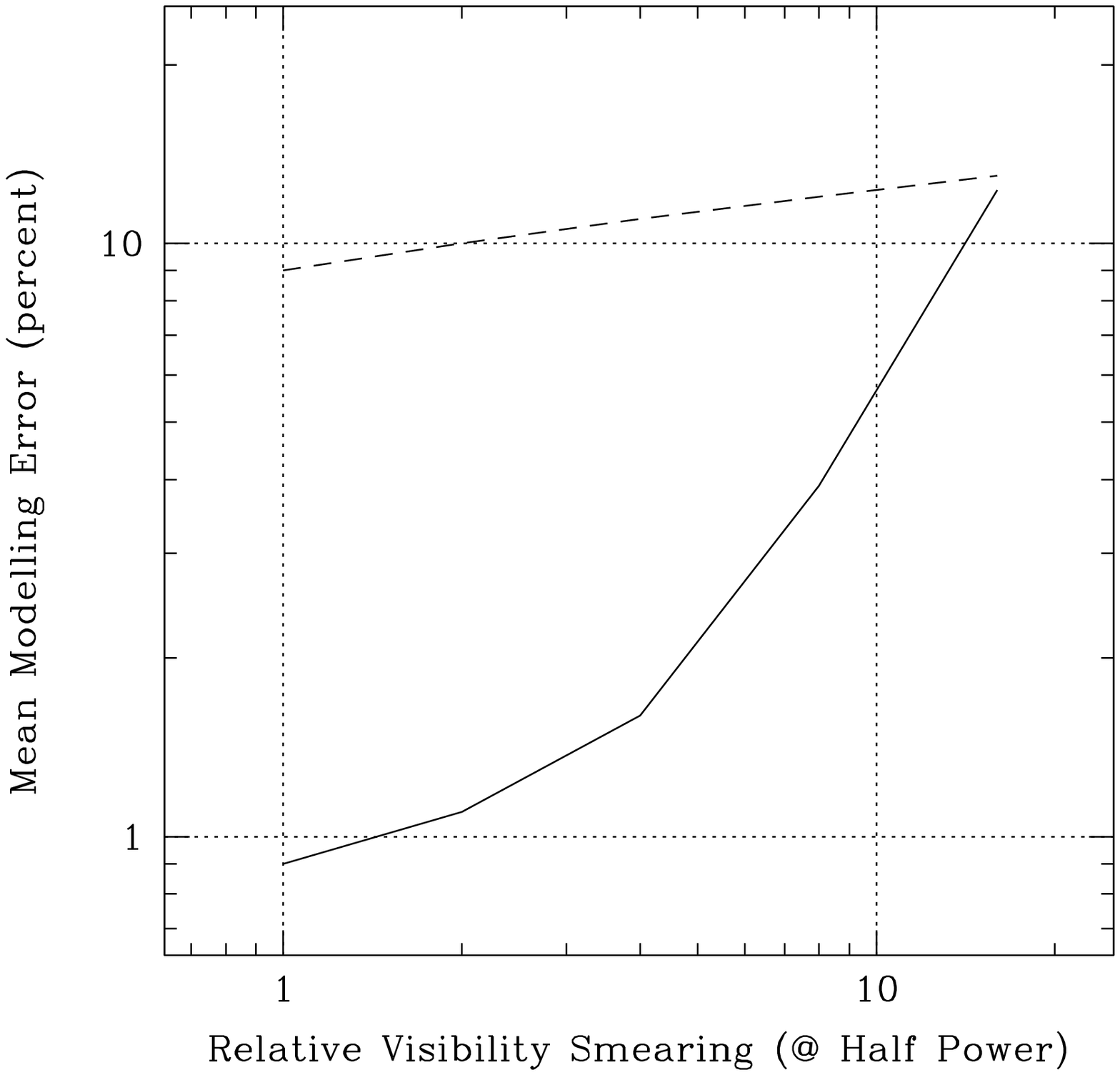}}
\caption{Mean source modelling error as function of the relative
  visibility smearing scale on the main beam flank. The reference
  smearing scale is taken to be 10\% of the synthesised beam at half power
  of the main beam. The solid curve is a high precision model that
  likely represents a best case scenario, while the dashed curve is a
  simple model that is constrained to have only positive components.}
\label{fig:moderr}
\end{figure}

When the same source is observed significantly off-axis, by 15 arcmin
(a half beamwidth) in both RA and Dec, using a channel bandwidth and
integration time that keep smearing effects in this configuration
below 10\% of the synthesised beam at the main beam edge: $\Delta \nu
/ \nu = 10^{-3}$ and $\tau = 10$ sec, the modelling errors increase as
seen in Figure~\ref{fig:modeloff}. The mean visibility errors increase
to 0.9\% and 9\%, for the unconstrained and constrained cases. The
unconstrained modelling error increases with larger visibility
smearing timescales, reaching 12\% for integration times of $\tau \sim
160$ sec (while $\Delta \nu / \nu$ is fixed at $10^{-3}$). When the
positivity constraint is invoked, this effect dominates over smearing
errors until $\tau \sim 120$ sec. The magnitude of the modelling error
increases approximately linearly with the smearing scale as illustrated in
Figure~\ref{fig:moderr}. Background sources dominated by a compact
flat spectrum core would be more likely to be modelled accurately, at
least when the dominant component is placed exactly on an image
pixel. However, the vast majority of random field sources encountered
at GHz frequencies are not of this type. The resulting visibility
modulation is given approximately by,
\begin{equation}
\label{eqn:sigm}
 \sigma_M = 0.7 \epsilon_M S_{RMS}(\Omega_M) \bigg({ \tau_S \over
   \tau^\star}\bigg) \bigg({ \Delta\nu_S \over
   \Delta\nu^\star}\bigg)
\end{equation}
where a linear scaling with self-cal solution timescales and frequency
intervals $(\tau_S, \Delta\nu_S)$ is introduced relative to the
nominal intervals $(\tau^\star, \Delta\nu^\star)$ that keep smearing
effects below 10\% of the synthesised beamwidth in eqns.~\ref{eqn:tsm}
and \ref{eqn:bsm}.  A typical attenuation of 0.7 for the brightest
field source on the main beam flank is also assumed. We will also
consider the case where a high precision source model is available
from other sources, in which case the residual modelling error,
$\epsilon_M^\star$, need not scale with the smearing scale,
\begin{equation}
\label{eqn:sigms}
 \sigma_M^\star = 0.7 \epsilon_M^\star S_{RMS}(\Omega_M).
\end{equation}
This circumstance might arise, for example, if a previous high quality
model of the field were available, or the field were fully sampled
with the overlapping primary beams of a phased array feed and
visibility smearing effects were explicitly included when undertaking
data comparison during self-calibration. The number of independent
time and frequency samples at this off-axis distance are given by,
\begin{equation}
\label{eqn:mtm}
 M_{TM} = {\tau_T B \over 2 d 1.38\times 10^4}.
\end{equation}
and
\begin{equation}
\label{eqn:mfm}
 M_{FM} = {\Delta\nu_T B \over 2 \nu d }.
\end{equation}
Significant research on better source modelling methods
\citepads[eg.][]{2010arXiv1008.1892Y, 2011arXiv1101.2830Y} will be
vital to allowing such high precisions to be achieved.

\subsection{Self-cal parameters}\label{sec:scparms}

\begin{figure*}
\resizebox{\hsize}{!}{\includegraphics{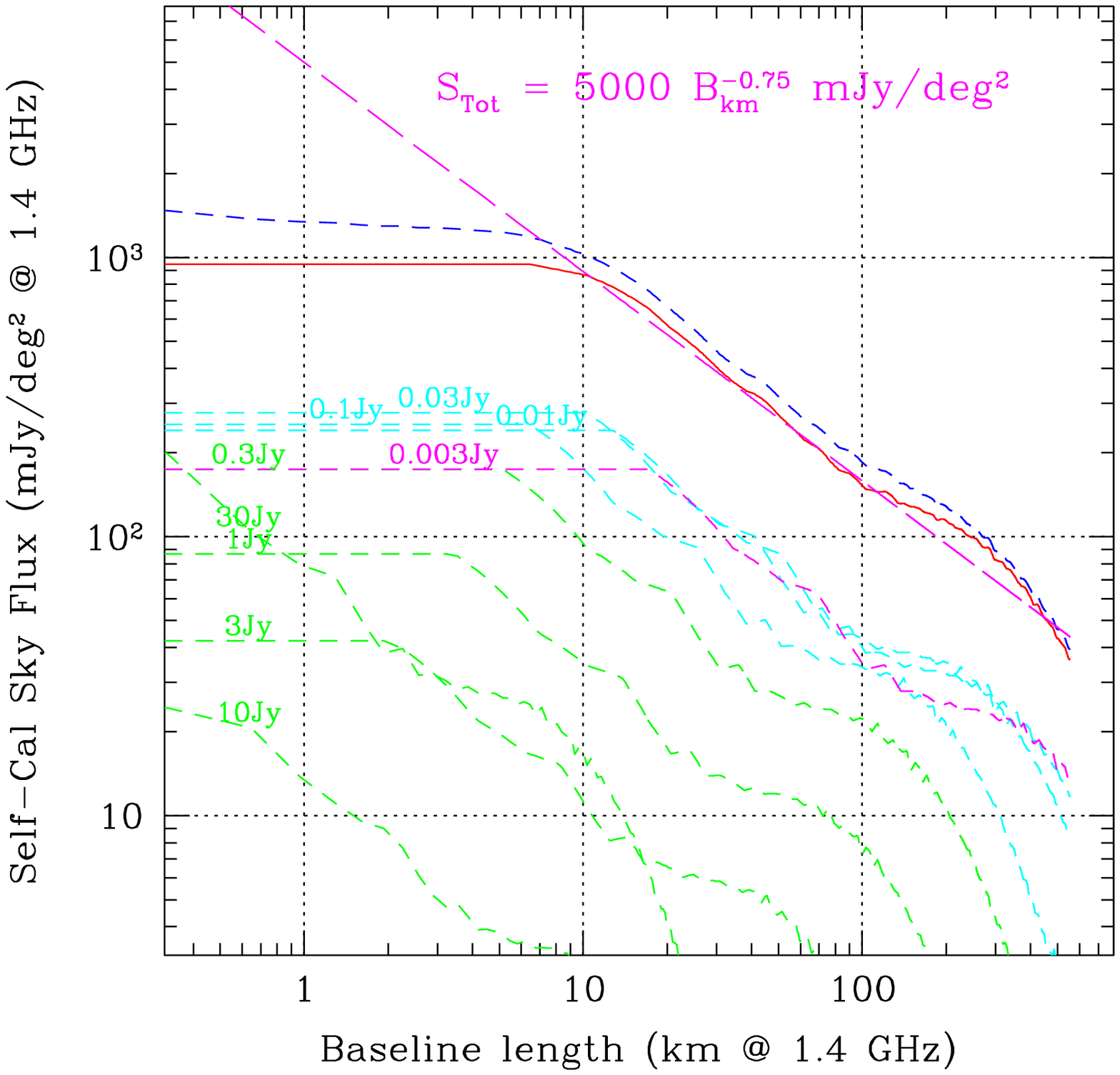},\includegraphics{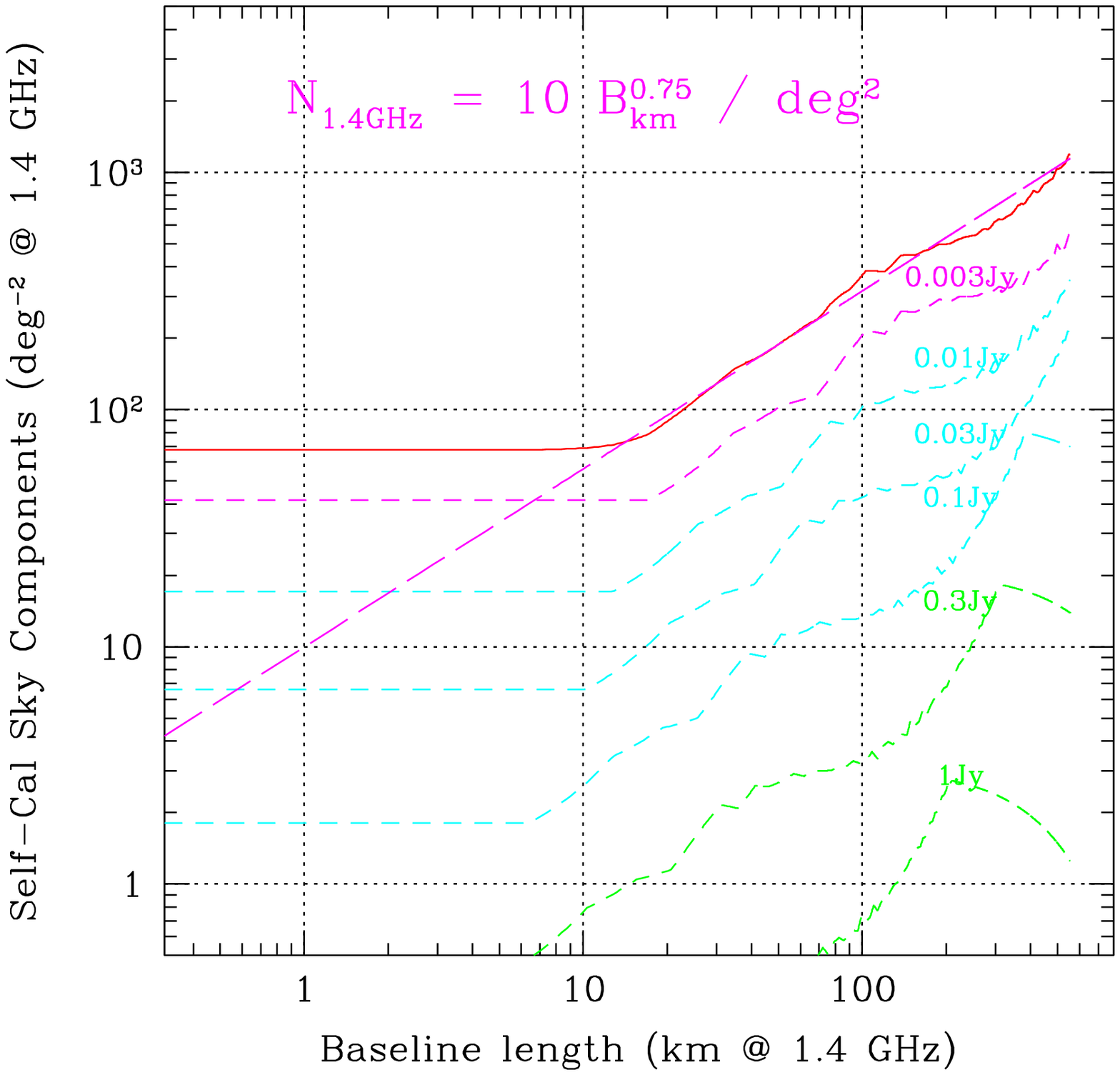}}
\caption{Mean integrated flux density per unit area (left) due to
  the extragalactic sky and the corresponding number density of
  flux-weighted source components (right) as function of angular
  scale. The total is given by the solid curve, while individual bins
  from Table 1 are plotted as short-dashed curves.}
\label{fig:selfcals}
\end{figure*}

The signal-to-noise ratio obtained within a self-cal solution interval
depends crucially on both the integral source flux density available
and the number of flux-weighted components over which it is
distributed as demonstrated by eqn.~\ref{eqn:phi}.  The NVSS
statistics of Table~\ref{tab:nvss} and the Cygnus A power spectrum of
Figure~\ref{fig:powspec} were used to calculate the integrated flux
density per unit area together with the sum of flux-weighted component
number shown in Figure~\ref{fig:selfcals}. The solid line in the
figure gives the total flux per square degree due to sources between 3
and 300 mJy at 1.4 GHz (for which the likelihood of occurrence in a
one degree field is at least unity) while the dark dashed line
includes brighter sources as well. Contributions to the integral are
quite evenly split between the 10, 30 and 100 mJy bins, while
contributions to component number are dominated by the faintest flux
bins. As noted previously, source populations are likely to be
dominated by luminous radio galaxies (like Cygnus A) above about 30
mJy, while edge-darkened radio galaxies are more likely to dominate
below this flux, implying that some modification of our power spectrum
template might be appropriate for the faintest flux bins. A
representative expression for the integrated flux density per unit
area is,
\begin{eqnarray}
\label{eqn:stot}
 S_{Tot} & = & 920\ \Omega_{deg} \bigg({\nu \over {\rm
     1.4\ GHz}}\bigg)^{-0.8} \qquad (B < B_R) \\
\nonumber & = & 920\ \Omega_{deg} \bigg({\nu \over {\rm
     1.4\ GHz}}\bigg)^{-0.8} 
  \bigg({B \over B_R}\bigg)^{-0.75} \hfil (B > B_R)  \quad {\rm
    mJy\ deg^{-2}}
\end{eqnarray}
where the total brightness scales roughly as $\nu^{-0.8}$, and the
baseline length is normalised to the break at about $B_R = 10
(\nu/{\rm 1.4\ GHz})^{-1}$ km, below which the total flux
saturates. The corresponding expression for flux-weighted major
component number is,
\begin{eqnarray}
\label{eqn:nc}
 N_{C} & = & 66\ \Omega_{deg}  \qquad (B < B_R) \\
\nonumber & = & 66\ \Omega_{deg} 
  \bigg({B \over B_R}\bigg)^{0.75} \hfil (B > B_R)  \quad {\rm
     deg^{-2}}.
\end{eqnarray}

\begin{figure*}
\resizebox{\hsize}{!}{\includegraphics{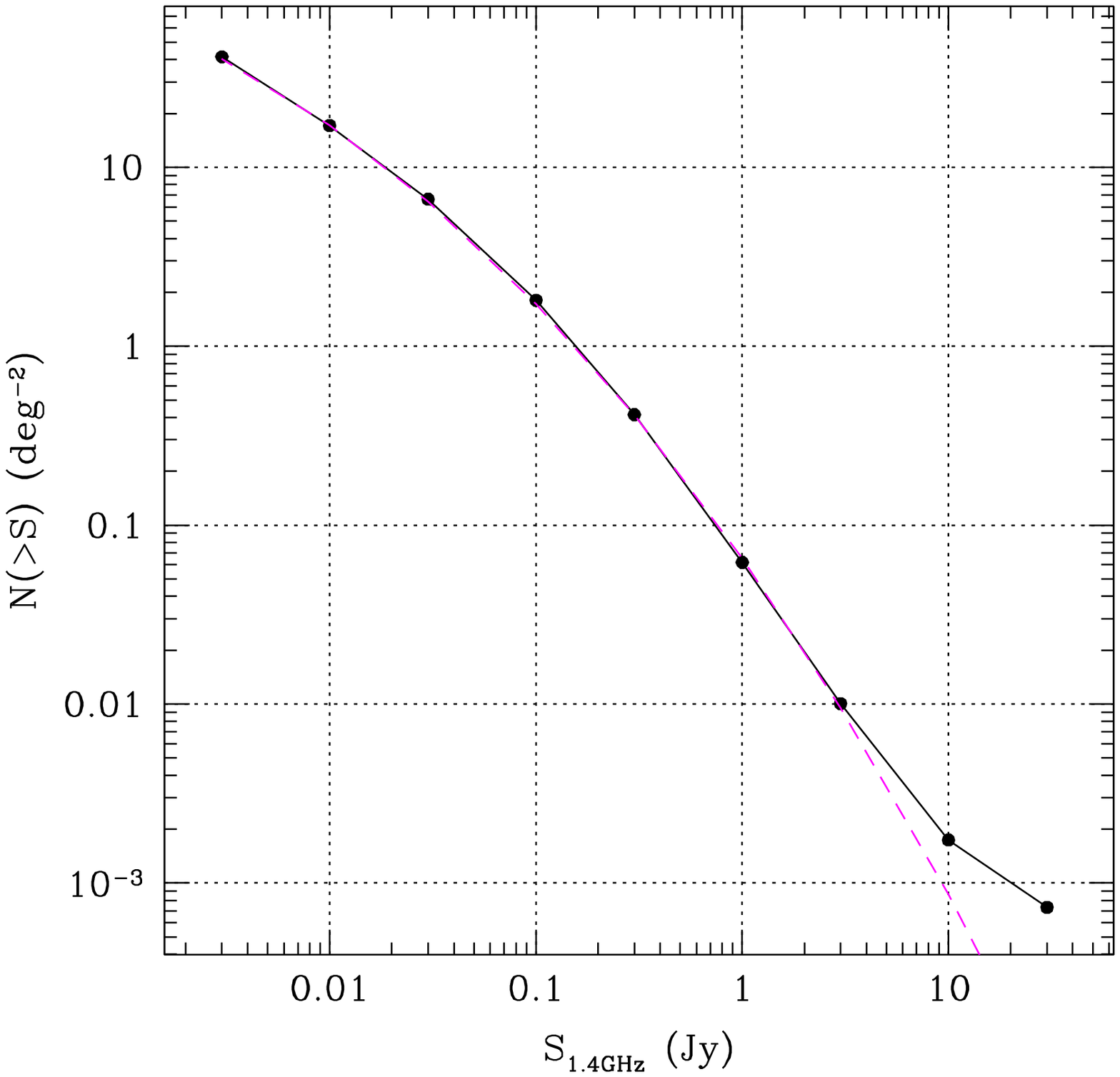},\includegraphics{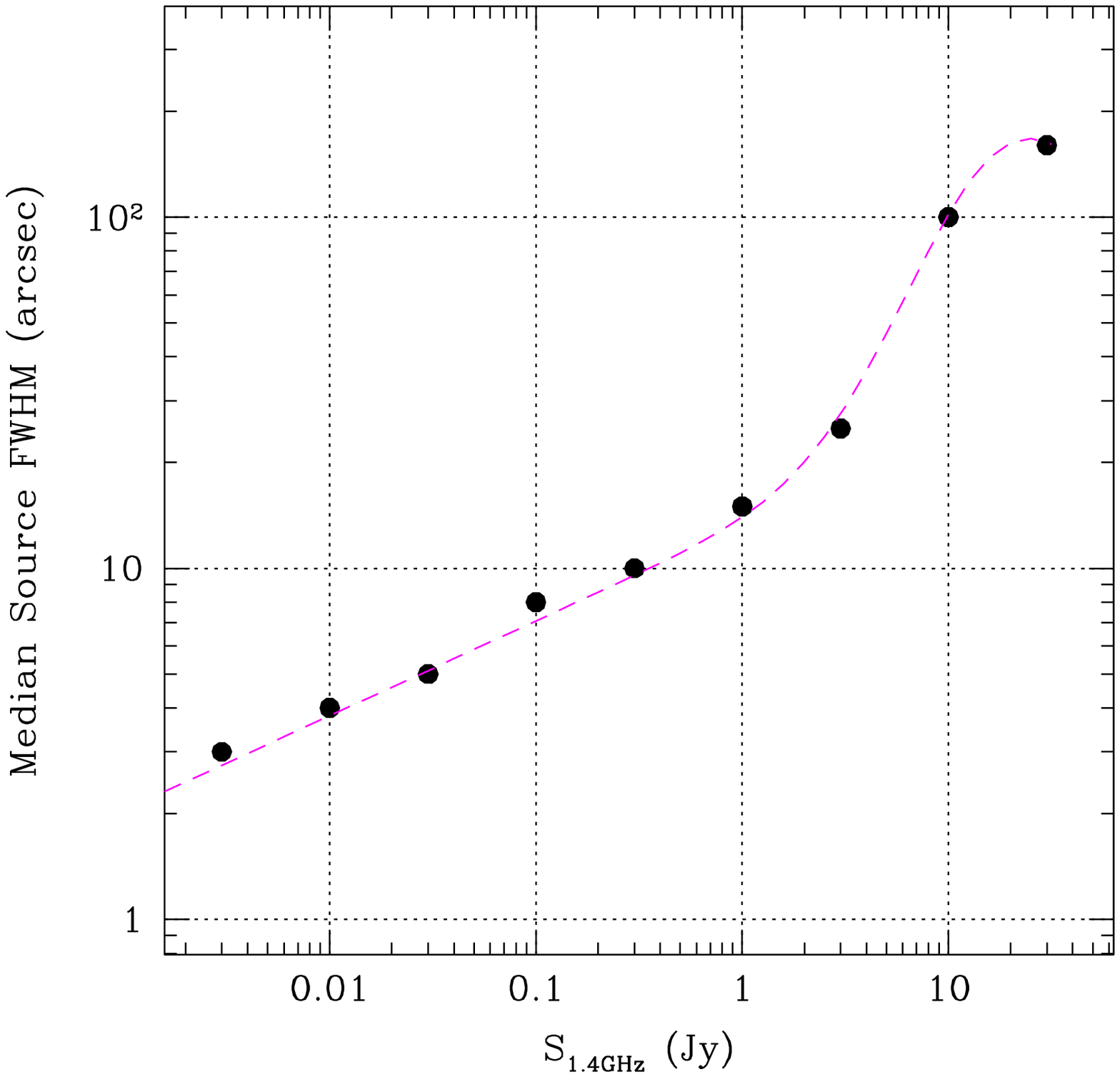}}
\caption{NVSS 1.4 GHz source counts (left) per 0.5 dex bin are
  shown as the filled points connected by a solid line. The analytic
  approximation over the range $0.003 <S_{1.4} < 3.$ Jy that is given
  by eqn.~\ref{eqn:nofs} is overlaid as the dashed curve. The median
  source size from Table~\ref{tab:nvss} overlaid with an analytic
  approximation is shown in the right hand panel.}
\label{fig:modsofs}
\end{figure*}

Other important quantities needed to assess the magnitude of many
contributions to the visibility noise budget are $S_{Max}(\Omega)$, the
maximum source brightness that can be expected by chance in a field of
a given solid angle together with its overall angular size,
$\psi$ in arcsec. The NVSS source counts of Table~\ref{tab:nvss} can be
approximated very well with the arbitrary analytic form,
\begin{equation}
\label{eqn:nofs}
{\rm log}(N(>S)) = 2.24 - {[{\rm log}(S_{1.4 GHz})+5]^{2.4}\over 13.9}
\end{equation}
which is overlaid on the count data in Figure~\ref{fig:modsofs} and
agrees within 0.02 dex over the range $0.003 <S_{1.4} < 3.$ Jy. The
median source size, $\psi$, of Table~\ref{tab:nvss} is also plotted in
Figure~\ref{fig:modsofs} together with an analytic approximation,
\begin{eqnarray}
\label{eqn:pofs}
{\rm log}(\psi) & = & 0.31 + 0.27\ [{\rm log}(S_{1.4 GHz})+3] \\ 
\nonumber & + & 0.74\ {\rm exp}\lbrace-2[{\rm log}(S_{1.4 GHz})-1.3]^2\rbrace
\end{eqnarray}
that agrees with the data to within 0.05 dex. Equation~\ref{eqn:nofs}
can be inverted to yield the flux of the brightest 1.4 GHz source
occurring in a solid angle $\Omega_{deg}$, as,
\begin{equation}
\label{eqn:smaxp}
S_{1.4 Max} = 1.4 \cdot
{\rm antilog}_{10}(\lbrace 13.9 [2.24-{\rm log}(1/\Omega_{deg})]\rbrace^{0.417}-5) {\rm Jy},
\end{equation}
where the leading factor of 1.4 accounts for the median flux of
sources within the 0.5 dex bins used in the source counts. This can be
scaled to other observing frequencies and baseline lengths via,
\begin{eqnarray}
\label{eqn:smax}
 S_{Max} & = & S_{1.4 Max} \bigg({\nu \over {\rm
     1.4\ GHz}}\bigg)^{-0.8} \qquad (B < B_\psi) \\
\nonumber & = & S_{1.4 Max} \bigg({\nu \over {\rm
     1.4\ GHz}}\bigg)^{-0.8} 
  \bigg({B \over B_\psi}\bigg)^{-0.75} \hfil (B > B_\psi)  \quad {\rm
    Jy},
\end{eqnarray}
where the characteristic baseline $B_\psi$, that resolves the source
is given by,
\begin{equation}
\label{eqn:bpsi}
B_\psi = 5\ \bigg({\psi \over {\rm
     10\ arcsec}}\bigg)^{-1} \bigg({\nu \over {\rm
     1.4\ GHz}}\bigg)^{-1} \quad {\rm
    km}.
\end{equation}

\subsection{External calibration}\label{sec:excal}

There are instances where self-calibration may not be practical and
others where it may not even be necessary. It is useful to define the
base noise level due to main beam sources under these
circumstances. For a gain error of magnitude, $\phi_C$, with a
correlation timescale, $\tau_C$ and fractional correlation bandwidth
$(\Delta\nu/\nu)_c~=~\delta_C$, we can write the visibility noise
associated with a source in the main beam as,
\begin{equation}
\label{eqn:sigcal}
 \sigma_{Cal} = \phi_C S_{RMS}(\Omega_M),
\end{equation}
together with the number of independent time and frequency samples in
a full track observation as,
\begin{equation}
\label{eqn:mtc}
 M_{TC} = {\tau_T / \tau_C},
\end{equation}
and
\begin{equation}
\label{eqn:mfc}
 M_{FC} = {\Delta\nu_T \over \nu \delta_c }. 
\end{equation}
In the event that the main beam field of view approaches 200 deg$^2$,
the large field version of the visibility fluctuations,
$S_{RMS}^\prime$, must be considered yielding,
\begin{equation}
\label{eqn:sigcalp}
 \sigma_{Cal}^\prime = \phi_C S_{RMS}^\prime(\Omega_M).
\end{equation}

\section{Some practical examples}\label{sec:exam}

With all of the quantities and scaling relationships defined in
\S~\ref{sec:cal}, it becomes possible to characterise the calibration
challenges faced by specific antenna systems. Although the principles
we outline are completely general, the source statistics and
properties we model are based on data obtained near 1.4 GHz and will
be most relevant to frequencies between about 100 MHz and 5
GHz. Outside of this range, there may be significant differences in
the source population (eg.  \citetads{2012MNRAS.423L..30S},
\citetads{2011MNRAS.412..318M}).

In the subsections that follow, the attributes of several existing and
planned arrays listed in Table~\ref{tab:tparms} will be considered,
both in terms of visibility noise contributions at the basic self-cal
solution interval, as well as the final image noise contributions in
an observation of a random field that is tracked for $\tau_T = 12$
hours in the case of a dish array or $\tau_T = 4$ hours for the case
of aperture arrays. For the full-track case we will consider both a
standard continuum mode observation, in which we assume that a total
fractional bandwidth of $\Delta\nu_T/\nu = 0.1$ is ultimately averaged
together as well as a standard spectral line mode observation in which
a spectral resolution of $\Delta\nu_T/\nu = 10^{-4}$ is utilised in
the image. For the self-cal solution interval, it will be important to
achieve a signal-to-noise ratio that is sufficient to allow
convergence to be achieved. We will define this by requiring a
residual calibration error, as given by eqn.~\ref{eqn:phi}, of
$\phi(\tau_S) < 0.5$ rad. This yields a visibility solution noise
limit,
\begin{equation}
\sigma_{Sol} < 0.5\ S_{Tot}(N-3)^{0.5}/N_C^{0.5}.
\label{eqn:sigsol}  
\end{equation}
For the full-track
continuum imaging it will be of relevance to compare the final image
noise contributions with the source confusion
level. \citetads{2012ApJ...758...23C} have recently estimated the
magnitude of such confusion due to the faint source population to be,
\begin{equation}
\sigma_{Cfn} \approx 1.2~ \mu{\rm Jy}~ \bigg( {\nu \over 3.02~ {\rm
    GHz}}\bigg)^{-0.7} \bigg( {\theta \over 8''}\bigg)^{10/3}.
\label{eqn:sigcfn}  
\end{equation}
For the purposes of illustration we will consider several examples of
main beam source modelling errors, one in which significant effort has
been expended to achieve a nominal precision $\epsilon_M = 0.01$, a
cruder version $\epsilon_M^\prime = 0.1$ and finally a high precision
variant, $\epsilon_M^\star = 0.001$. To simplify the interpretation,
we will not apply additional scaling of these values with the smearing
factors of eqn.~\ref{eqn:sigm}. Recall from Sec.~\ref{sec:synbm} that
a 1\% precision was found challenging to achieve for marginally
resolved background sources with a realistic degree of complexity and
that a 10\% error was often encountered in practise with simplistic
modelling approaches. Achieving 0.1\% modelling precision represents a
major technical challenge that would rely both on more extensive
information of the field under study than is typically available as
well as the algorithm complexity to utilise that information
effectively. The full track cases will also be compared with the
baseline noise level due to gain calibration errors of main beam
sources to illustrate the consequence of not applying self-calibration
at all. A nominal gain error of $\phi_C~=~0.2$, corresponding to
10 degrees of phase, a correlation timescale of $\tau_C ~=~15$ minutes and
a correlation fractional bandwidth of $\delta_C~=~0.1$ will be assumed
throughout for illustration purposes. All of the relevant parameters
are collected and defined in Table~\ref{tab:defns}.

\begin{table*}
\caption{Measured and Assumed Telescope Parameters.}
\label{tab:tparms}
\centering
\begin{tabular}{l c c c c c c c c c c c c c}
\hline\hline
Telescope&$N$&$d$&$B_{Max}$&$B_{Med}$&$\eta_F$&$\epsilon_S$&$P$&$\tau_P$&$\epsilon^\prime_P$&
$\tau^\prime_P$&$\epsilon_Q$&$\epsilon_B$&$l_C$\\
 & &(m)&(km)&(km)& & &(arcsec)&(min)& &(min)& & &(m)\\
\hline
JVLA-D Config.&27&25&1.0&0.17&0.1&0.02&10&15& & &0.055&0.05&8.2\\
JVLA-C Config.&27&25&3.4&0.57&0.1&0.02&10&15& & &0.055&0.05&8.2\\
JVLA-B Config.&27&25&11.1&1.85&0.1&0.02&10&15& & &0.055&0.05&8.2\\
JVLA-A Config.&27&25&36.4&6.07&0.1&0.02&10&15& & &0.055&0.05&8.2\\
ATA           &42&6.1&0.32&0.08&0.7&0.01&90&15& & &0.04&0.01&3.0\\
ASKAP&36&12&6.0&0.63&0.2/2\tablefootmark{a}&0.02/10\tablefootmark{b}&10&15&0.01&1&0.04/100\tablefootmark{a}\tablefootmark{b}&0.05/10\tablefootmark{a}&6.0\\
MeerKAT       &64&13.5&8.0&0.50&0.2&0.01&10&15& & &0.04&0.01&7.0\\
SKA1-Survey   &96&15&20&1.0&0.2/2\tablefootmark{a}&0.01&10&15&0.01&1&0.04/10\tablefootmark{a}&0.01/10\tablefootmark{a}&7.0\\
SKA1-Dish   &250&15&20$\times$5\tablefootmark{c}&1.0&0.2&0.01&10&15& & &0.04&0.01&7.0\\
\hline
LOFAR HBA Core&48&30.8&3.5&0.25&0.5&0.1& & &0.01&1&0.01& & \\
LOFAR HBA Ext&64&30.8&121.&1.0&0.5&0.1& & &0.01&1&0.01& & \\
MWA          &128&4.4&3.0&0.3&0.5&0.1& & &0.01&1&0.1& & \\
SKA1-Low Core&35&180&5.0&0.5&0.5&0.1& & &0.01&1&0.01& & \\
SKA1-Low Ext &50&180&100.&2.5&0.5&0.1& & &0.01&1&0.01& & \\
\hline
\end{tabular}
\tablefoot{
\tablefoottext{a}{Phased array feeds provide improved aperture
  illumination and reduced beam modulation.}
\tablefoottext{b}{A parallactic angle mount provides stationary
  near-in sidelobes and main beam.}
\tablefoottext{c}{Analysis is presented for $B_{Max} = 20$~km, while
  the array is projected to extend to $B_{Max} = 100$~km.}
}
\end{table*}

\begin{table*}
\caption{Parameter Definitions.}
\label{tab:defns}
\centering
\begin{tabular}{l c l}
\hline\hline
Variable&Defining Equations& Comments \\
\hline
$\tau^\star$&\ref{eqn:timesm}&Nominal self-cal solution timescale\\
$\Delta\nu^\star$&\ref{eqn:bandsm}&Nominal self-cal solution bandwidth\\
$\phi(\tau_S)$&\ref{eqn:phi}&Self-cal residual error on solution timescale\\
$\sigma_{Map}$&\ref{eqn:sens}&Image plane natural noise \\
$\sigma^\prime_{Map}$&\ref{eqn:farsens}&Self-cal related image plane noise \\
$\Omega_M$&\ref{eqn:om}& Main beam solid angle\\
$\Omega_S$&\ref{eqn:oms}& Near-in sidelobe solid angle\\
$S_{Tot}(\Omega)$&\ref{eqn:stot}& Integral source brightness in solid angle\\
$N_C(\Omega)$&\ref{eqn:nc}& Flux-weighted number of source components in solid angle\\
$S_{RMS}(\Omega)$&\ref{eqn:srms}& RMS source brightness in solid angle\\
$S_{RMS}^\prime(\Omega)$&\ref{eqn:srmsp}& Large field RMS source brightness in solid angle\\
$S_{Max}(\Omega)$&\ref{eqn:smax}& Maximum source brightness in solid angle\\
$\sigma_{Sol}$&\ref{eqn:sigsol}&Self-cal solution noise required for convergence \\
$\sigma_{Cfn}$&\ref{eqn:sigcfn}&Source confusion noise \\
$\sigma_{Cal}$&\ref{eqn:sigcal}&Gain calibration noise \\
$\sigma_{Cal}^\prime$&\ref{eqn:sigcalp}&Large field gain calibration noise \\
$SEFD$&\ref{eqn:sefd}&System equivalent flux density \\
$\sigma_T$&\ref{eqn:vlasigt}&Thermal noise \\
\hline
$\phi_C$&\ref{eqn:sigcal}&Main beam gain calibration error \\
$\epsilon_F$&\ref{eqn:epf}&Far sidelobe attenuation relative to on-axis \\
$\epsilon_S$&\ref{eqn:sigs}&Near-in sidelobe attenuation relative to on-axis \\
$\epsilon_P^\prime$&\ref{eqn:sigpp}&Electronic pointing error\\
$\epsilon_Q$&\ref{eqn:sigq}&Main beam shape asymmetry\\
$\epsilon_B$&\ref{eqn:sigb}&Main beam shape modulation with frequency\\
$\epsilon_M$&\ref{eqn:sigm}&Source modelling error\\
\hline
$\sigma_N$&\ref{eqn:sign}&Nighttime far sidelobe visibility modulation \\
$\sigma_D$&\ref{eqn:sigd}&Daytime far sidelobe visibility modulation \\
$\sigma_S$&\ref{eqn:sigs}&Near-in sidelobe visibility modulation  \\
$\sigma_P$&\ref{eqn:sigp}&Main beam mechanical pointing visibility modulation\\
$\sigma^\prime_P$&\ref{eqn:sigpp}&Main beam electronic pointing visibility modulation\\
$\sigma_Q$&\ref{eqn:sigq}&Main beam asymmetry visibility modulation\\
$\sigma_B$&\ref{eqn:sigb}&Main beam modulation visibility modulation\\
$\sigma_M$&\ref{eqn:sigm},\ref{eqn:sigms}&Source modelling error visibility modulation\\
\hline
$P$&\ref{eqn:sigp}&Mechanical pointing precision\\
$\tau_P$&\ref{eqn:mtp}&Timescale for mechanical pointing modulation\\
$\tau^\prime_P$&\ref{eqn:mtpp}&Timescale for electronic pointing modulation\\
$\tau_C$&\ref{eqn:mtc}&Timescale for calibration error\\
$\delta_C$&\ref{eqn:mfc}&Fractional bandwidth for calibration error\\
\hline
$M_{TF}$&\ref{eqn:mtf}&Far sidelobe independent time solutions\\
$M_{FF}$&\ref{eqn:mff}&Far sidelobe independent frequency solutions\\
$M_{TS}$&\ref{eqn:mts}&Near-in sidelobe independent time solutions\\
$M_{FS}$&\ref{eqn:mfs},\ref{eqn:mfsp}&Near-in sidelobe independent frequency solutions\\
$M_{TP}$&\ref{eqn:mtp}&Main beam mechanical pointing independent time solutions\\
$M_{FP}$&\ref{eqn:mfp}&Main beam mechanical pointing independent frequency solutions\\
$M^\prime_{TP}$&\ref{eqn:mtpp}&Main beam electronic pointing independent time solutions\\
$M^\prime_{FP}$&\ref{eqn:mfpp}&Main beam electronic pointing independent frequency solutions\\
$M_{TQ}$&\ref{eqn:mtq}&Main beam asymmetry independent time solutions\\
$M_{FQ}$&\ref{eqn:mfq}&Main beam asymmetry independent frequency solutions\\
$M_{TB}$&\ref{eqn:mtb}&Main beam modulation independent time solutions\\
$M_{FB}$&\ref{eqn:mfb},\ref{eqn:mfbp}&Main beam modulation independent frequency solutions\\
$M_{TM}$&\ref{eqn:mtm}&Source modelling error independent time solutions\\
$M_{FM}$&\ref{eqn:mfm}&Source modelling error independent frequency solutions\\
$M_{TC}$&\ref{eqn:mtc}&Calibration error independent time solutions\\
$M_{FC}$&\ref{eqn:mfc}&Calibration error independent frequency solutions\\
\hline
\end{tabular}
\end{table*}

Of further interest is an answer to the question: ``What is the
ultimate limit to image noise contributions that follows from multiple
long duration tracks of the same field?''. The answer follows from
considering which of the error terms are correlated over multiple
observing tracks and which are not. The thermal contribution
$\sigma_T$, will clearly be uncorrelated from one track to the
next. For each of the other contributions we have considered, this is
not guaranteed. If the same model is being employed to determine the
self-calibration variables for all data tracks, then residual errors
due to model imperfections will be identical at the corresponding hour
angles of subsequent tracks and the $\sigma_M$ component will not
average down. In the case of pointing errors, $\sigma_P$, which have not
been modelled, it will depend upon the nature of these errors. If they
are the result of an incomplete pointing model or diurnal temperature
variations they may well approximately repeat on subsequent days. If
instead they are due to more random variations such as wind buffeting,
they will average down. The contributions due to main beam frequency
modulation, $\sigma_B$, as well as unmodelled sources both within
near-in, $\sigma_S$, and far sidelobes, $\sigma_N$, are intrinsic to
the sky-telescope system and as such will repeat exactly for multiple
tracks and will not average away. An exception to this is the case of
far side-lobe pickup of the Sun, $\sigma_D$, which will have a
different relative location and hence different visibility response on
subsequent observing dates. The nominal initial calibration error,
$\phi_C$, is not likely to be correlated between observing sessions if
it is dominated by ionspheric or tropospheric phase fluctuations.

\subsection{The JVLA} 

Let's first consider the JVLA, with its $N=27$ $(alt,az)$ mounted dishes
of $d=25$ m. The logarithmic three-arm ``Y'' of the JVLA has a high
central concentration of baselines in each of its configurations
\citepads[see Fig. 25-4 of][]{1989ASPC....6..477H} so that more than
half of all baselines are within $B_{Med} = B_{Max}/6$. The four
standard configurations, designated (A, B, C, D) are defined by
$B_{Max} =$ (36.4, 11.1, 3.4, 1.03) km.  The recently upgraded
Cassegrain receivers provide continuous frequency coverage with a
series of approximately octave band feeds between about 1 and 50 GHz
with a system equivalent flux density, 
\begin{equation}
SEFD = {2 k_B T_{Sys} \over (A \eta_A) }
\label{eqn:sefd}  
\end{equation}
in terms of the Boltzman constant, $k_B$, the
system temperature, $T_{Sys}$, the physical area of a dish, $A$, and
the aperture efficiency, $\eta_A$. Online JVLA documentation suggests,
\begin{equation}
SEFD \approx 250 + 3.4\bigg( {\nu \over 1~ {\rm
    GHz}} - 9\bigg)^2 \quad {\rm Jy},
\label{eqn:vlasefd}  
\end{equation}
that applies to $1 < \nu < 15$ GHz, yielding a visibility thermal
noise per polarisation product,
\begin{equation}
\sigma_T = { SEFD  \over (\tau \Delta\nu)^{0.5}} \quad {\rm Jy}.
\label{eqn:vlasigt}  
\end{equation}
Pointing errors of the VLA 25m dishes have an RMS magnitude of about
$P = 10$ arcsec under good nighttime conditions, that is likely
correlated over timescales of about $\tau_P = 15$ minutes. The
pointing errors have a larger amplitude under daytime and high wind
conditions, sometimes exceeding an arcminute, but can also be very
effectively reduced under good conditions by a reference pointing
strategy to better than about 5 arcsec. Beam squint of the JVLA main
beam has been accurately measured \citepads{2008A&A...486..647U} and
would contribute $\epsilon_Q = 0.055$ if not modelled in an
observation. The magnitude of main beam modulation with frequency of
the VLA has not been documented, but the similarity in feed geometry
to that of the ATCA suggests that a comparable value of $\epsilon_B =
0.05$ will likely apply when this modulation is not explicitly
accounted for in the self-calibration process. The
primary-subreflector separation is, $l_C = 8.2$ m. The near-in
sidelobe level of the main beam is about 2\%, so $\epsilon_S = 0.02$,
if such sources, including their apparent time variability are not
accounted for in the self-cal model of the field. In view of the
significant azimuthal variation in the sidelobe properties, and their
rotation with parallactic angle, accounting for these effects adds
substantial complexity to the processing
\citepads[eg.][]{2008A&A...487..419B}. The far sidelobe contributions
under daytime, $\sigma_D$, and nighttime, $\sigma_N$, conditions
follow directly from the dish diameter and eqn.~\ref{eqn:epf}, which
has been calibrated with $\eta_F = 0.1$ from
\citetads{2002.Dhawan} and eqns. \ref{eqn:sigd} and \ref{eqn:sign}.
The magnitude of the residual source modelling errors will depend on
the methods and effort that have been employed to minimise these
effects.

\begin{figure*}
\resizebox{\hsize}{!}{\includegraphics[bb=40 180 535 660]{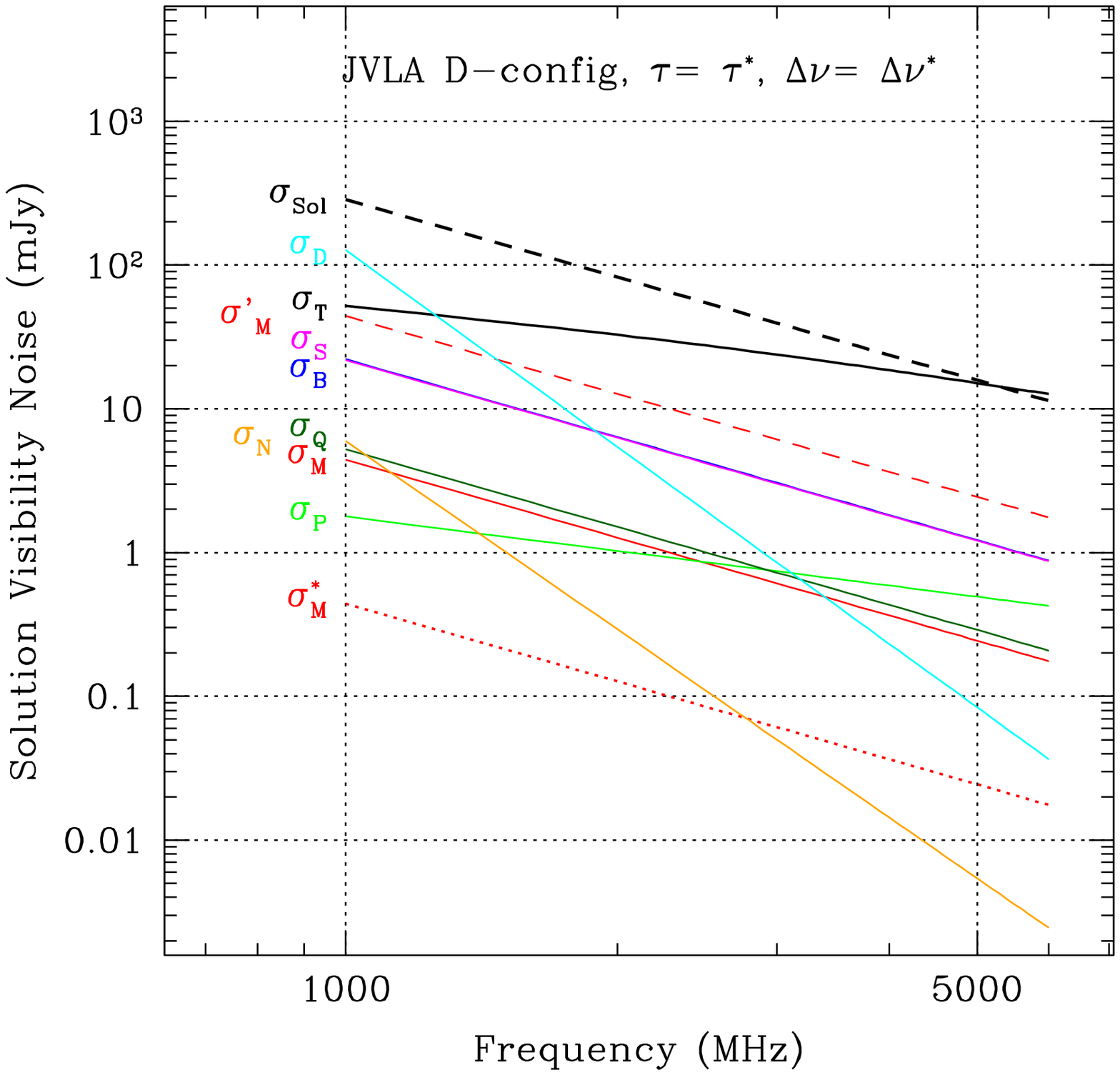},\includegraphics[bb=40 180 535 660]{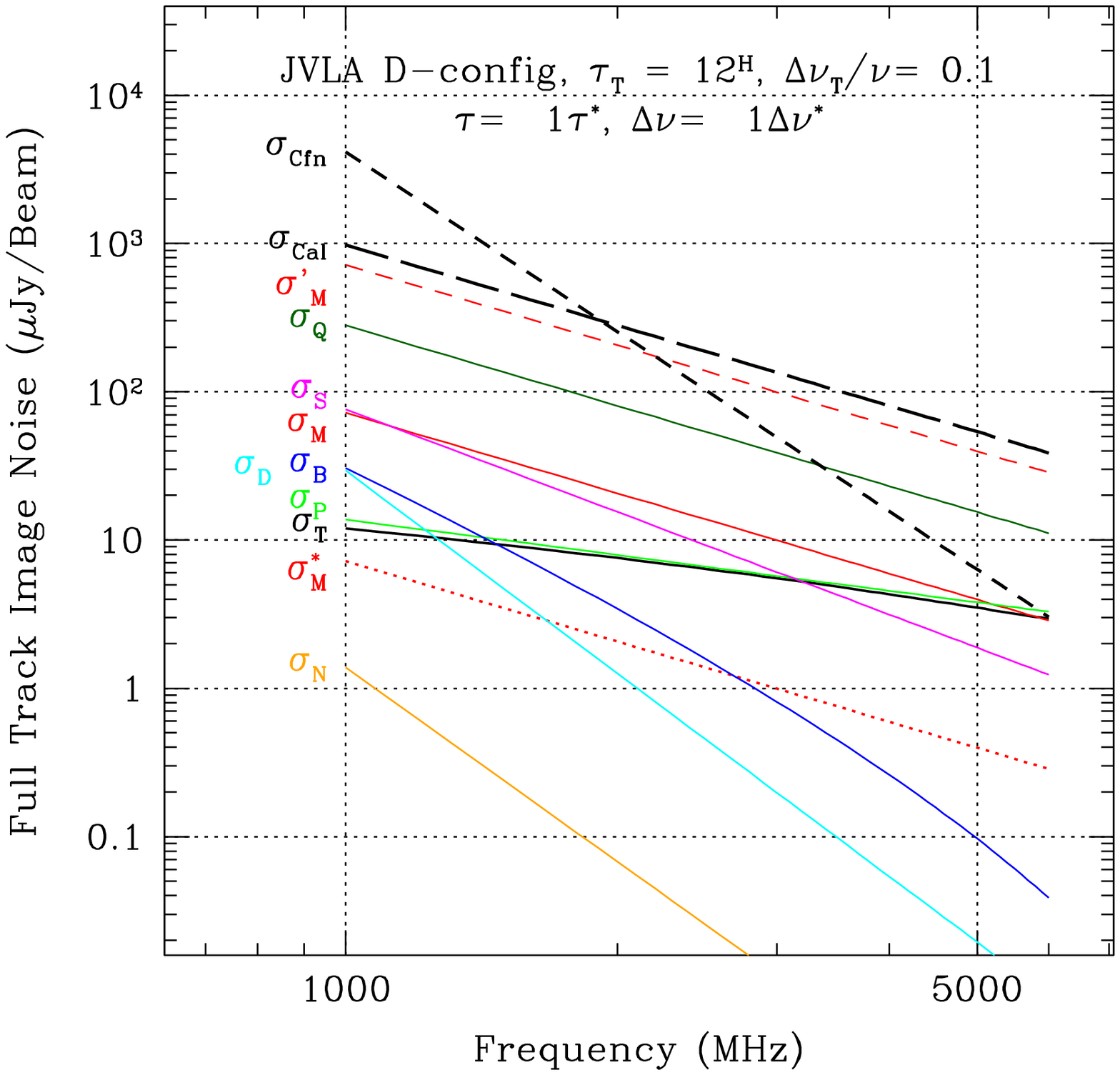},\includegraphics[bb=40 180 535 660]{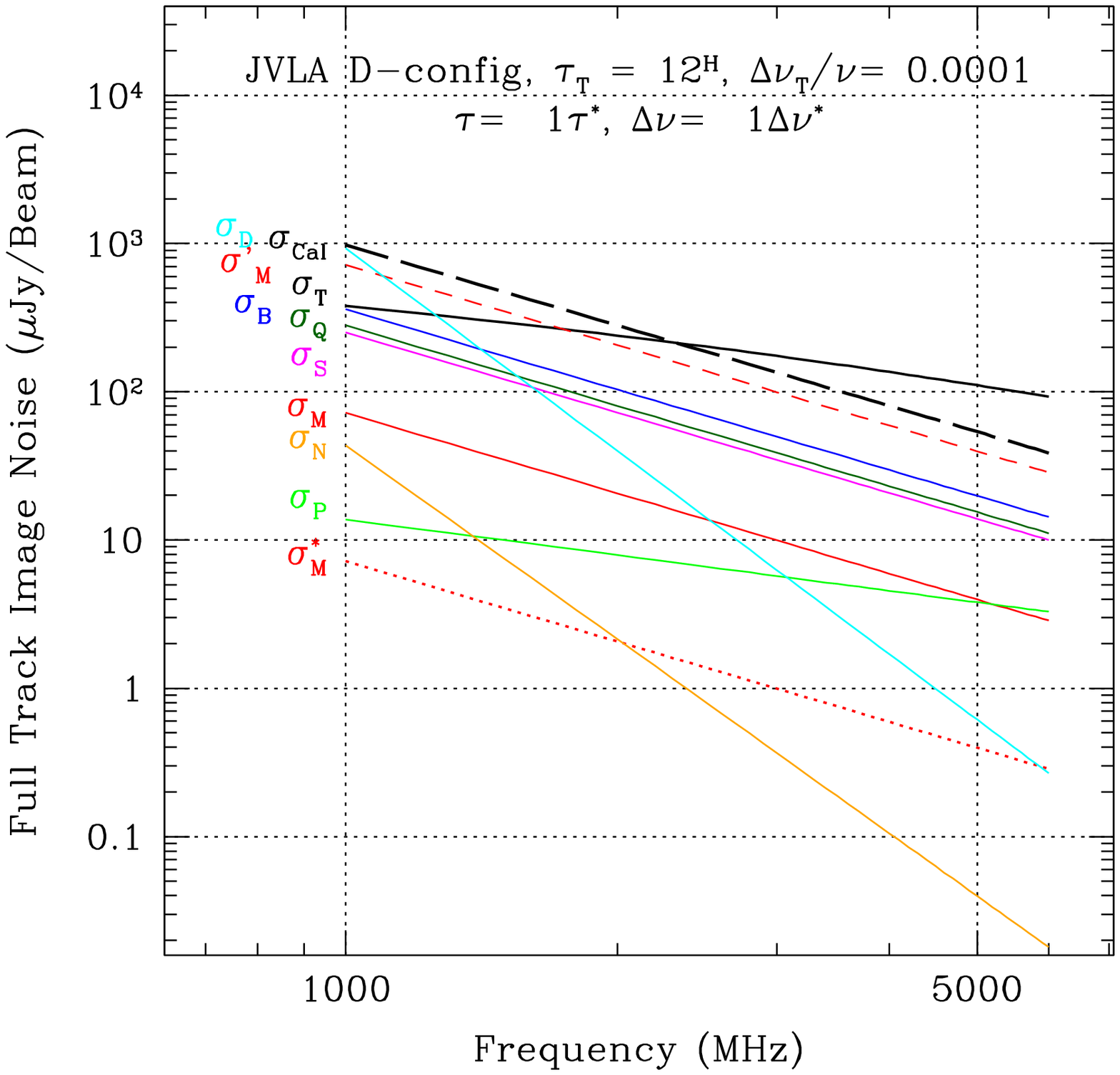}}
\caption{Noise budget for the JVLA-D configuration for a single
  solution interval (left) for a full track continuum observation
  (centre) and for a full track spectral line observation
  (right). Individual noise contributions are as defined in the text.}
\label{fig:vlad}
\end{figure*}

The noise budget for the JVLA in its D configuration is shown in
Fig.~\ref{fig:vlad}. On the solution timescale appropriate to keep
smearing effects below 10\% of the synthesised beamwidth the thermal
noise, $\sigma_T$, is below that which is required for self-cal
convergence $\sigma_{Sol}$. This implies that full-field self-cal can
be performed with minimal smearing effects. As will become apparent
below, this condition is often not met in practise by telescope
systems, so that more extensive data averaging becomes necessary. The
thermal noise is seen to dominate over all noise contributions except
for daytime observations below about 1.4~GHz, where visibility
modulation by the Sun becomes comparable. A full track continuum
observation is dominated by source confusion noise, $\sigma_{Cfn}$,
below about 1.5 GHz. However, the calibration noise, $\sigma_{Cal}$,
vastly exceeds the thermal noise and becomes dominant overall at
higher frequencies, implying a clear need for self-calibration. The
next largest contributions come from source modelling errors (in the
case, $\sigma_M^\prime$, where only 10\% precision is achieved),
followed by beam squint $\sigma_Q$, near-in sidelobe $\sigma_S$ and
main beam modulation $\sigma_B$. Sky-limited continuum performance
above about 2 GHz will rely on achieving a high source modelling
precision and explicit beam squint compensation. In the case of a full
track spectral line obervation, the thermal noise is seen to dominate
above 2--3 GHz, although self-calibration will be necessary at lower
frequencies. Solar modulations will have a negative impact on daytime
spectral line performance at frequencies below about 1.5
GHz. Ultra-deep spectral line observations will ultimately become
limited by the stationary error contributions as noted above. The most
important of these are $\sigma_M^\prime$, $\sigma_S$, $\sigma_B$ and
$\sigma_Q$. Even so, it should be possible to improve upon the single
track sensitivity by about a factor of several in long
integrations, provided the source modelling precision is better than
about 3\% and most importantly, that RFI is
negligible.

\begin{figure*}
\resizebox{\hsize}{!}{\includegraphics[bb=40 180 535 660]{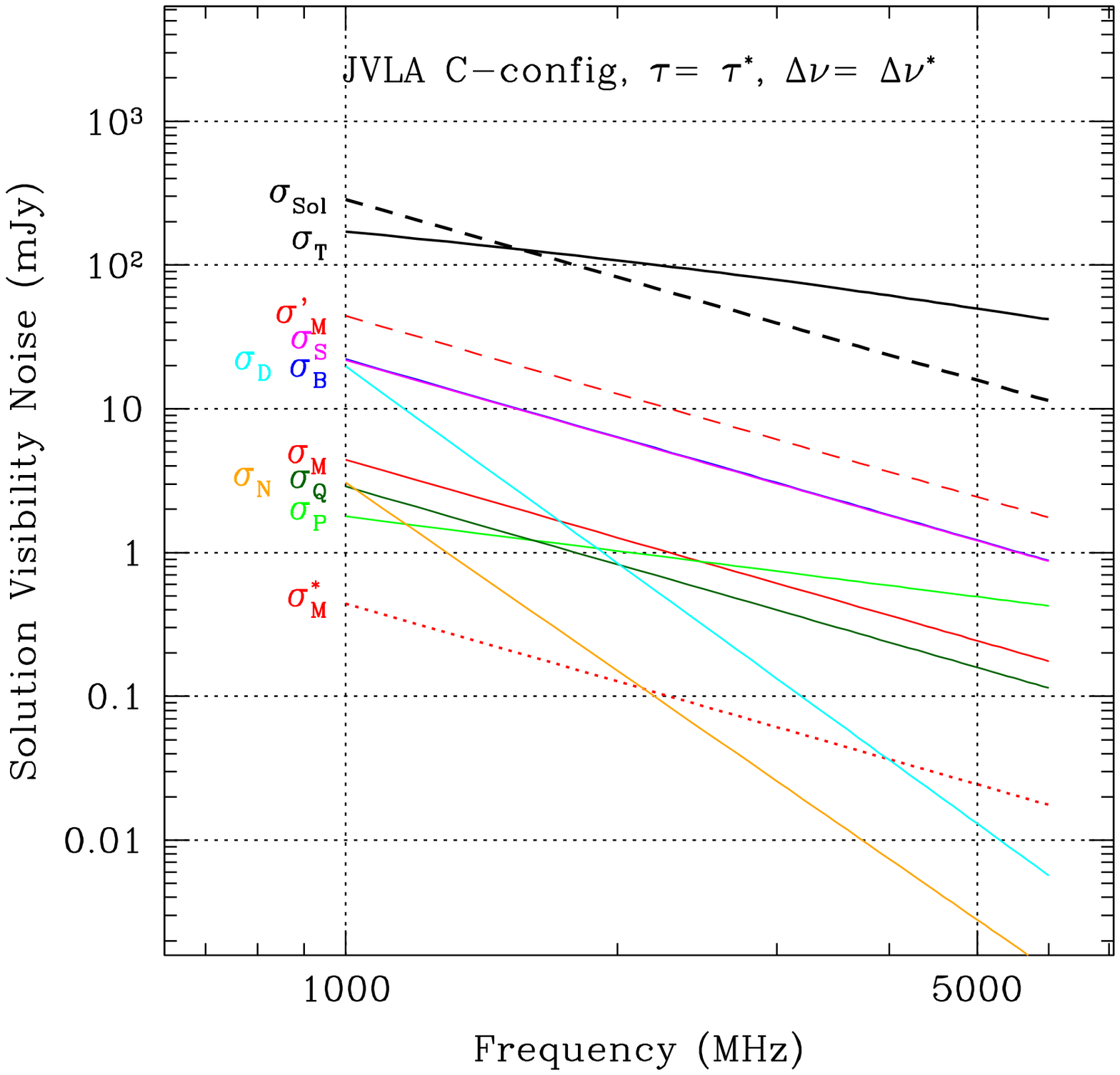},\includegraphics[bb=40 180 535 660]{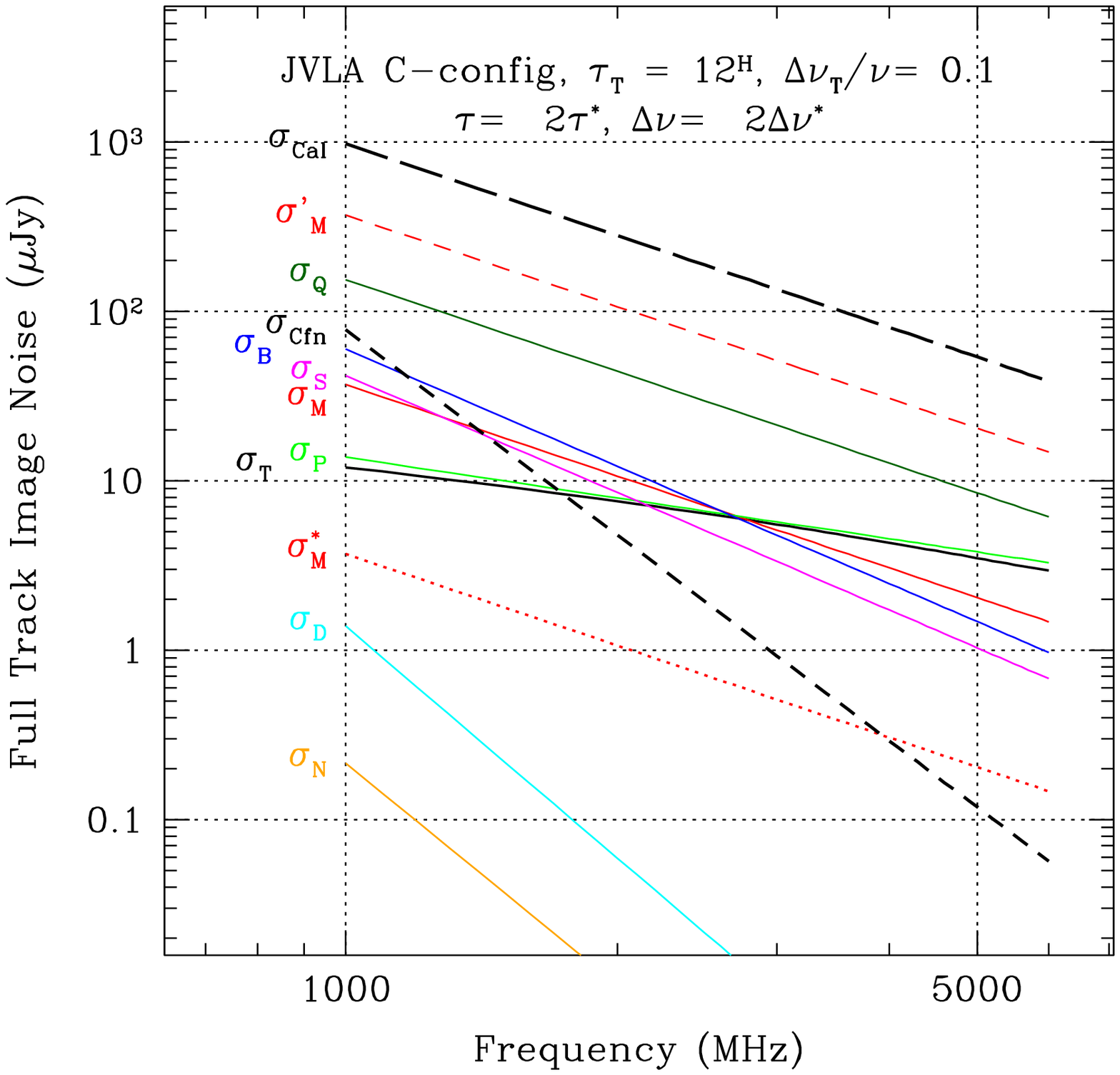},\includegraphics[bb=40 180 535 660]{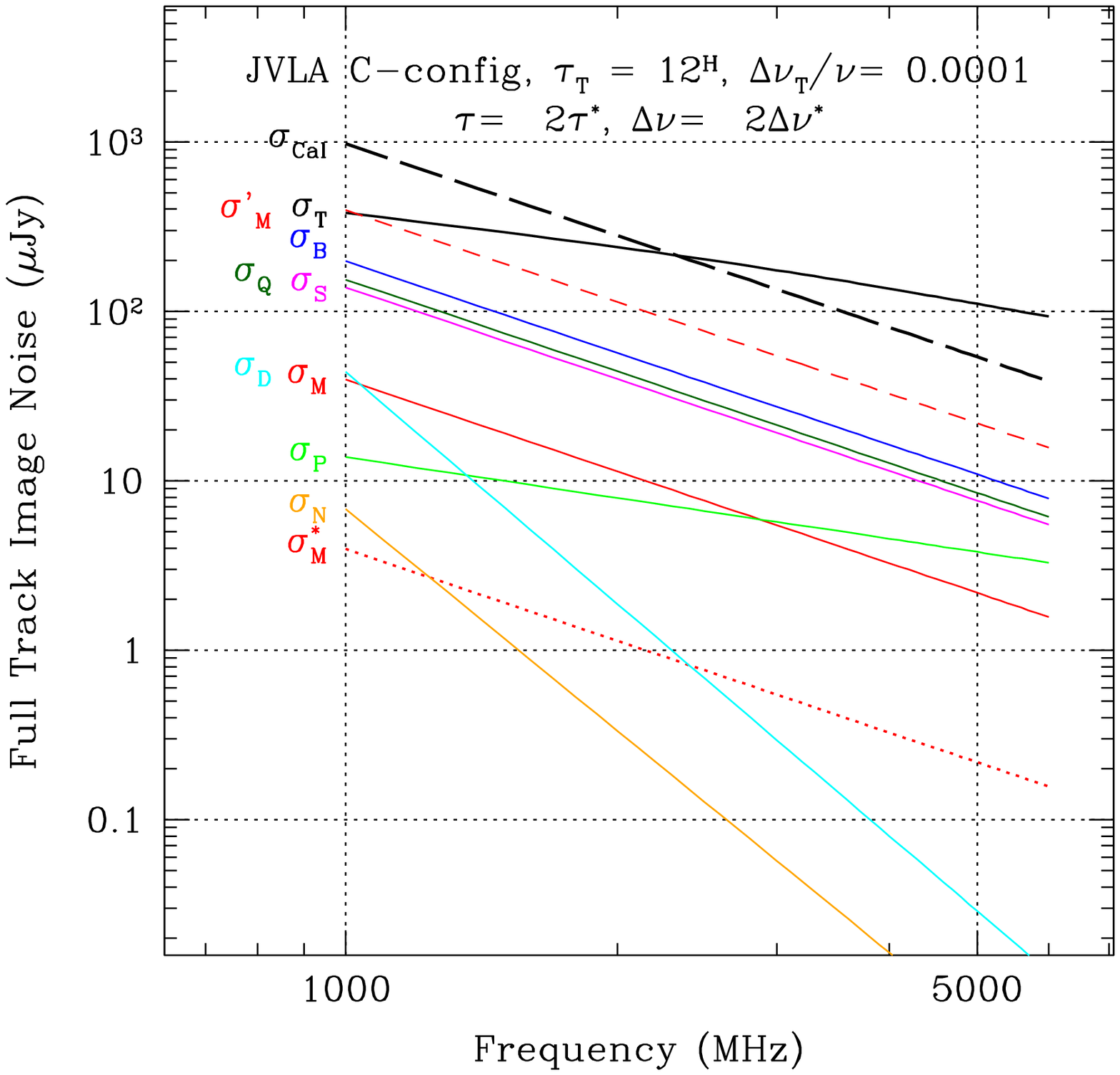}}
\caption{Noise budget for the JVLA-C configuration as in
  Fig.~\ref{fig:vlad}. }
\label{fig:vlac}
\end{figure*}

The same type of analysis applied to the JVLA-C configuration is given
in Fig.~\ref{fig:vlac}. Here we see that while the thermal noise
dominates over other contributions on the desired solution timescale,
it exceeds what is required for self-cal convergence $\sigma_{Sol}$,
so that more extensive data averaging, by about a factor of two in
both domains, becomes necessary. This will imply more extensive time
and bandwidth smearing for off-axis sources, which may limit the
source modelling precision which can routinely be acheived. The full
track continuum observation will require self-calibration to reach the
thermal noise. Substantial effort needs to be expended on achieving a
source modelling precision approaching 1\%, otherwise this contribution
will dominate the noise budget below 3 GHz. The next most serious
effect is that of beam squint. As noted previously, active modelling of
this effect has proven effective at improving image dynamic range at
the expense of computational complexity
\citepads{2008A&A...486..647U}. By addressing both of these issues
satisfactorily it should become possible to reach the thermal
noise. Reaching the source confusion level $\sigma_{Cfn}$ in deep
integrations will be considerably more challenging, since several
additional error contributions would need to be overcome.  The full
track spectral line observation is thermal noise limited above 2--3
GHz, while self-calibration with at least moderate source modelling
precision ($<10\%$) will be required at lower frequencies. Even deep
integrations should still be in this regime, provided that RFI remains
negligible.

\begin{figure*}
\resizebox{\hsize}{!}{\includegraphics[bb=40 180 535 660]{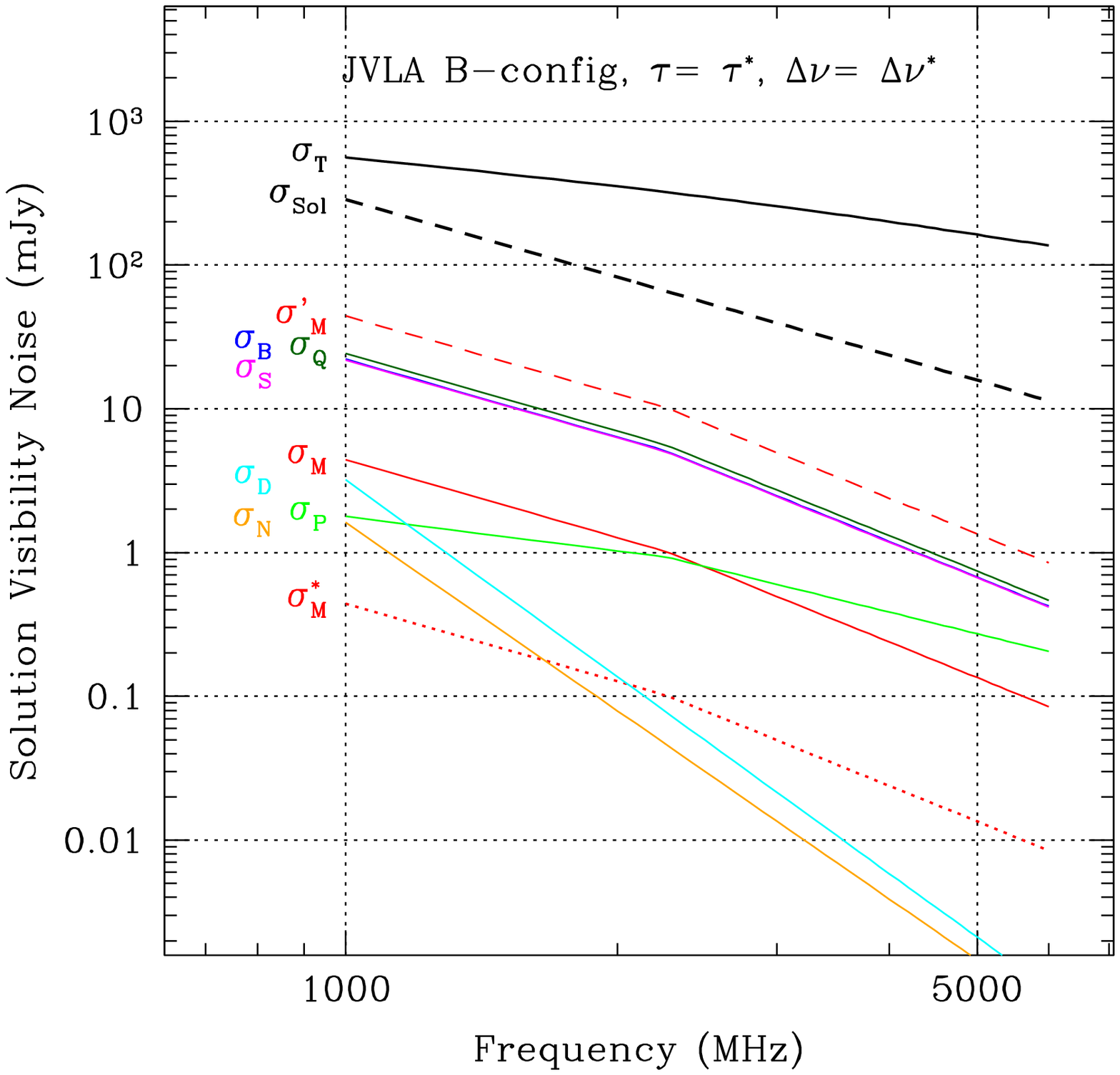},\includegraphics[bb=40 180 535 660]{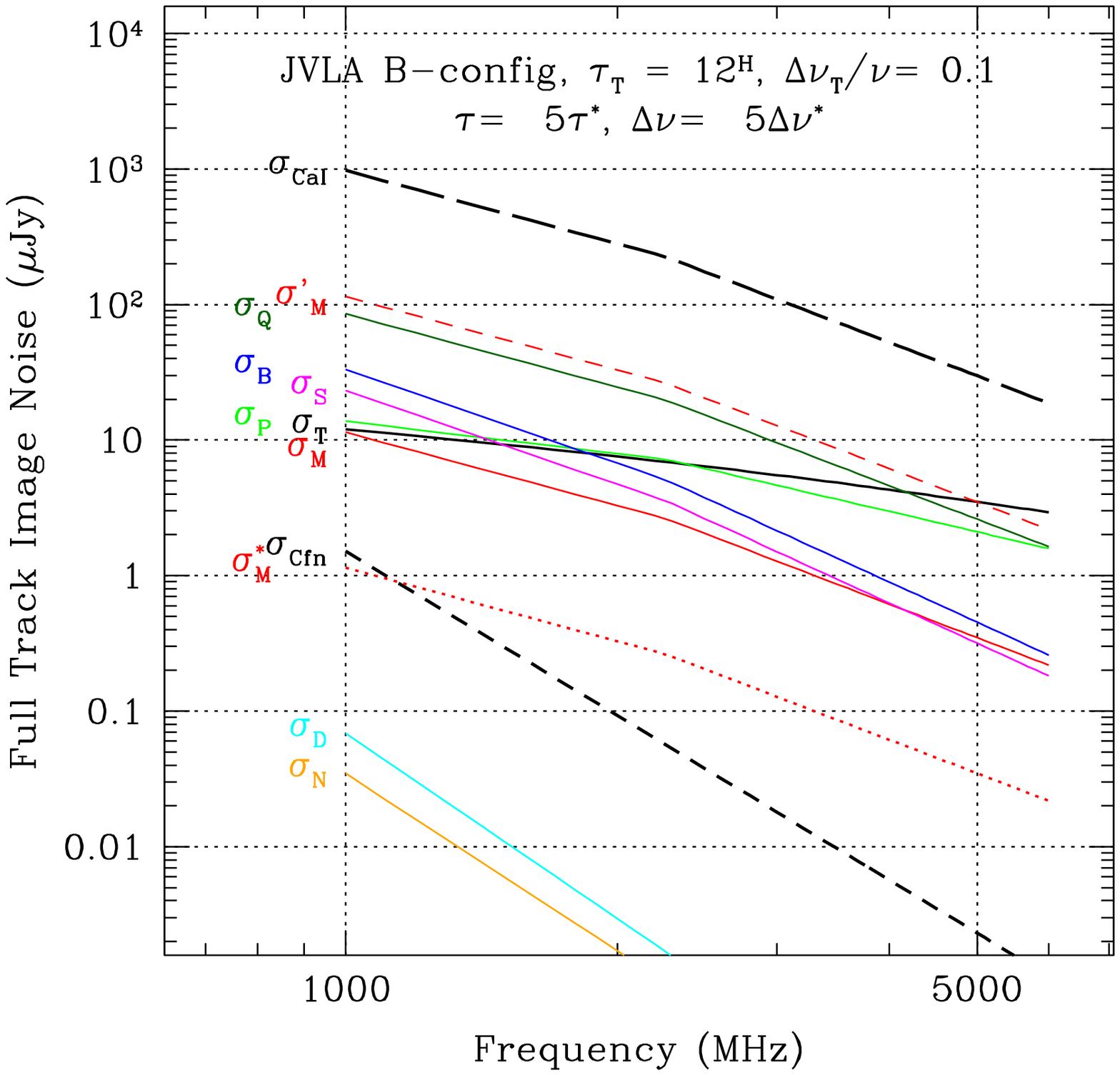},\includegraphics[bb=40 180 535 660]{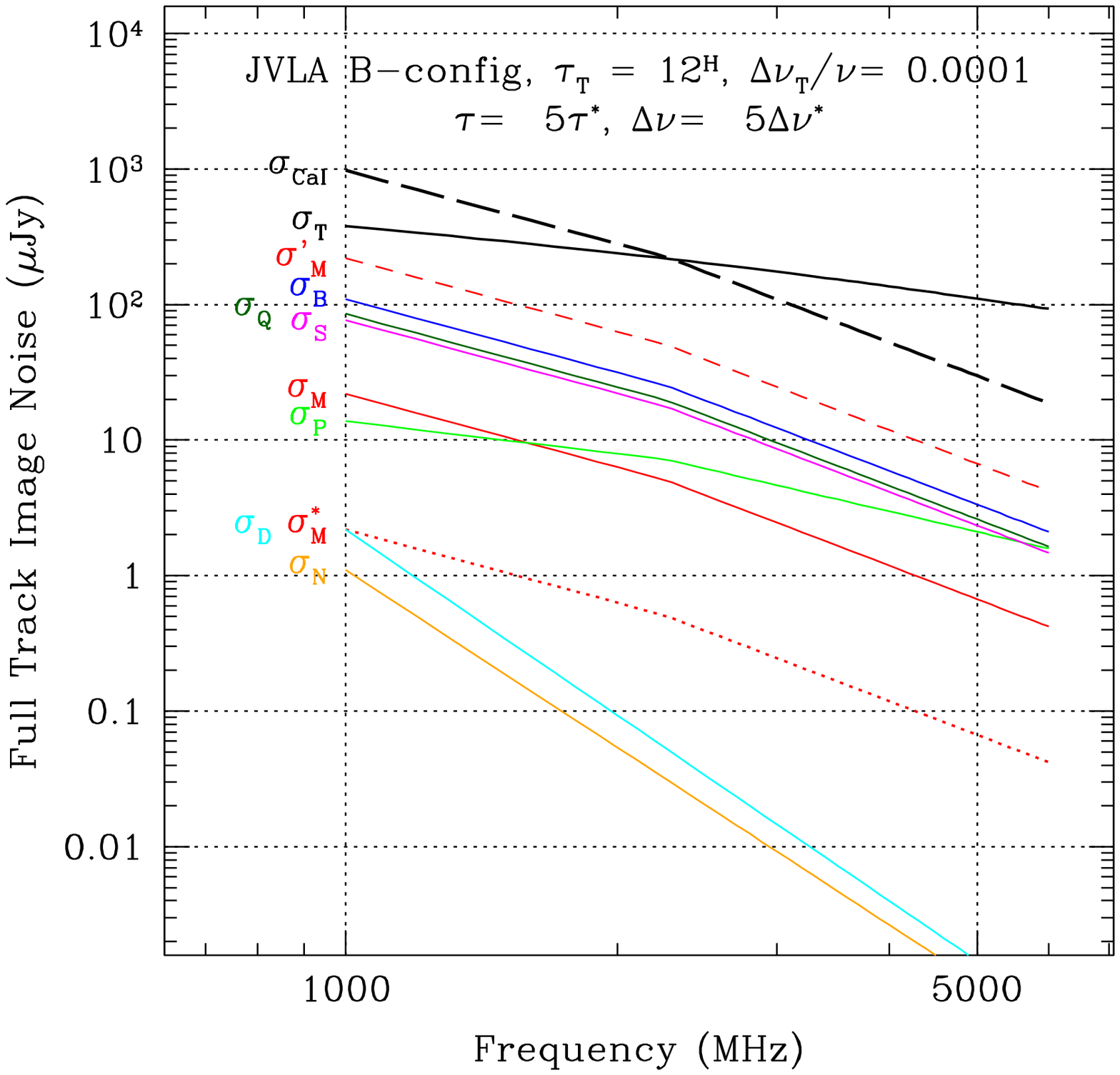}}
\caption{Noise budget for the JVLA-B configuration as in
  Fig.~\ref{fig:vlad}.}
\label{fig:vlab}
\end{figure*}

\begin{figure*}
\resizebox{\hsize}{!}{\includegraphics[bb=40 180 535 660]{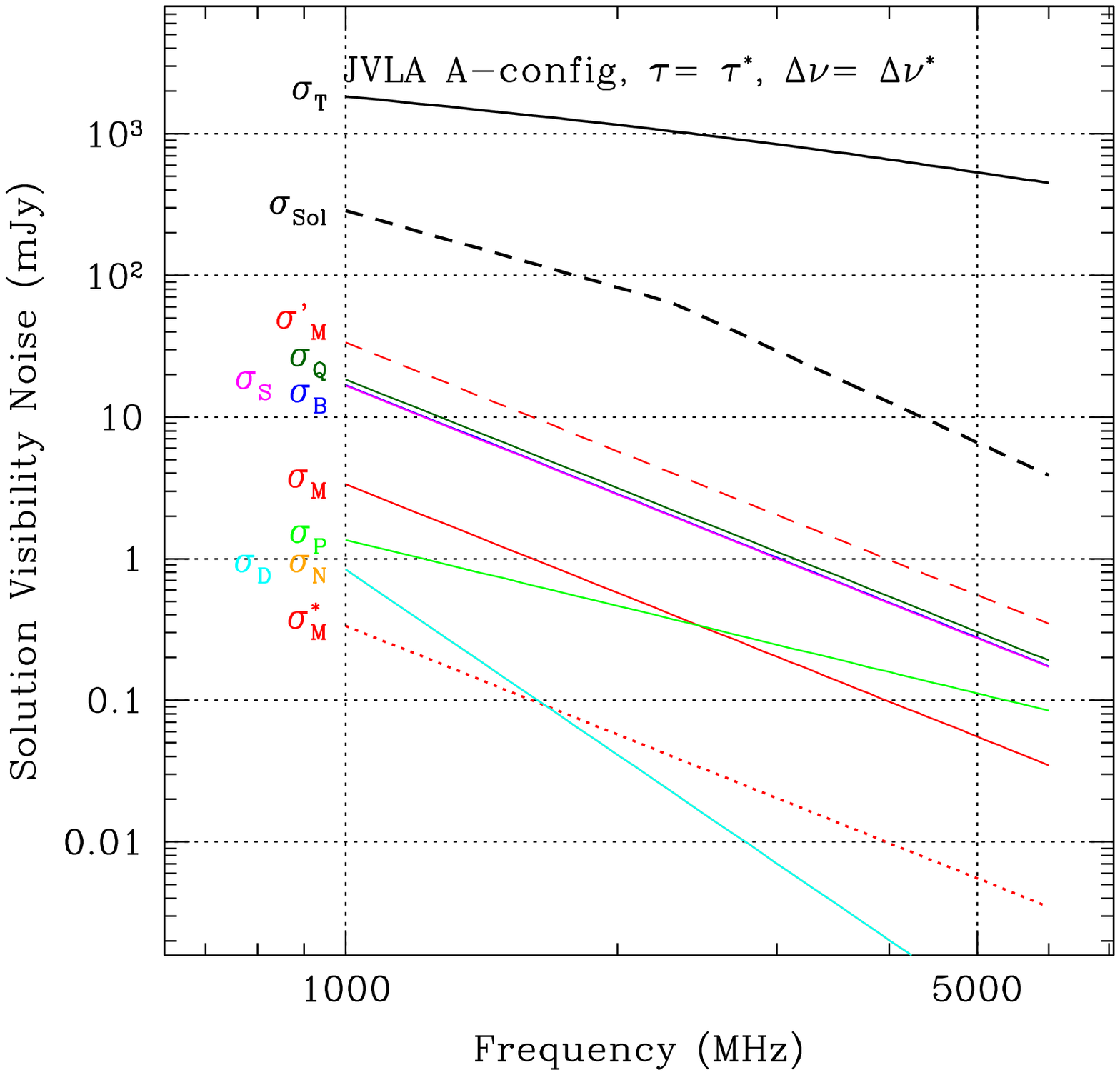},\includegraphics[bb=40 180 535 660]{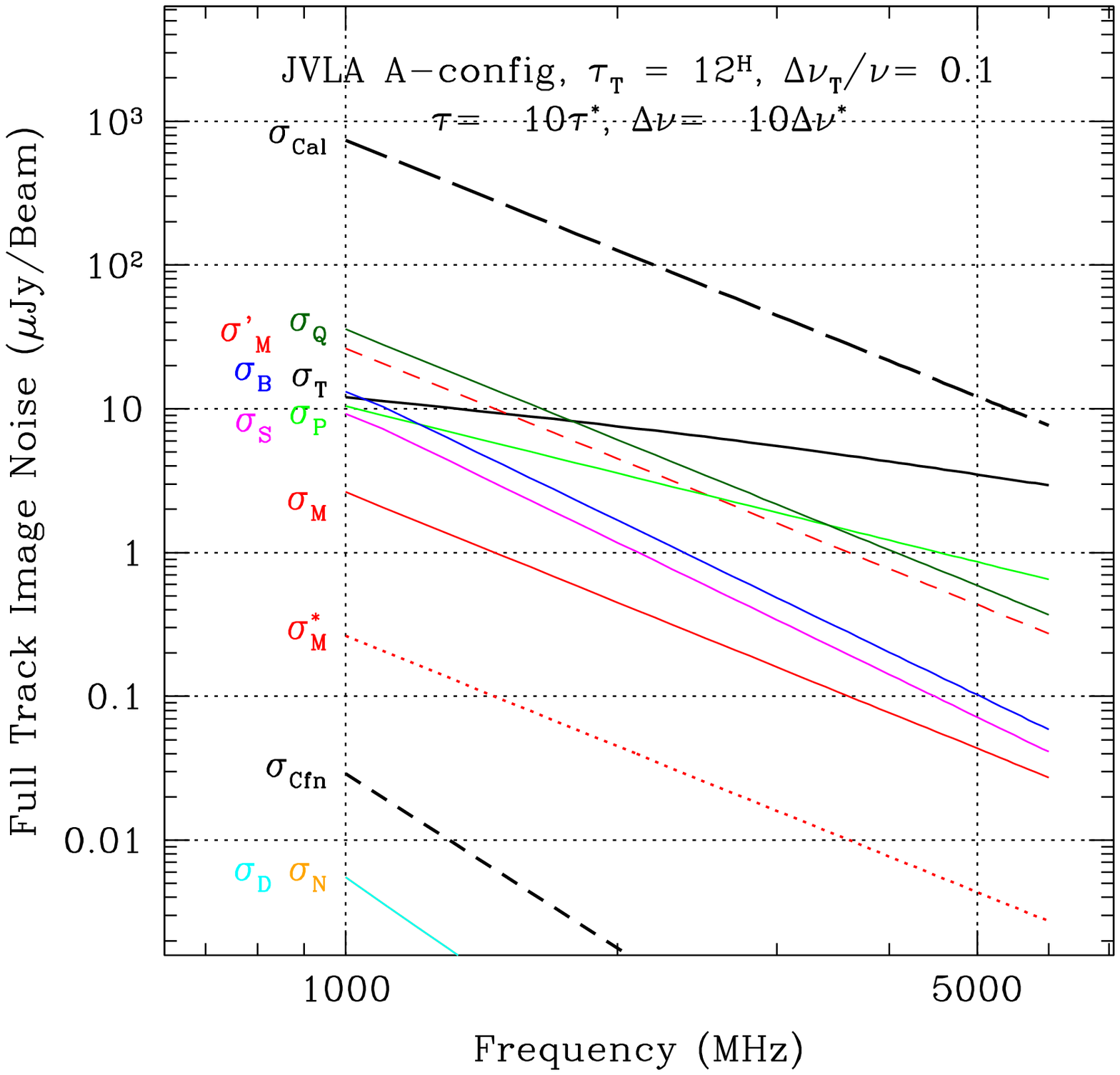},\includegraphics[bb=40 180 535 660]{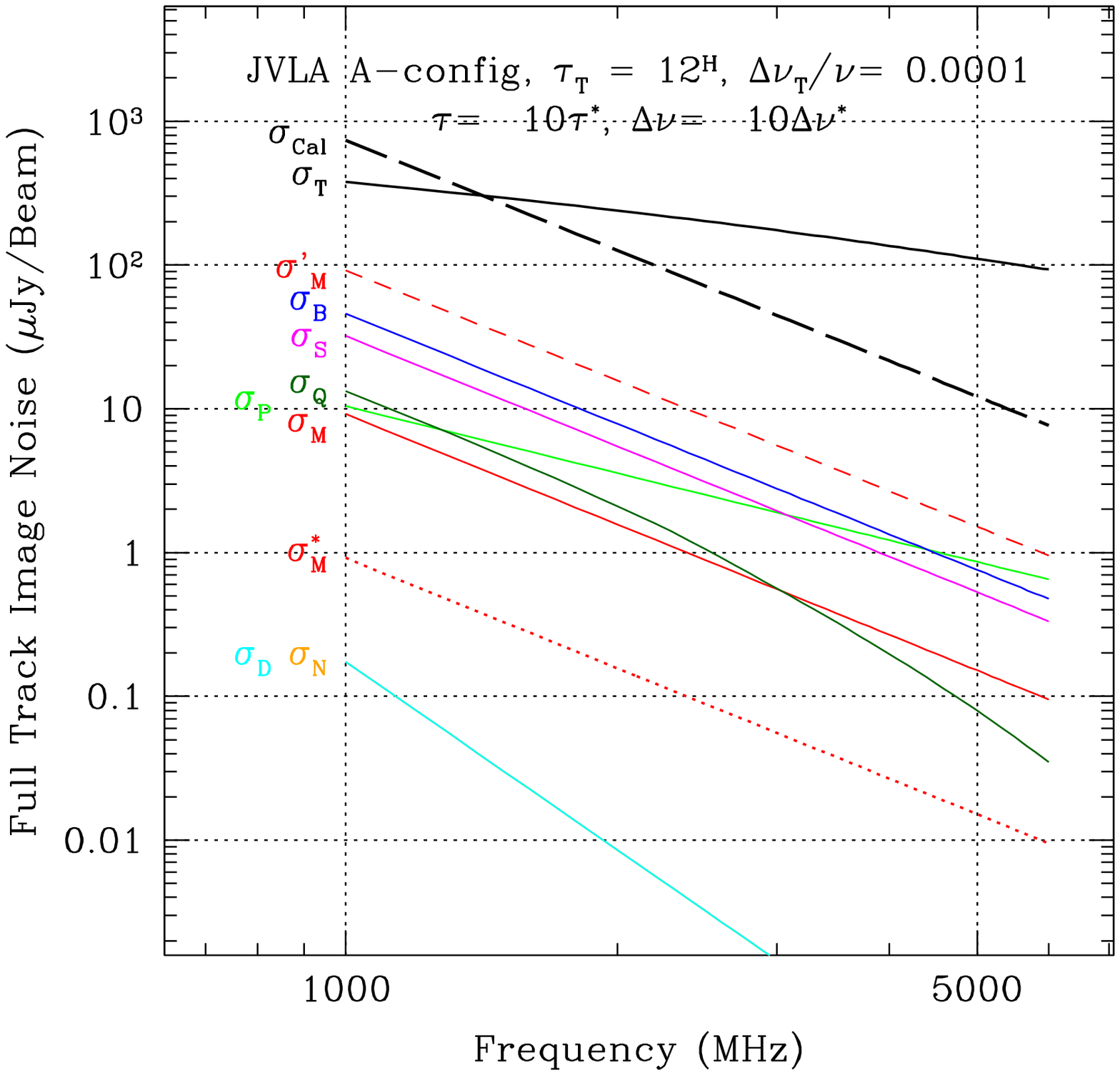}}
\caption{Noise budget for the JVLA-A configuration as in
  Fig.~\ref{fig:vlad}.}
\label{fig:vlaa}
\end{figure*}

Very similar patterns are seen in the more extended B and A
configurations as shown in Figs.~\ref{fig:vlab} and
\ref{fig:vlaa}. The more extended configurations require smaller time
and frequency averaging intervals to support the iterative full-field
source modelling inherent to self-calibration, but the finite
sensitivity per visibility does not allow this to be achieved while
providing enough signal-to-noise. Excess averaging in both domains by
about a factor of five (and ten) are required for the B (and A)
configuration(s). This will severely impair the ability to achieve
high levels of modelling precision on the main beam flanks for the
simplest imaging algorithms. More complex wide-field imaging and
modelling strategies will be mandatory under these circumstances, such
as employment of multiple distributed phase centres. While such
methods are well-understood, they impose a significant computational
overhead. The full track continuum performance relies on
self-calibration and beam squint effects will also need to be
addressed to reach the thermal noise. Source modelling precision
better than about 1\% will be required, together with explicit
modelling of the full range of main beam and near-in sidelobe effects
($\sigma_Q$, $\sigma_P$, $\sigma_B$ and $\sigma_S$) if deep
integrations are to remain in that regime. The challenges in all these
areas are greatest at the lowest frequencies and for the less extended
configuration.

\subsection{ATA}

The Allen Telescope Array (\citetads{2009IEEEP..97.1438W},
\citetads{5722985}) consists of
$N = 42$ off-axis $(alt,az)$ mounted dishes of $d=6.1$~m. The ATA
configuration is very compact with $B_{Max} = 323$~m and $B_{Med} =
81$~m. The dishes are fed by an ultra-wideband log-periodic feed from
a Gregorian focus over the frequency range 0.5 to 10 GHz. A model for
the system temperature performance of the system as function of
frequency presented by \citetads{2009IEEEP..97.1438W} provides,
\begin{equation}
T_{Sys} = 19.7 + 4\nu_G^{-0.5} +9.5\nu_G^{0.5}+0.8\nu_G + 3\nu_G^{-2/7} 
 \quad {\rm K},
\label{eqn:atatsys}  
\end{equation}
in terms of the frequency in GHz, $\nu_G$, at an assumed aperture
efficiency, $\eta_A = 0.6$. Typical values of receiver temperature and
illumination efficiency tabulated by \citepads{5722985} suggest
significantly higher values of $T_{Sys}/\eta_A$ between about 70 and
270~K.  The values of $T_{Sys}$ given by eqn.~\ref{eqn:atatsys} range
between about 37 and 60 K, with corresponding SEFDs between 5800 and
9400 Jy. The pointing accuracy of the dishes under good nighttime
conditions is $P = 90$ arcsec with a likely correlation timescale
$\tau_P = 15$ minutes. The beam squint for this offset design has been
measured \citepads{5722985} to be about 4\% FWHM at 1.4 GHz, yielding
$\epsilon_Q = 0.04$.  Major asymmetries in the main beam shape, at the
20\% level, are also documented graphically by \citetads{5722985},
suggesting that a beam squash contribution to $\epsilon_Q$ may be
significantly larger. The peak near-in sidelobe levels measured
at 2300 MHz are about 1\%, yielding $\epsilon_S = 0.01$, with
substantial azimuthal variation in sidelobe intensity. While beam and
sidelobe modulation with frequency have not been documented directly,
they might be expected to track the gain modulation with frequency
intrinsic to the log-periodic feed, which has an RMS fluctuation level
of about 1 dB, or $\epsilon_B^\prime \approx 0.25$ and correlation
bandwidth, $\Delta\nu/\nu \approx 0.03$. This particular gain
modulation is due to the quantised, resonant element wavefront
sampling of the wide-band feed and is in addition to any multi-path
induced gain modulation with frequency as discussed
previously. Multi-path beam modulation should have a much lower
amplitude, of perhaps $\epsilon_B \approx 0.01$ (twice the GBT value).
Since the feed is mounted on the edge of the primary reflector
surface, the primary-subreflector separation of about $l_C \approx3$~m
is likely the most relevant cavity dimension for an estimate of
modulation periodicity. For the purpose of our analysis, we will
consider both of these possibilities for main beam modulation
$\epsilon_B$. \citetads{5722985} have also carried out far sidelobe
measurements of the ATA dishes at 2300 MHz which are in the range of
-30 to -40 dB over much of the sky. This implies a value of $\eta_F =
0.7$ for use in eqn.~\ref{eqn:epf} which is significantly higher than
that which applies to the upgraded JVLA.

\begin{figure*}
\resizebox{\hsize}{!}{\includegraphics[bb=40 180 535 660]{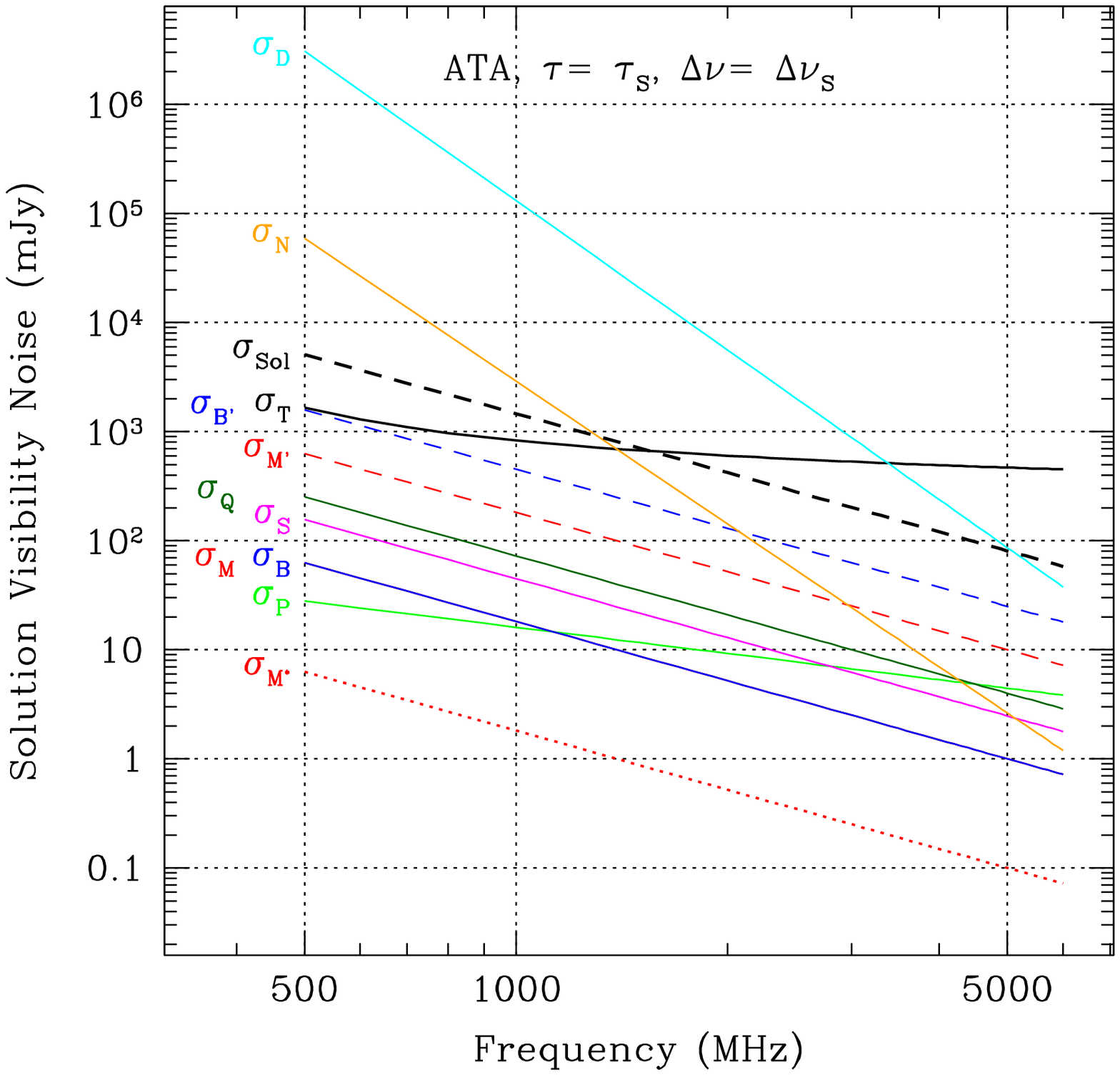},\includegraphics[bb=40 180 535 660]{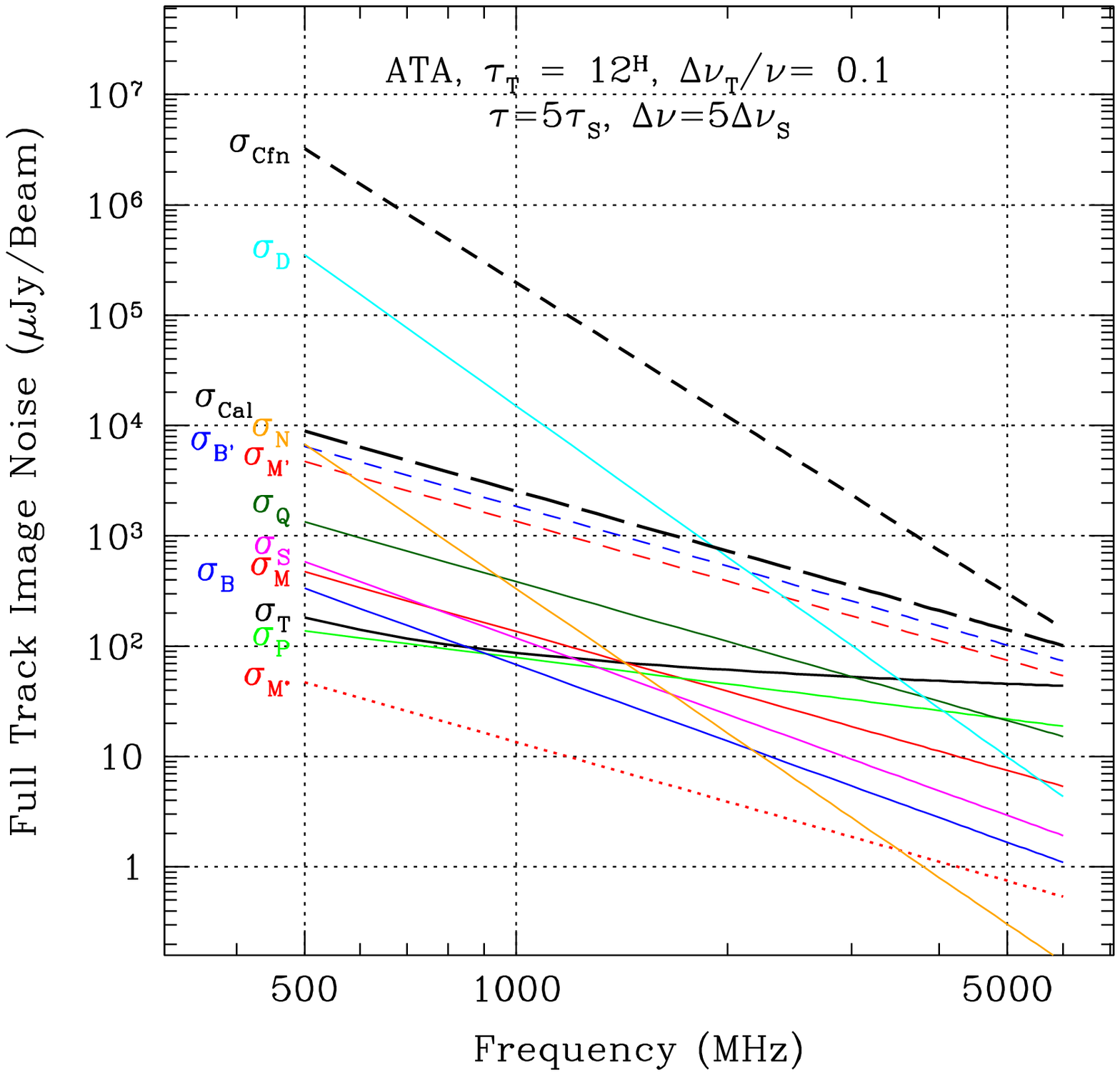},\includegraphics[bb=40 180 535 660]{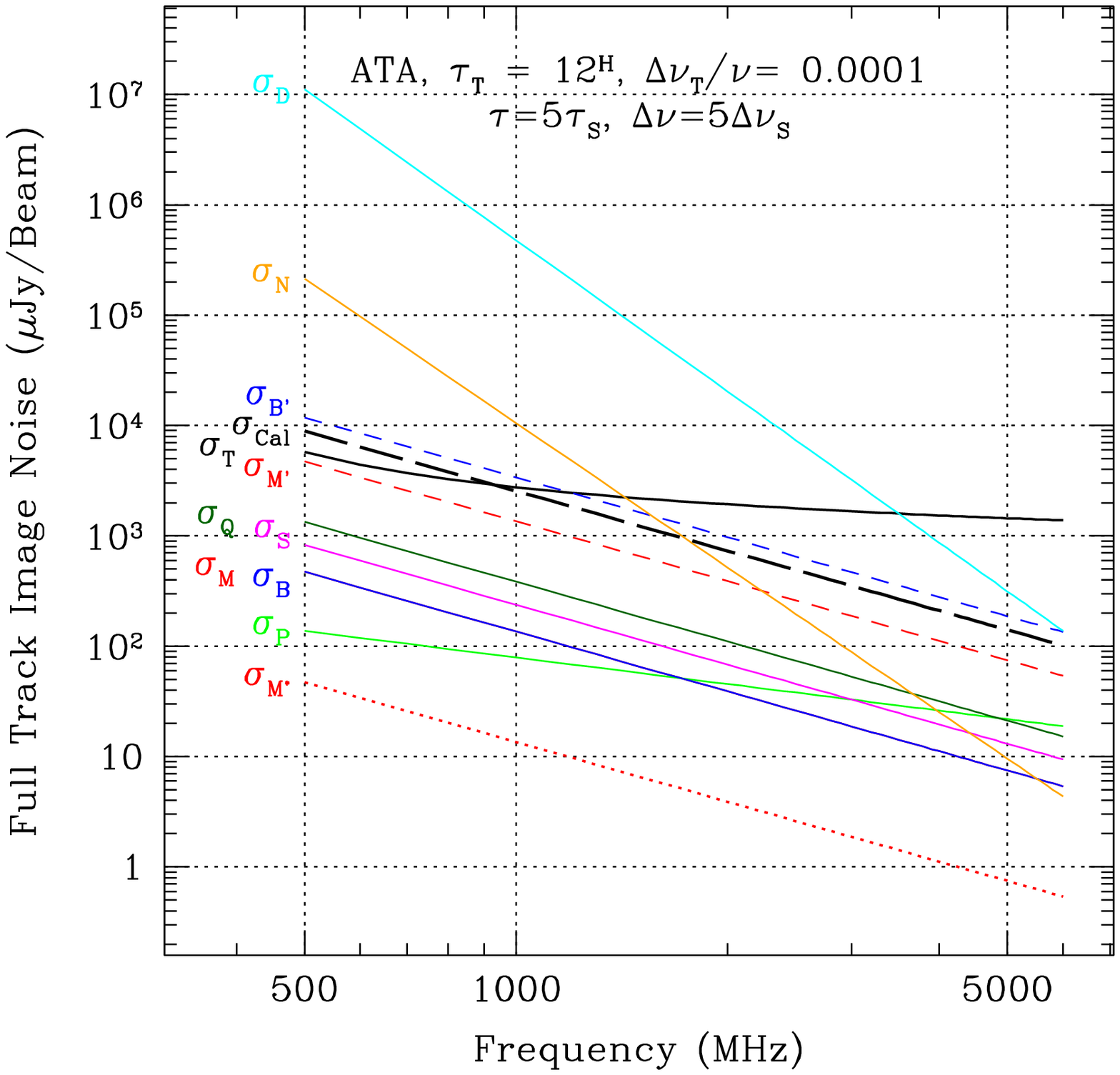}}
\caption{Noise budget for the ATA telescope as in
  Fig.~\ref{fig:vlad}.}
\label{fig:ata}
\end{figure*}

The noise budget analysis for the ATA telescope is shown in
Fig.~\ref{fig:ata}. What is striking is that visibility fluctuation
levels are predicted to be completely dominated by the far sidelobe
response on the self-cal solution timescale. During daytime observing
this is true up to 3 GHz, while even for nighttime observing this will
be the case below 1.4 GHz. This is a consequence of both the small
dish size and the very compact array configuration.  Substantial
visibility averaging (by about a factor of five in both domains) would
diminish these contributions to the point that the self-cal
signal-to-noise criterion could be met. A full track continuum
observation would be source confusion noise limited, but many of the
noise contributions under consideration would exceed the thermal noise
below about 1 GHz. Above 1 GHz, the potential main beam modulation
associated with gain variations of the wide-band log-periodic feed
becomes the most important contributor for nighttime observations. The
full track spectral line case is dominated by far sidelobe effects
below a few GHz. The nominal calibration noise exceeds the thermal
noise below about 1 GHz, implying that self-cal may not be required
for spectral line applications at higher frequencies. In that case, it
may be possible to simply ignore the far sidelobe response
contributions, since the visibility gridding function will provide
some 40dB of suppression for such distant sources. Any sidelobes of
the instantaneous synthesised beam response will of course not be
suppressed during imaging which may become a limitation in this
case. Deep spectral line integrations would encounter a range of
issues that would necessitate extensive modelling and 
self-calibration to overcome.

\subsection{ASKAP}

Let us now consider what challenges will be faced with the ASKAP
array, with its $N = 36$, three axis $(alt,az,pol)$ mounted dishes of
$d=12$~m. The ASKAP configuration is highly centrally concentrated
with $B_{Max} = 6$~km and $B_{Med} = 630$~m. The dishes are fed with a
prime focus PAF (phased array feed) that covers the frequency range
700 to 1800 MHz. Current performance estimates are for $T_{Sys}/\eta_A
= 55$~K across this band, yielding an SEFD of 1340 Jy for the 12m dishes.

Mechanical pointing errors of the ASKAP dishes under normal operating
conditions have not yet been determined, but will be assumed to be
about $P = 10$ arcsec under good nighttime conditions and correlated
over timescales of about $\tau_P = 15$ minutes. What must also be
assessed is the electronic pointing error that might be associated with
gain fluctuations of the individual antenna elements that are combined
to produce the main beam in a PAF. While actual values have not yet
been determined, current estimates of the element gain variability are
better than $\epsilon_P^\prime = 0.01$ with variability timescales of
$\tau_P^\prime = 1$ minute.

The main beam modulation phenomenon that afflicts single-pixel-fed
radio telescopes has been demonstrated to be diminished by between 20
and 30 dB in a PAF system \citepads{2010iska.meetE..43O}
employing Vivaldi antenna elements. The reason for this improvement
appears to be related to a much better and physically more extended
power matching to the incident radiation field, such that reflectivity
from the feed is minimised. While similar measurements are still
required for the chequerboard antenna array of the ASKAP phased array
feeds, it is expected that they will have similar performance in this
regard. We will assume a conservative 10 dB reduction in the frequency
modulation, to $\epsilon_B \approx 0.005$.

The phased array feed also provides the potential to strongly suppress
beam squash through optimisation of the individual polarisation beams
and their relative alignment. Furthermore, as noted previously, the
ability to track parallactic angle with the ``polarisation'' axis
allows any residual beam squash pattern to be fixed on the sky during
tracking so that the apparent brightness of sources on the beam flank
remains unchanged. For illustration purposes we will assume a 10 dB
reduction due to each of these contributions to
beam squash from a base level that would apply to an $(alt,az)$ mounted
single pixel feed of 4\%, giving $\epsilon_Q = 0.0004$. 

The near-in sidelobe pattern of a phased-array feed illuminated
reflector is expected to have a comparable near-in sidelobe level to
that of the single-pixel-fed equivalent
\citepads{2010iska.meetE..43O}, so is likely to have a peak amplitude
of some 2\%. However, unlike the single-pixel-fed counterpart, this
pattern is not expected to display an amplitude modulation with
frequency. As noted above for the main beam modulation, the phased
array feed illumination pattern is expected to be very similar over
frequency rather than modulated at the 5\% level. Moreover, the
polarisation rotation axis of the ASKAP telescope mount will permit
all of the telescope sidelobes, including the radial spokes associated
with the prime focus support struts, to remain stationary on the sky
during source tracking. The time and frequency stability of the
near-in sidelobe response would enable straightforward modelling of
sources at these large offsets. Furthermore, it is planned that all 36
digital main beams formed by the PAF will be processed simultaneously
via a joint self-cal source model for the entire field of view. The
implication is that source information for modelling the field of each
main beam and its near-in sidelobes is provided by a beam in which
that source is essentially on-axis, with minimal smearing effects or
main beam attenuation. Provided that the relevant time and frequency
smearing effects for each beam are explicitly included during data
comparison, it should permit very low values of both $\epsilon_S$
and $\epsilon_M$ to be obtained. For illustration purposes, we will
assume only a conservative 10 dB reduction in this component to
$\epsilon_S = 0.002$.

For the far sidelobe contributions under daytime, $\sigma_D$, and
nighttime, $\sigma_N$, conditions we will assume $\eta_F = 0.1$ since
we expect excellent illumination quality in this prime focus system,
comparable to what currently applies to the Cassegrain JVLA and the
prime focus WSRT together with eqns.~\ref{eqn:epf}, \ref{eqn:sigd} and
\ref{eqn:sign}. This represents an improvement by perhaps a factor of 2
relative to a single pixel fed dish of this size.

\begin{figure*}
\resizebox{\hsize}{!}{\includegraphics[bb=40 180 535 660]{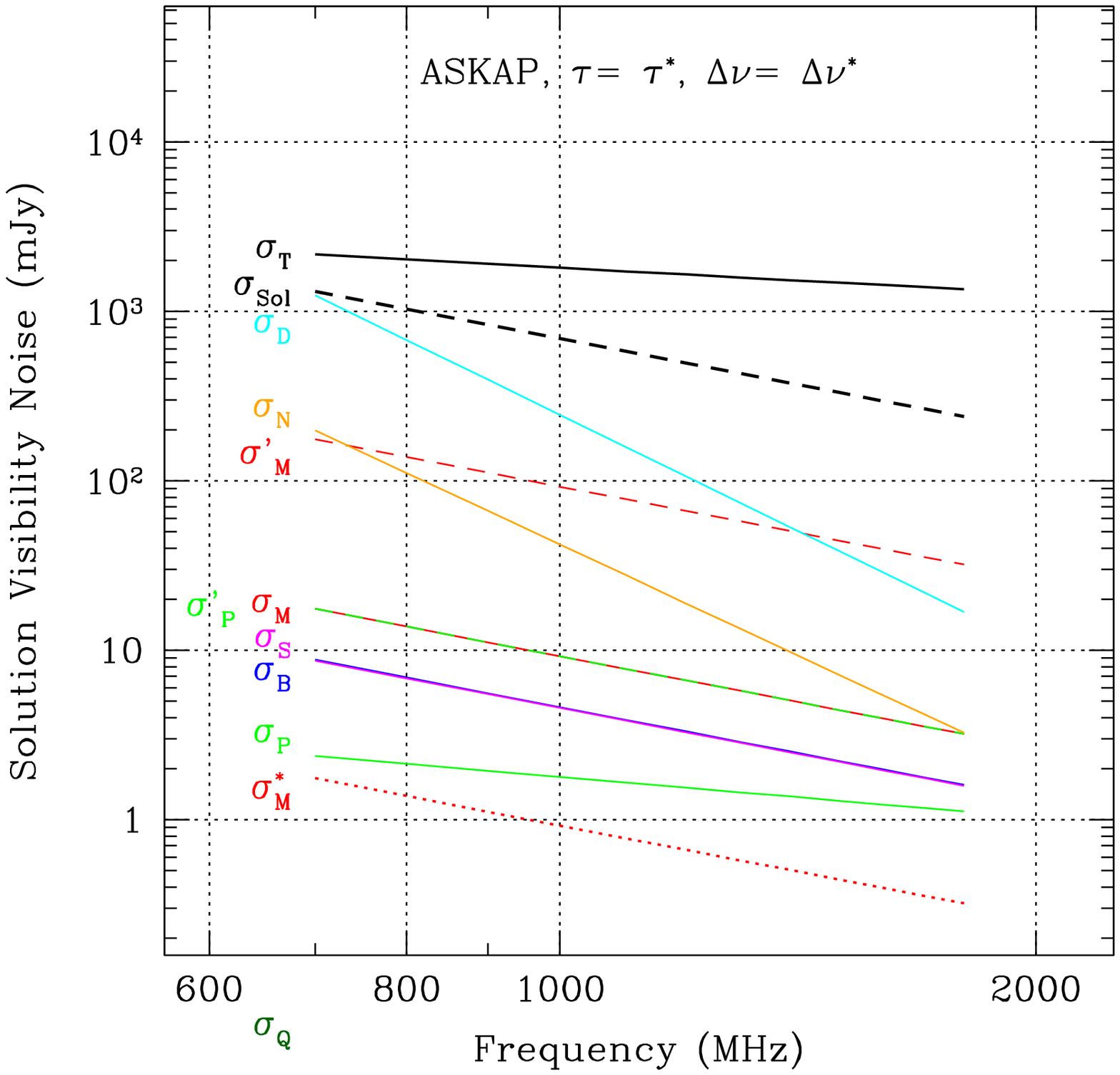},\includegraphics[bb=40 180 535 660]{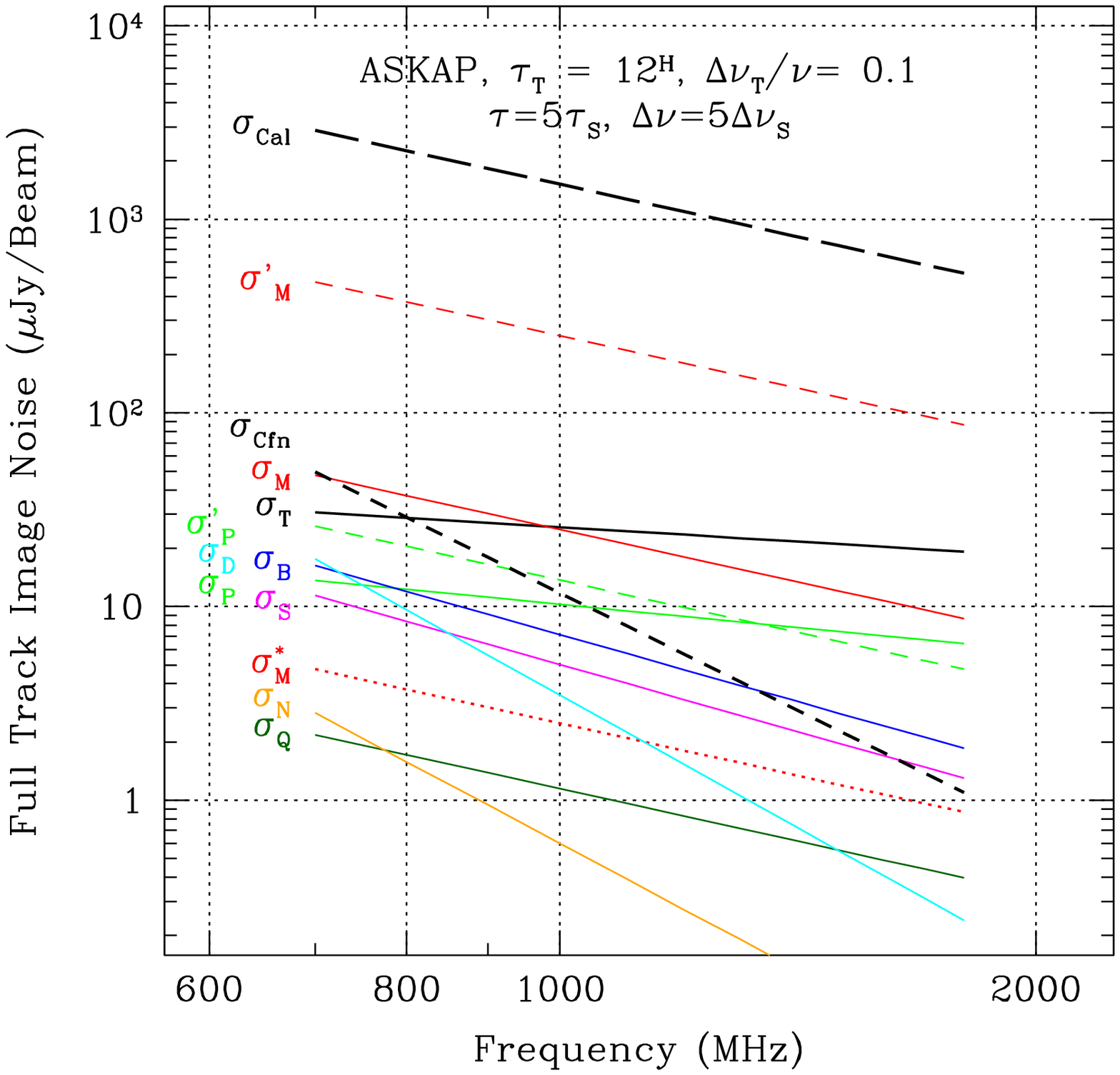},\includegraphics[bb=40 180 535 660]{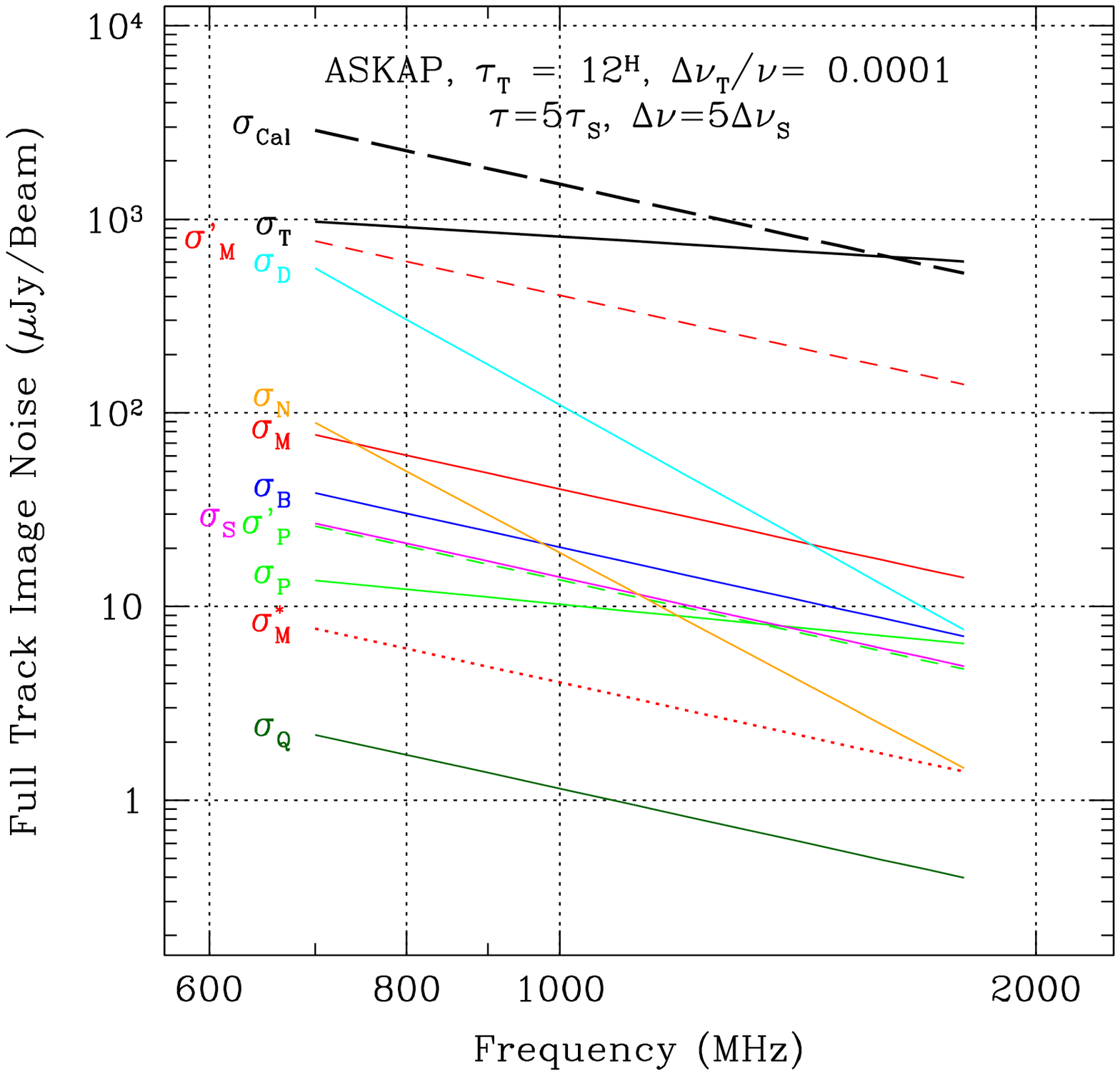}}
\caption{Noise budget for the ASKAP telescope as in
  Fig.~\ref{fig:vlad}.}
\label{fig:askap}
\end{figure*}

The noise budget analysis is presented in
Fig.~\ref{fig:askap}. Thermal noise is seen to dominate on the
self-cal solution interval, although the signal-to-noise requirement
implies that additional data averaging (by about a factor of five in
both domains) is needed. The full track continuum case demonstrates
that self-calibration will be essential and that source modelling errors will be the single most important factor
to address for this array design. For single track observations, these
must be kept below about 1\% to achieve thermal noise limited
performance, but for deep integrations above 1 GHz, it will be
necessary to achieve even better precision in
order to reach the source confusion level. As noted above, the
simultaneous imaging of the multi-beam field-of-view offers good
prospects for achieving high source modelling precision. It is also of
interest to compare the contributions of electronic,
$\sigma_P^\prime$, and mechanical pointing, $\sigma_P$, to the noise
budget in this case. While these contributions have different
frequency scalings, they have a similar magnitude for the parameters
assumed here. It is also worth noting that the electronic component is
not expected to be correlated over multiple observing sessions, so
will average down further in a deep observation. The full track
spectral line case should permit routine thermal noise limited
performance to be achieved, although self-calibration is required
across the band. Deep spectral line integrations will
become dependent on the quality of source modelling.

\subsection{MeerKAT}

MeerKAT is currently planned to consist of $N = 64$ off-axis $(alt,az)$
mounted dishes of $d=13.5$~m \citepads{2012AfrSk..16..101B}. A single
Gregorian focus feed covering about 1 to 1.74 GHz is planned for the
first phase of deployment with a specified total sensitivity of about
220 m$^{2}$/K, corresponding to $T_{sys}/\eta_A \approx 42$ K and an
SEFD of 810 Jy. While the telescopes have not yet been fabricated, we
will assume that relatively good pointing accuracy of about $P \approx
10$ arcsec may be realised under good conditions with a correlation
timescale $\tau_P \approx 15$ minutes. Beam squint for the MeerKAT
dishes is expected to be somewhat larger than that of the GBT, given
the smaller size of the subreflector of about 4 versus 7.5~m, so may
be about $\epsilon_Q \approx 0.01$ in practise. Beam squash
simulations are not available, so we will assume that values
comparable to those achieved at the GBT will apply of $\epsilon_Q
\approx 0.04$. Peak near-in sidelobe levels will depend on the degree
of edge illumination, but might be about $\epsilon_S \approx
0.01$. Similarly, the multi-path induced beam modulation effects might
also be slightly larger than those of the GBT, $\epsilon_B \approx
0.01$, while the dominant resonant ``cavity'' dimension in the system
will probably be the primary-subreflector separation of about $l_C
\approx 7$~m. The recent study of \citetads{2009.Cortes} provides an
indication of far sidelobe levels from similar optical designs and
dish dimensions of about -40 dB at 1.4 GHz. This would suggest $\eta_F
\approx 0.5$ for use in eqn.~\ref{eqn:epf}. However, the
\citetads{2009.Cortes} study used the non-optimum illumination of an
ultra-wideband feed and is strongly influenced by subreflector
spillover. An octave band Gregorian feed design is more likely to
achieve superior illumination quality of perhaps $\eta_F \approx
0.2$. The MeerKAT array configuration will have $B_{Max} = 8$~km, but
be highly centrally concentrated with $B_{Med} \approx 500$~m.

\begin{figure*}
\resizebox{\hsize}{!}{\includegraphics[bb=40 180 535 660]{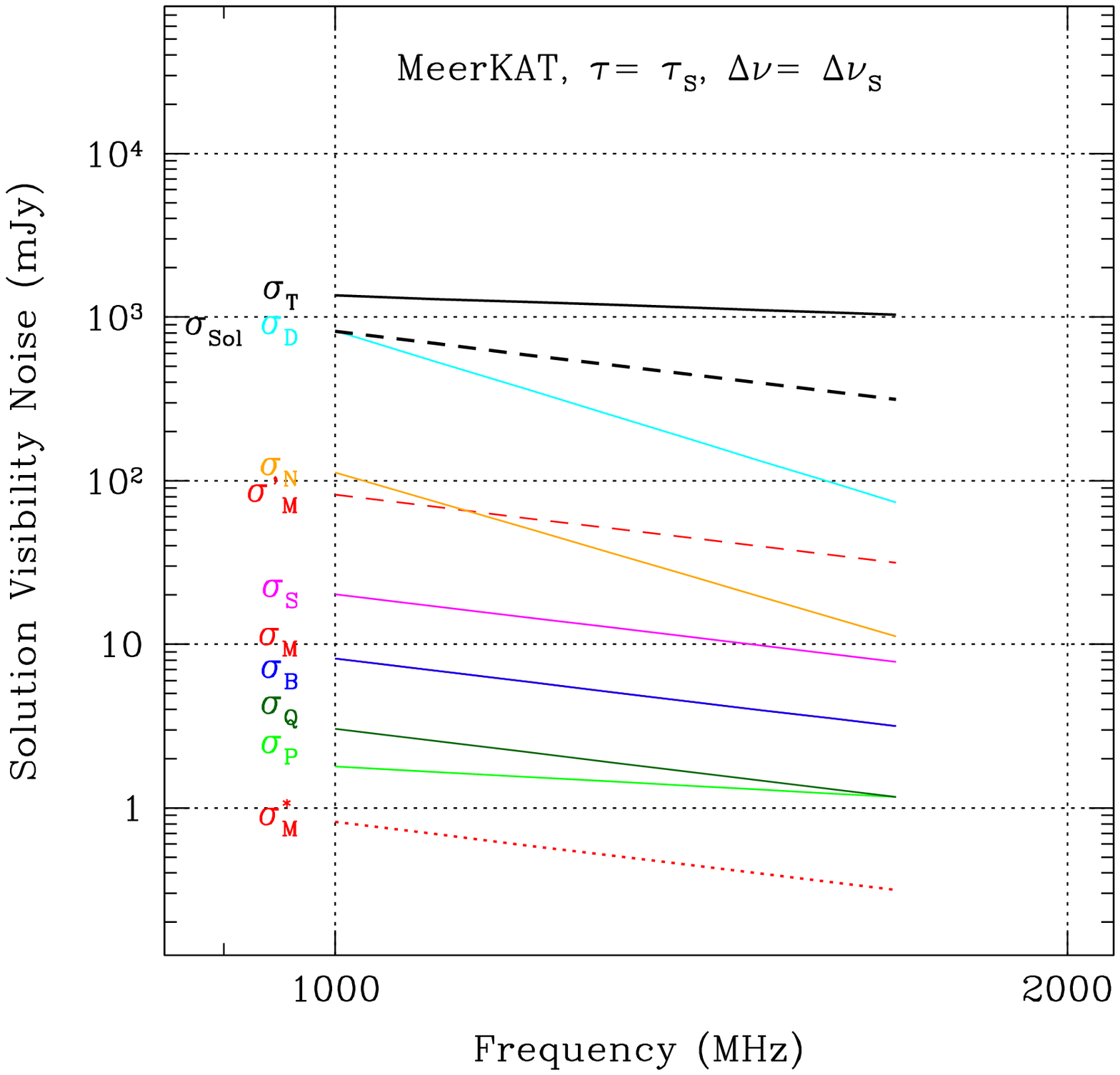},\includegraphics[bb=40 180 535 660]{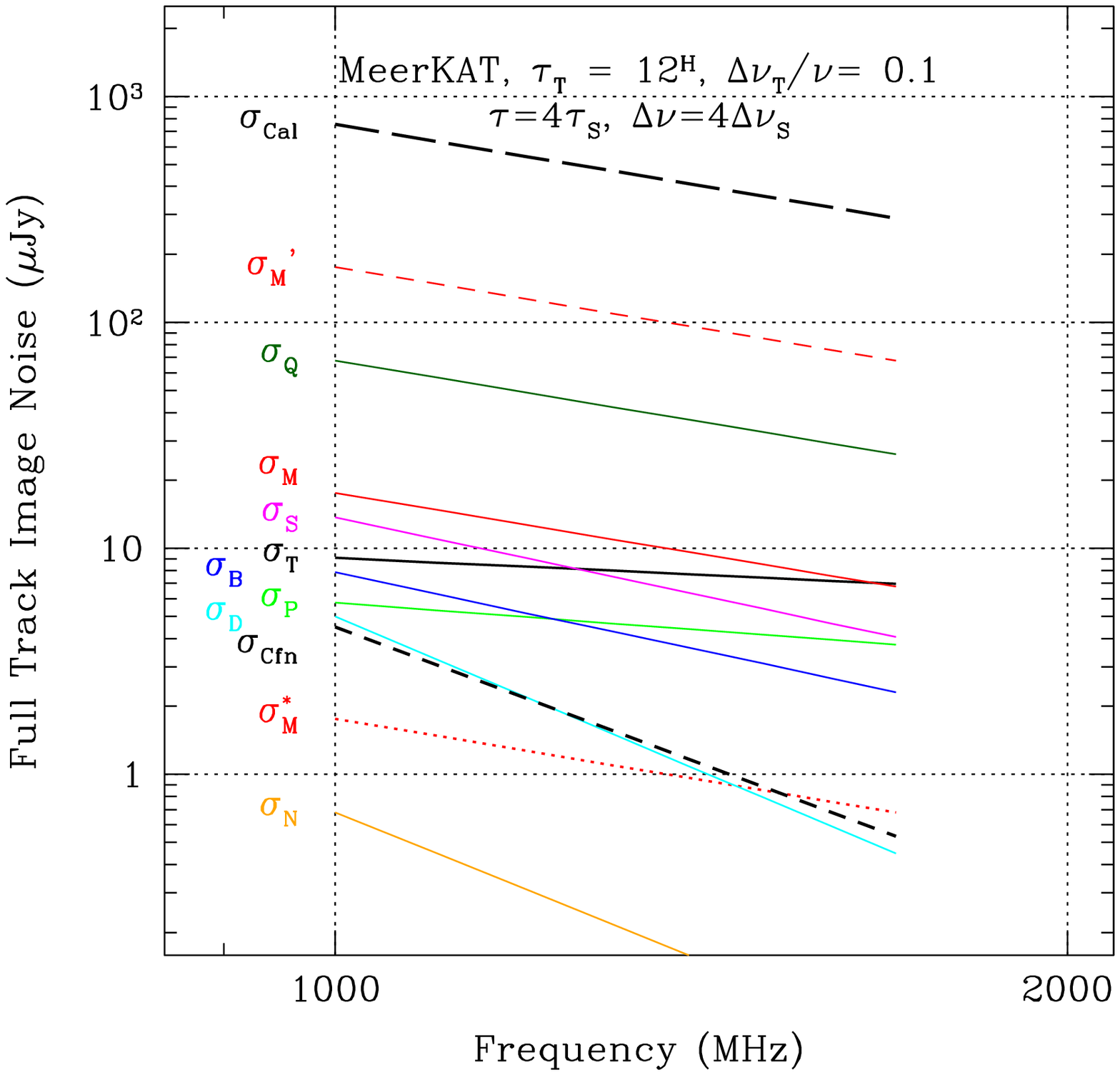},\includegraphics[bb=40 180 535 660]{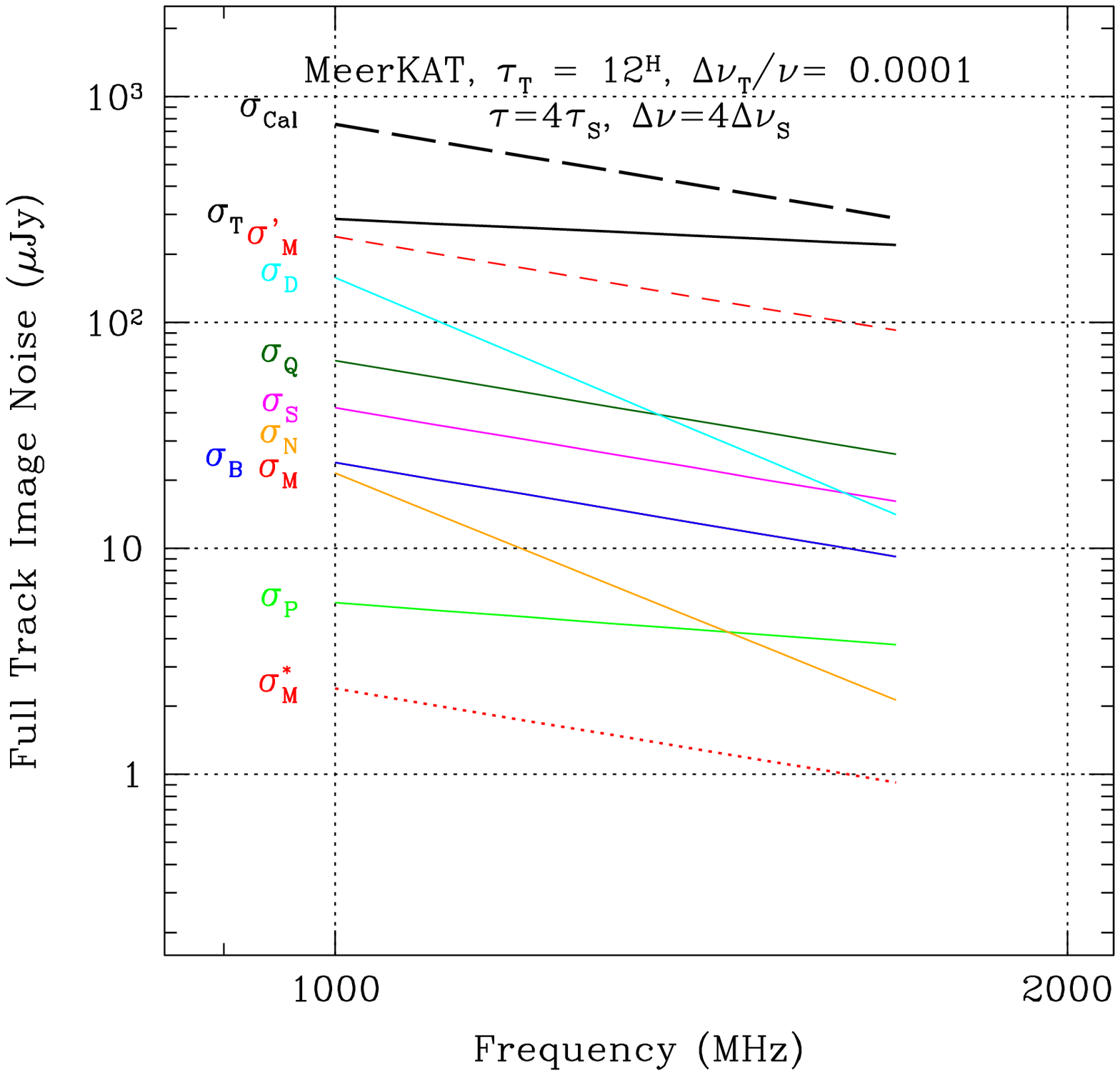}}
\caption{Noise budget for the MeerKAT telescope as in
  Fig.~\ref{fig:vlad}.}
\label{fig:meerkat}
\end{figure*}

Inspection of the noise budget analysis in Fig.~\ref{fig:meerkat}
shows that more substantial data averaging (by about a factor of four)
is required to meet the $\sigma_{Sol}$ criterion in the self-cal
solution interval. For the full track continuum case, self-calibration
is clearly required and the subsequent noise budget
is completely dominated by source modelling errors. Reaching the
thermal noise in a single track will require achieving better than
about 1\% modelling precision, which will be challenging. Beam
squash effects constitute the next challenge, with these contributions
exceeding the thermal noise by factors of several. Explicit modelling
of the time variable response in the elliptical polarisation beams
will be essential to achieving thermal noise limited
performance. Comparison with the JVLA plots for similar configurations
in Figs.~\ref{fig:vlac} and \ref{fig:vlab}, suggests that it will be
about twice as challenging for MeerKAT to achieve thermal noise
limited performance as the JVLA.

Achieving even lower noise levels in deep integrations will require a
significant further reduction in the source modelling and beam squash
errors, together with explicit modelling of additional main beam and
near-in sidelobe issues ($\sigma_P$, $\sigma_B$ and $\sigma_S$).  The
full track spectral line analysis suggests that thermal noise limited
performance will be straightforward to achieve provided
self-calibration is employed, with deep integrations becoming
sensitive to the precision of source modelling.

\subsection{SKA1-Mid}

The Square Kilometre Array Organisation has recently announced the
intention that the SKA Phase 1 mid-frequency dish array be deployed as
two complimentary components: an SKA1-Survey instrument to be sited in
Australia and an SKA1-Dish array to be sited in South Africa.

\subsubsection{SKA1-Survey}

The SKA1-Survey instrument is projected to consist of $N = 96$ dishes
of $d = 15$~m covering the frequency band of 0.45 -- 3.0 GHz with
three overlapping phased array feed systems, each of octave
bandwidth. While the array configuration has not been determined in
detail, it might be expected to have $B_{Max} = 20$~km and $B_{Med} =
1$~km. The sensitivity specification calls for $T_{Sys}/\eta_A = 50$~K
above about 1~GHz, accepting the fact that the Galactic contribution
will become significant at lower frequencies. An estimate of the
frequency dependence of system performance is given by,
\begin{equation}
T_{Sys} \approx T_{Nom} + 60\lambda_m^{2.55} \quad {\rm K},
\label{eqn:ska1tsys}  
\end{equation}
in terms of a nominal temperature, $T_{Nom}$ (that excludes the
Galactic contribution) and the wavelength, $\lambda_m$ in
meters. Relevant values for the PAF systems might be $T_{Nom} = 37$~K
and $\eta_A = 0.8$. An antenna and mount design have not yet been
determined for the SKA, but for the present purpose we will assume
that an off-axis Gregorian-fed system with an $(alt,az)$ mount is
utilised. We will assume good mechanical pointing accuracy of about $P
\approx 10$ arcsec with a correlation timescale $\tau_P \approx 15$
minutes. Electronic pointing errors of $\epsilon_P^\prime = 0.01$
correlated over $\tau_P^\prime = 1$ minute will also be assumed. In
the absence of a polarisation mount axis, it will be necessary to
track the digital PAF beams which are offset from boresight across the
focal plane. The substantial intrinsic beam squash of about 4\% is
likely to be improved by about 10 dB with polarisation specific PAF
illumination to about $\epsilon_Q = 0.004$. The multi-path induced
beam modulation will be decreased from its nominal 1\% level in such
an off-axis design by at least 10 dB to $\epsilon_B \approx 0.001$. Near-in
sidelobes $\epsilon_S \approx 0.01$ will be assumed in this offset
design. Good quality PAF illumination should provide an improvement of
about a factor of two in the far sidelobe efficiency to $\eta_F =
0.1$.

\begin{figure*}
\resizebox{\hsize}{!}{\includegraphics[bb=40 180 535 660]{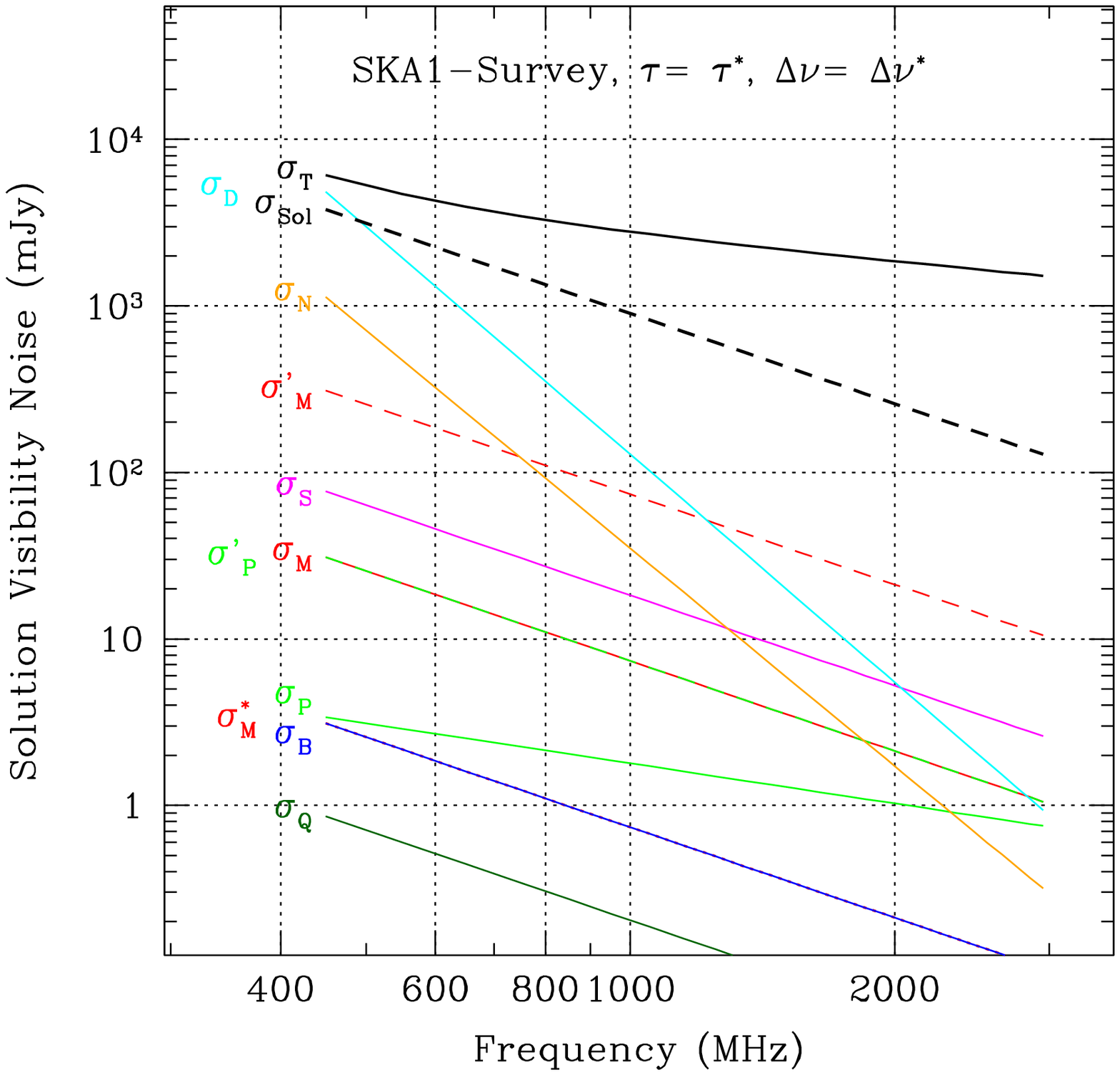},\includegraphics[bb=40 180 535 660]{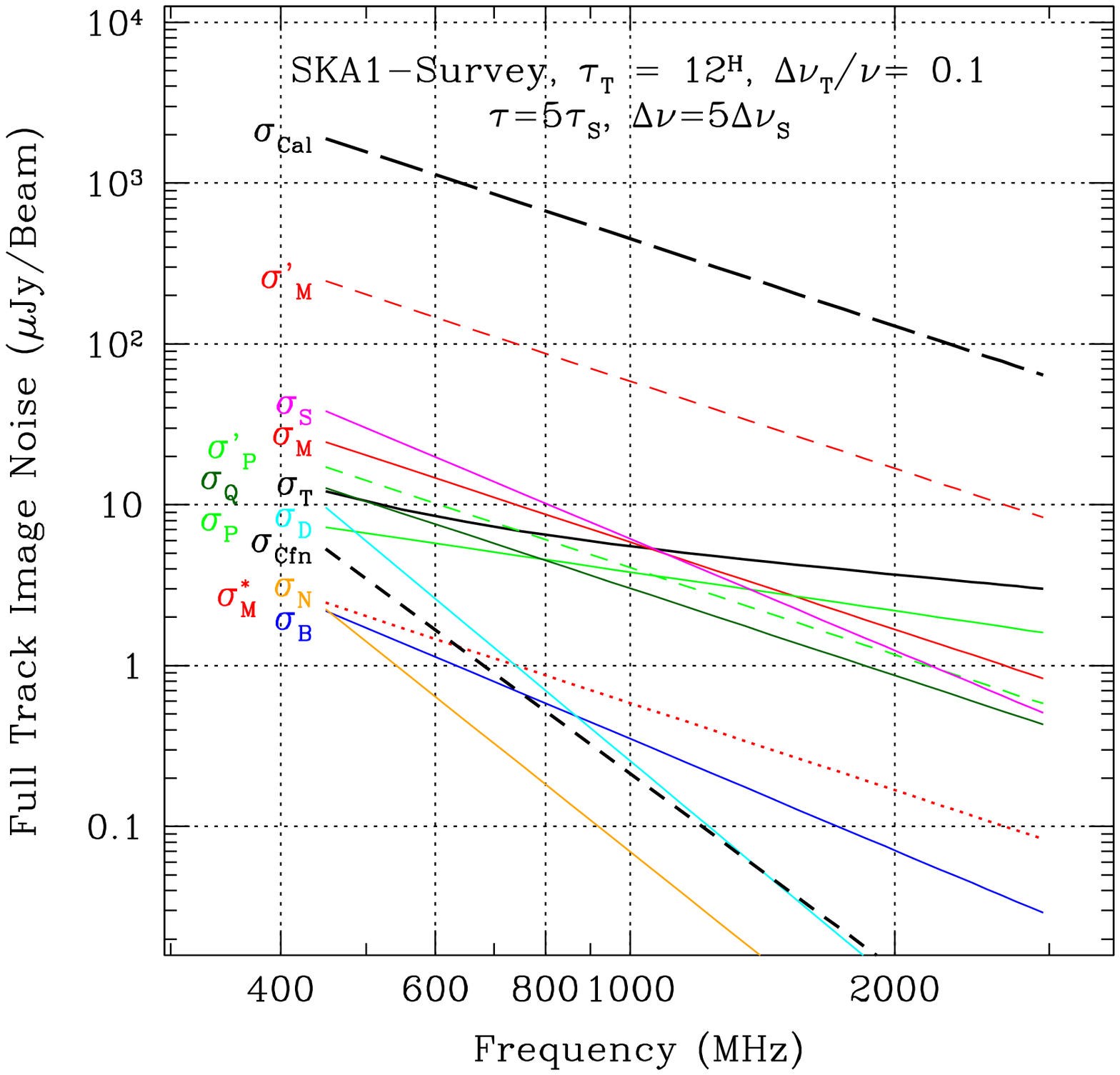},\includegraphics[bb=40 180 535 660]{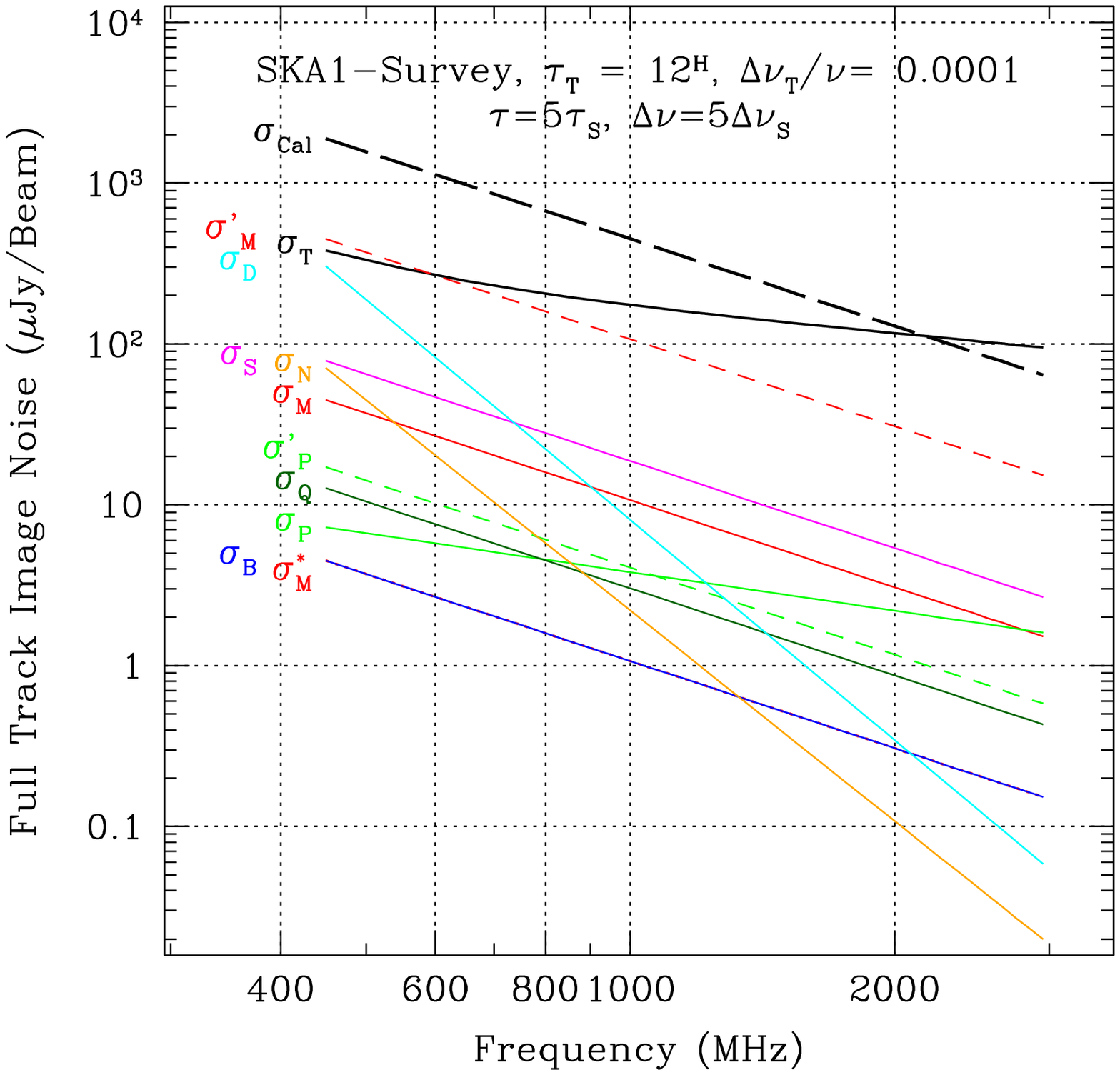}}
\caption{Noise budget for the SKA1-Survey telescope as in
  Fig.~\ref{fig:vlad}.}
\label{fig:ska1s}
\end{figure*}

The self-cal solution noise budget in Fig.~\ref{fig:ska1s}
demonstrates yet again that there is insufficient sensitivity for
self-calibration at the ``natural'' time and frequency interval for
straightforward, full beam imaging, necessitating some further data
averaging. The full track continuum case will require self-calibration
and significant attention to source modelling, particularly at the
lowest frequencies where a precision of $<$1\% will be required, with
the near-in sidelobes representing the next largest error
contribution. It is the absence of a polarisation axis which
introduces time variability into the near-in sidelobe response and
adds this complexity to the imaging problem. Obtaining very deep
integrations will rely on further precision in modelling sources both
in the time variable near-in sidelobes as well as in the main
beam. Full track spectral line observations are expected to be much
more straightforward, although they will require self-calibration.
Deep integations will become limited by source modelling precision.

\subsubsection{SKA1-Dish}

The SKA1-Dish array is projected to consist of $N = 250$ dishes of $d
= 15$~m covering the frequency band of 0.45 -- 3.0 GHz with three
overlapping single pixel feed systems, each of octave bandwidth.  The
array configuration is expected to extend out to $B_{Max} = 100$~km
but maintain a $B_{Med} \approx 1$~km. However, to simplify comparison
of the results with both SKA1-Survey as well as the more extended
configurations of the JVLA, we will only illustrate the case of
$B_{Max} = 20$~km.  The single pixel feed sensitivity specification
calls for $T_{Sys}/\eta_A = 44$~K above about 1~GHz. Relevant values
for use in eqn.~\ref{eqn:ska1tsys} might be $T_{Nom} = 28$~K and
$\eta_A = 0.7$. The mechanical pointing errors assumed above will
apply with $P \approx 10$ arcsec and a correlation timescale $\tau_P
\approx 15$ minutes, as well as the near-in sidelobe response of
$\epsilon_S \approx 0.01$. A beam squash of $\epsilon_Q \approx 0.04$
and multi-path modulation of $\epsilon_B \approx 0.01$ with $l_C
\approx 7$~m such as discussed for MeerKAT are also likely to be
relevant here. A far sidelobe response of $\eta_F = 0.2$ is assumed to
apply for the likely dish illumination quality.

\begin{figure*}
\resizebox{\hsize}{!}{\includegraphics[bb=40 180 535 660]{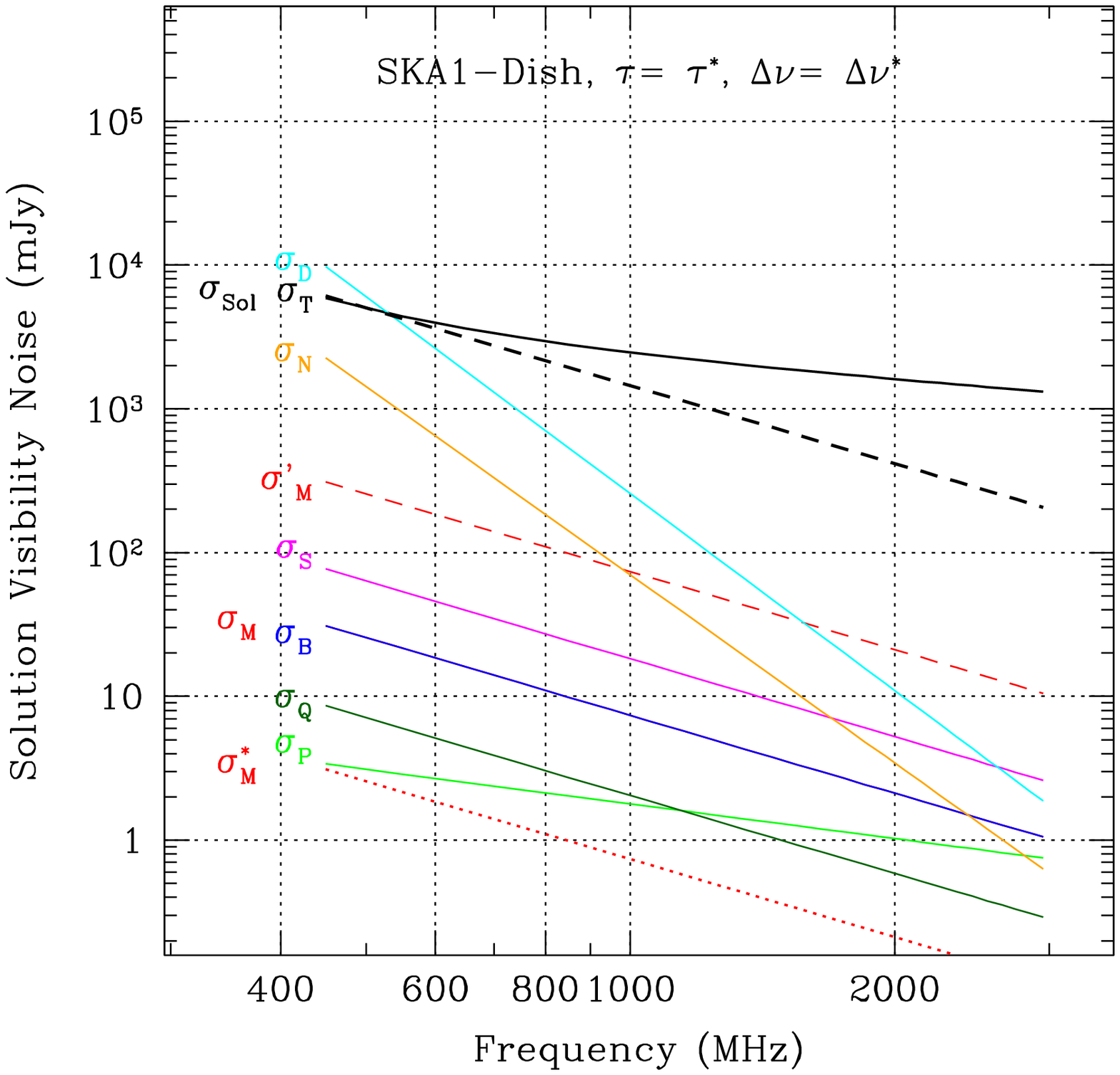},\includegraphics[bb=40 180 535 660]{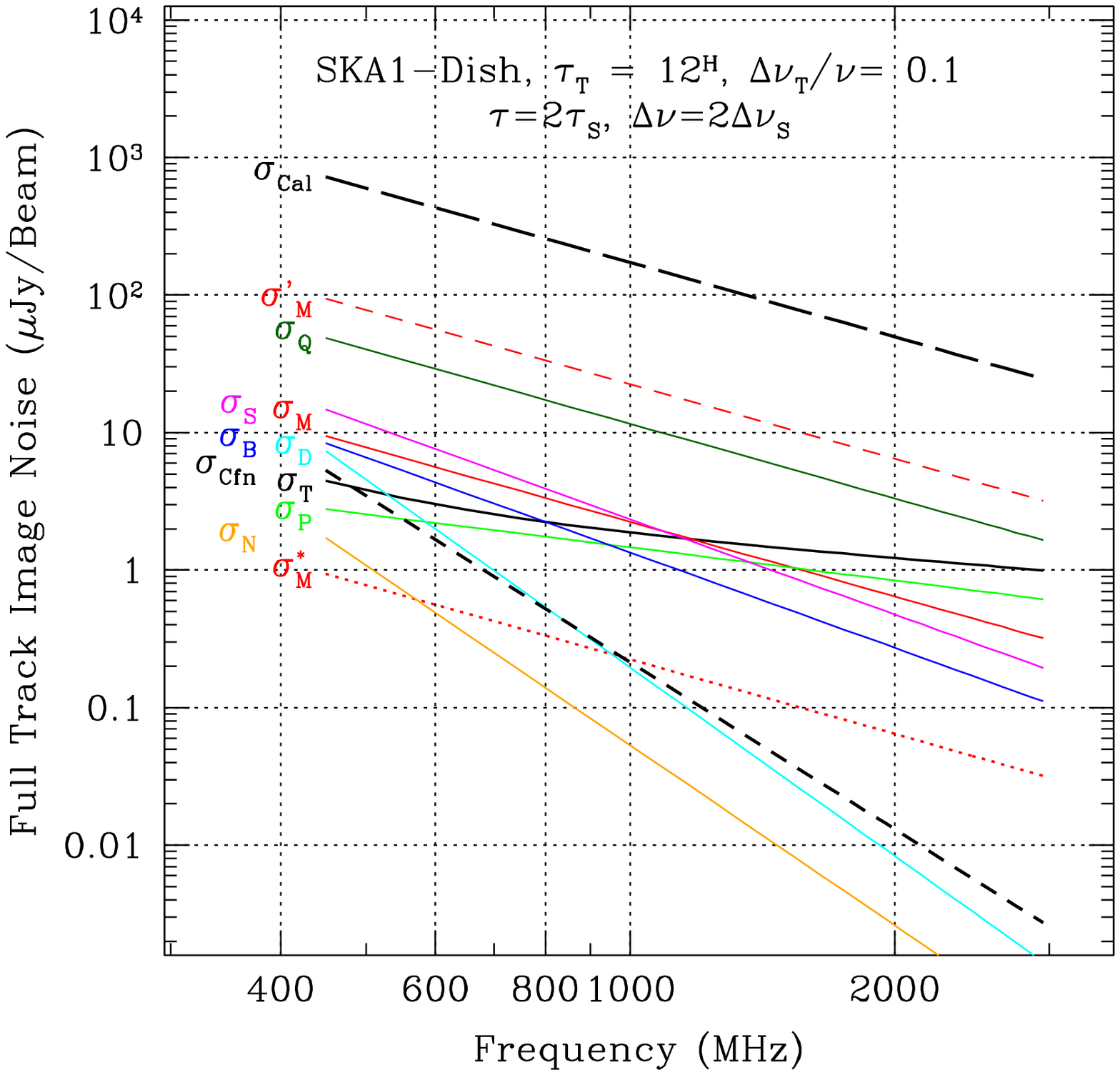},\includegraphics[bb=40 180 535 660]{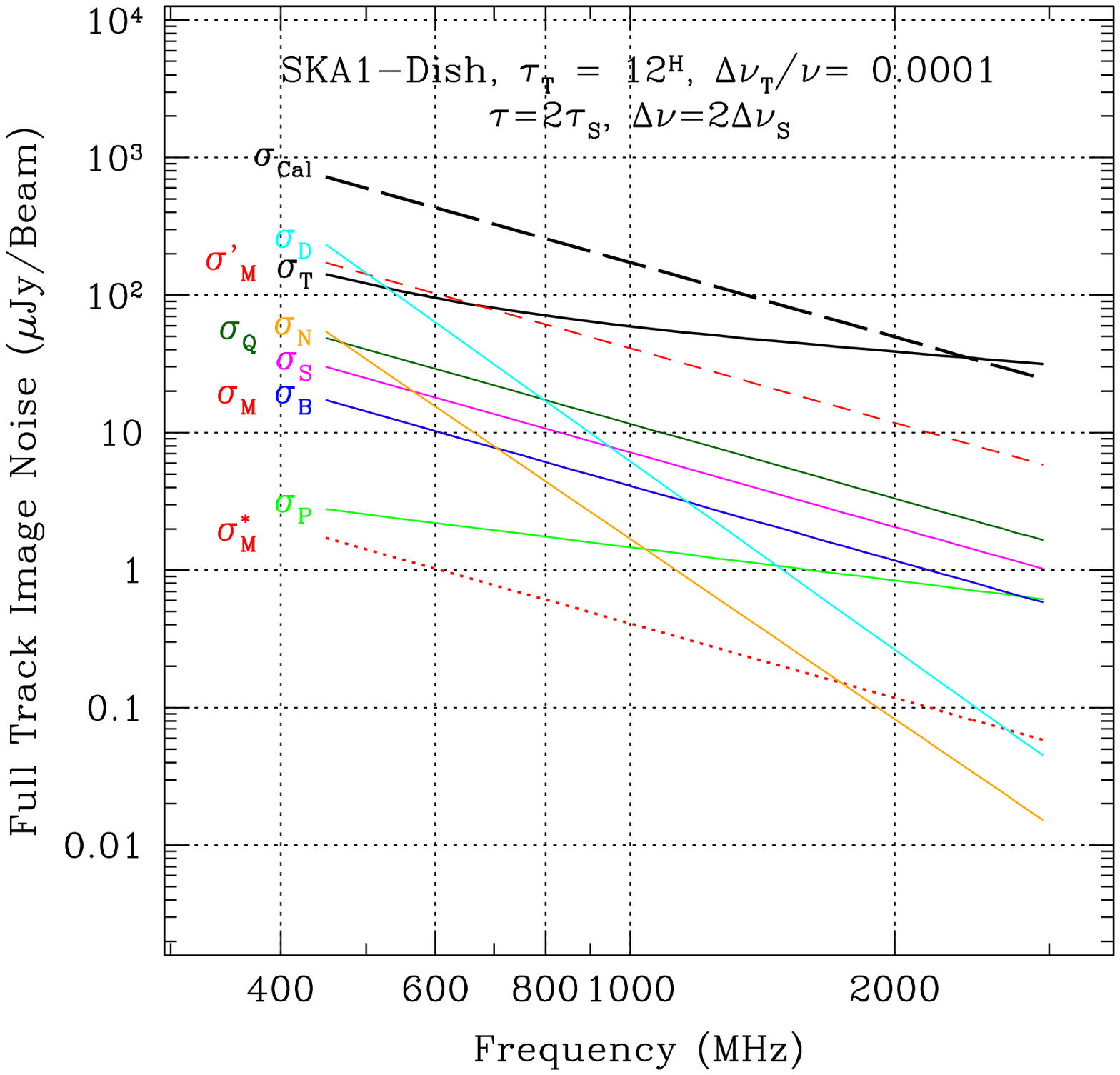}}
\caption{Noise budget for the SKA1-Dish telescope as in
  Fig.~\ref{fig:vlad}.}
\label{fig:ska1d}
\end{figure*}

The thermal noise is just sufficient to permit self-cal convergence on
the ``natural'' solution interval at the bottom end of the frequency
coverage, but exceeds that limit above about 600~MHz. A full track
continuum observation will require self-calibration and will need to
achieve high precision source modelling in the main beam,
varying between about 0.3 and 3\% from the bottom to the top of the
band. The beam squash effects must also be actively modelled in order
to permit the thermal noise limit to be achieved.  Deep integrations
would only approach the source confusion limit if a range of
additional effects were effectively modelled, including the near-in
sidelobes and main beam modulation. Full track spectral line
observations will require self-calibration, but are only moderately
sensitive to the source modelling precision.

\subsection{LOFAR high band}
The LOFAR array \citepads{2011JApA...32..589H} consists of stations
distributed over much of the Netherlands as well as in other European
countries. For the purpose of our analysis we will consider only the
LOFAR High Band that covers frequencies between 120 and 240 MHz and
the stations within NL. It will be useful to consider both the core
configuration of $N = 48$ stations of $d = 30.75$ m diameter for which
$B_{Max} = 3.5$~km and a core plus remote configuration with a total
of $N = 64$ stations and $B_{Max} = 121$~km. While the remote stations
have a larger physical size than those in the core, we will assume
that only the central $d = 30.75$ m of these outer stations is
utilised to simplify their combination with the core stations during
imaging. The configuration of stations is highly centrally
concentrated with a baseline density that declines roughly as $B^{-2}$
beyond about 100~m. For the core configuration we will assume $B_{Med}
= 250$~m while for the core plus remote configuration this might
increase to $B_{Med} = 1$~km. The SEFD of the LOFAR High Band core
stations is projected to vary approximately as,
\begin{equation}
SEFD \approx 2150 + 0.215\bigg( {\nu \over 1~ {\rm
    MHz}} - 175\bigg)^2 \quad {\rm Jy}.
\label{eqn:lofarsefd}  
\end{equation}
Mechanical pointing errors clearly do not apply to an aperture
array. There may well be an electronic pointing error associated with
gain fluctuations of the individual antenna elements that are combined
to produce the station beam. Actual values are not yet documented but
the impact will be assesed by assuming a station beam variability
amplitude of about $\epsilon_P^\prime = 0.01$ and timescale of
$\tau_P^\prime = 1$ minute. As was the case for the phased array feed,
we expect beam squint and beam squash to be minor, if there is
optimisation for these parameters. However, the minimisation of beam
squash for an aperture array will require active compensatation of the
geometric foreshortening of the stations for the duration of the
observing track. While this can be realised it will come at the
expense of sensitivity, since the ``lowest common denominator''
approach for the effective aperture would need to be applied. An
alternative to this straightforward strategy would be the use of the
``A-projection'' algorithm, as outlined previously, to attempt to
compensate for the changing, assymetric station beams during an
observation.  As noted above, this method only provides
approximate ``forward'' correction of the acquired data. Experience
on the applicability and precision of this method needs to be acquired
and documented.  A reference value of $\epsilon_Q = 0.01$ will be
assumed to track the impact of any residual unmodelled beam
assymetries. As noted at the outset of this section we will assume
that a full track observation has $\tau_T = 4$ hour duration to limit
the severity of geometric foreshortening.  Main beam modulation due to
multipath effects is clearly not relevant to the aperture array
geometry, eliminating the $\epsilon_B$ component. The near-in sidelobe
amplitude is also open to optimisation with the choice of digital
tapering of the station beam weights. However, a lower sidelobe will
come at the expense of station sensitivity. For illustration purposes,
we will assume a uniform aperture weighting to achieve maximum
sensitivity, although this yields a large value of $\epsilon_S =
0.1$. Measured HBA sidelobe levels are somewhat greater than 10\% and
are highly modulated with azimuthal angle. The magnitude of the
far-sidelobe response of the LOFAR stations is not yet documented. In
view of the discrete rather than continuous nature of the aperture
sampling, a relatively large value of $\eta_F \approx 0.5$ might be
appropriate.

\begin{figure*}
\resizebox{\hsize}{!}{\includegraphics[bb=40 180 535 660]{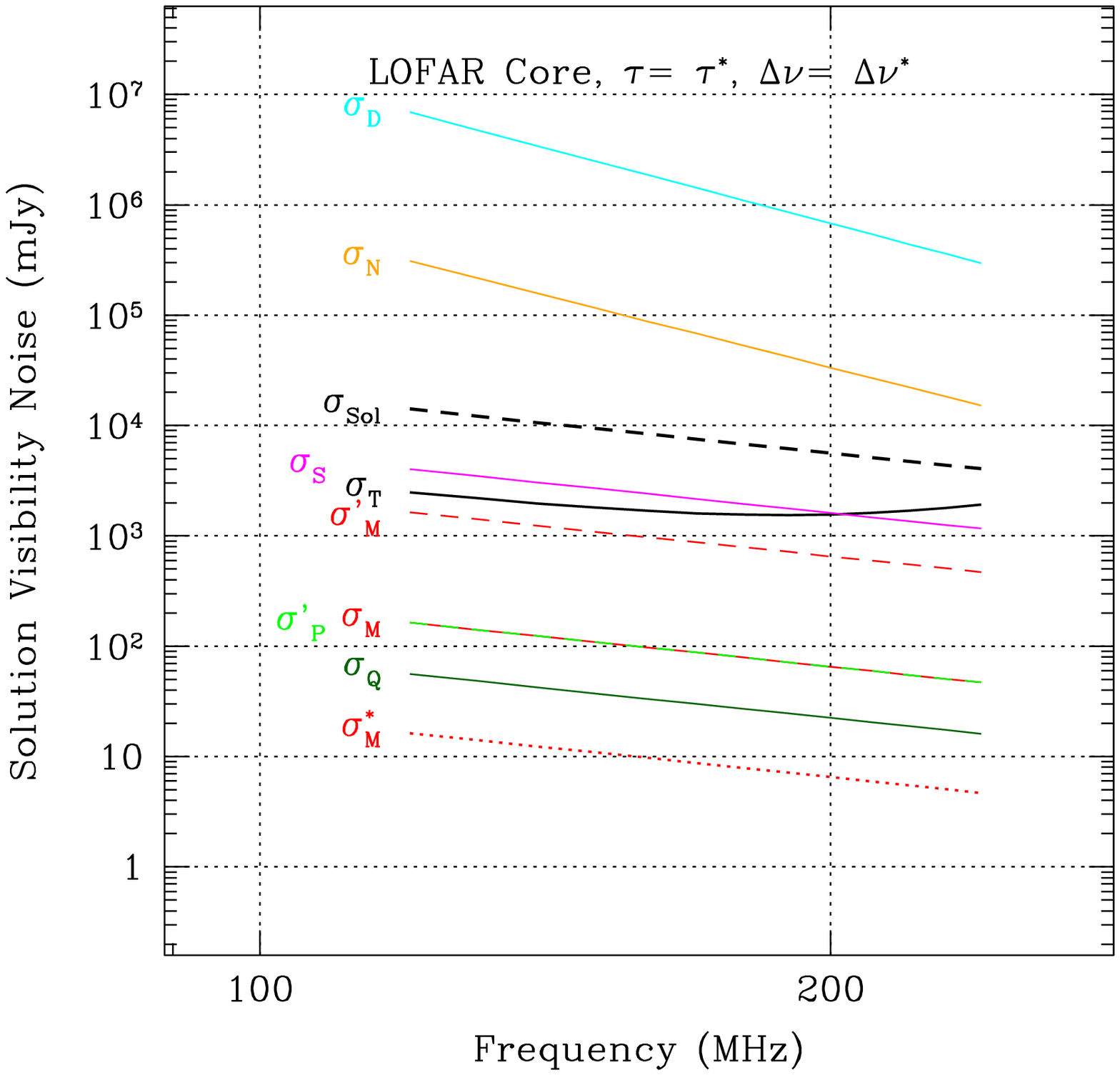},\includegraphics[bb=40 180 535 660]{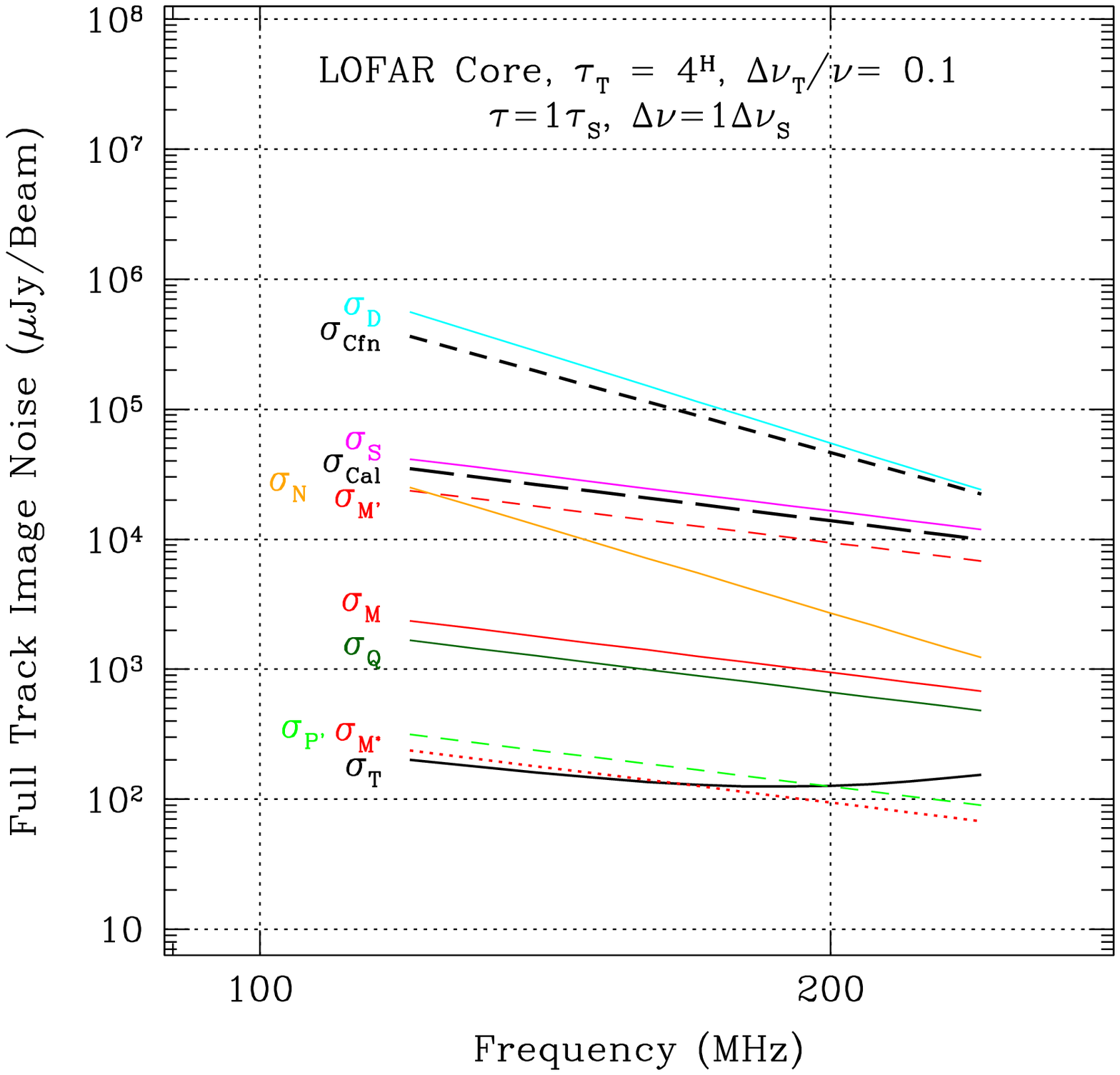},\includegraphics[bb=40 180 535 660]{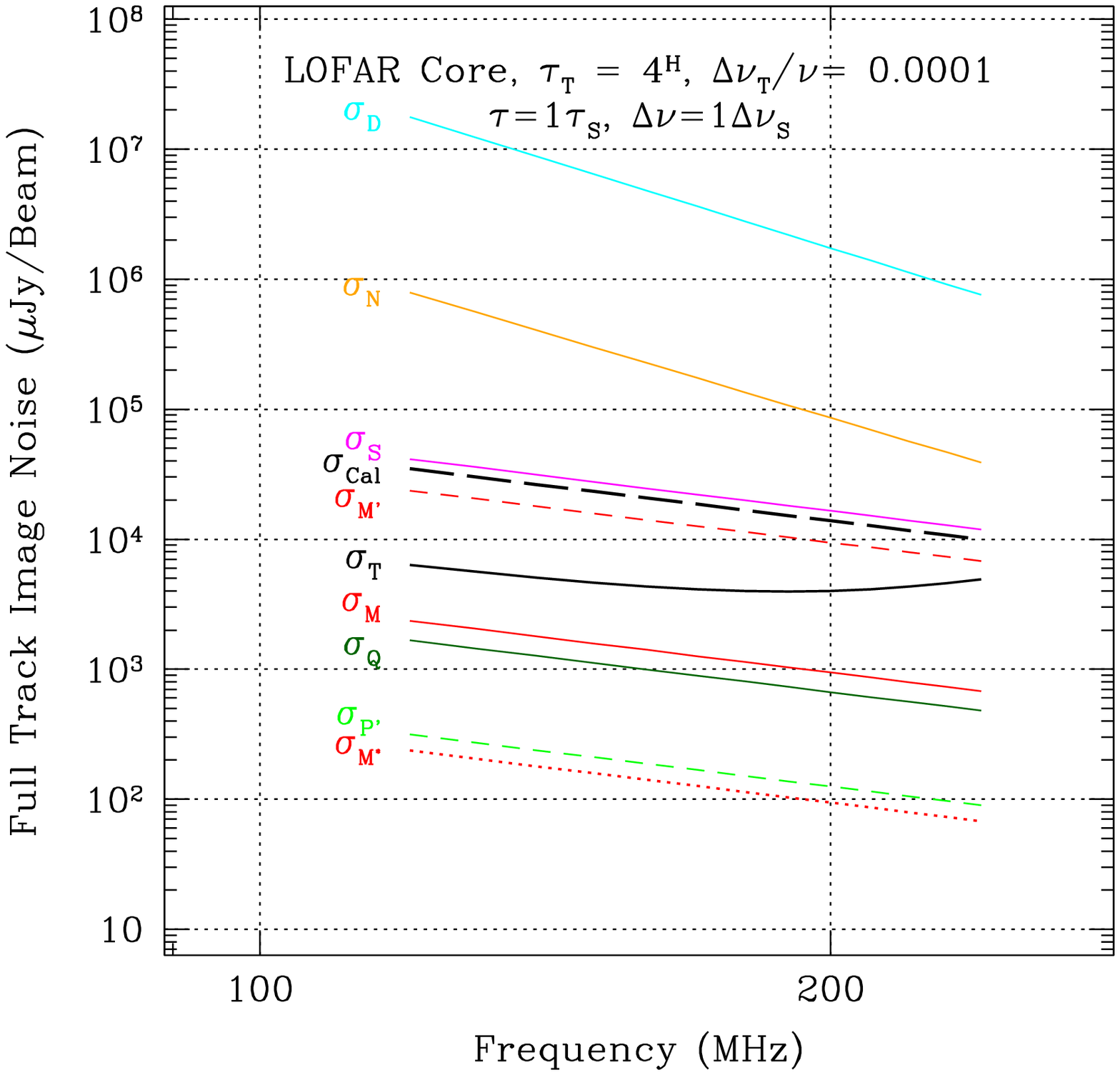}}
\caption{Noise budget for the LOFAR High Band Core telescope as in
  Fig.~\ref{fig:vlad}.}
\label{fig:lofarc}
\end{figure*}

The LOFAR Core configuration noise budget is shown in
Fig.~\ref{fig:lofarc}. At the nominal solution timescale, the thermal
noise easily satisfies the self-cal signal-to-noise requirement,
$\sigma_T < \sigma_{Sol}$. However, the far sidelobe response to both
the daytime, $\sigma_D$, and the nighttime sky $\sigma_N$, completely
dominate the noise budget by several orders of
magnitude. Self-calibration will not be possible without very
extensive modelling of the sky for a nominal solution interval
$(\tau^\star, \Delta \nu^\star)$. Comparison with Fig.~\ref{fig:sign}
demonstrates how deep it will be necesssary to go before these
contributions will no longer dominate the fluctuation level on a
solution interval. The first order of magnitude reduction will be
addressed by adequately modelling the 30 brightest sources (per $2\pi$
sr) with $S_{1.4GHz} > 30$~Jy. Unfortunately, even this may not be
sufficient below about $\nu = $180~MHz. The next flux bin, $10
<S_{1.4GHz} < 30$~Jy, contains an additional 70 sources per
hemisphere. With this reduction in fluctuation level (by about a
factor of 17) self-calibration should become practical over the entire
band. The other method for reducing the far sidelobe fluctuations
within a solution interval might be extensive data averaging in time
and frequency. As can be seen from eqns.~\ref{eqn:sign} and
\ref{eqn:sigd}, increasing each of these by an order of magnitude
would provide a factor of ten reduction in the fluctuation amplitude,
although at the expense of a major negative impact on the ability to
model sources on the main beam flank with precision. In practise a
hybrid approach might be employed whereby some number of the brightest
sources are modelled directly and the remainder are suppressed by a
degree of visibility averaging. The full-track noise budget plot
demonstrates that the LOFAR Core will be source confusion limited for
continuum observations after only short integrations. Of potential
concern is the fact that the thermal noise is exceeded by the majority
of other effects under consideration. 

The full-track spectral line observation demonstrates the challenges
that will be faced by the Epoch of Reionisation experiments
\citepads{2012MNRAS.425.2964Z}. Self-calibration will be necessary to
permit the thermal noise floor to be reached. However, far sidelobe
contributions vastly exceed the thermal noise and will require
extensive modelling to achieve the necessary 20 to 30 dB
reduction. This requirement is significantly more stringent than that
encountered on the self-cal solution timescale and will be very
challenging to meet in the lower half of the band, where 100's of
sources distributed over the entire sky must be modelled. Unmodelled
sources in the near-in sidelobes also lead to excess image noise that
exceed $\sigma_T$ by factors of several in each four hour track. As
noted previously, the near-in sidelobe pattern with uniform aperture
weighting is highly modulated with parallactic angle, implying that
this time variability will require active modelling. Since both
$\sigma_N$ and $\sigma_S$ are tied to the intrinsic telescope-sky
response they will not average down in long integrations. The ultimate
noise floor in a deep field observation will be entirely limited by
the quality of the sky modelling. It may well be deemed beneficial to
employ significant tapering of the station aperture to reduce the
near-in sidelobe level. Unfortunately, such aperture tapering may
have the undesirable side-effect that the far sidelobe response is
further enhanced, since the effective station diameter, $d$, is
reduced.

\begin{figure*}
\resizebox{\hsize}{!}{\includegraphics[bb=40 180 535 660]{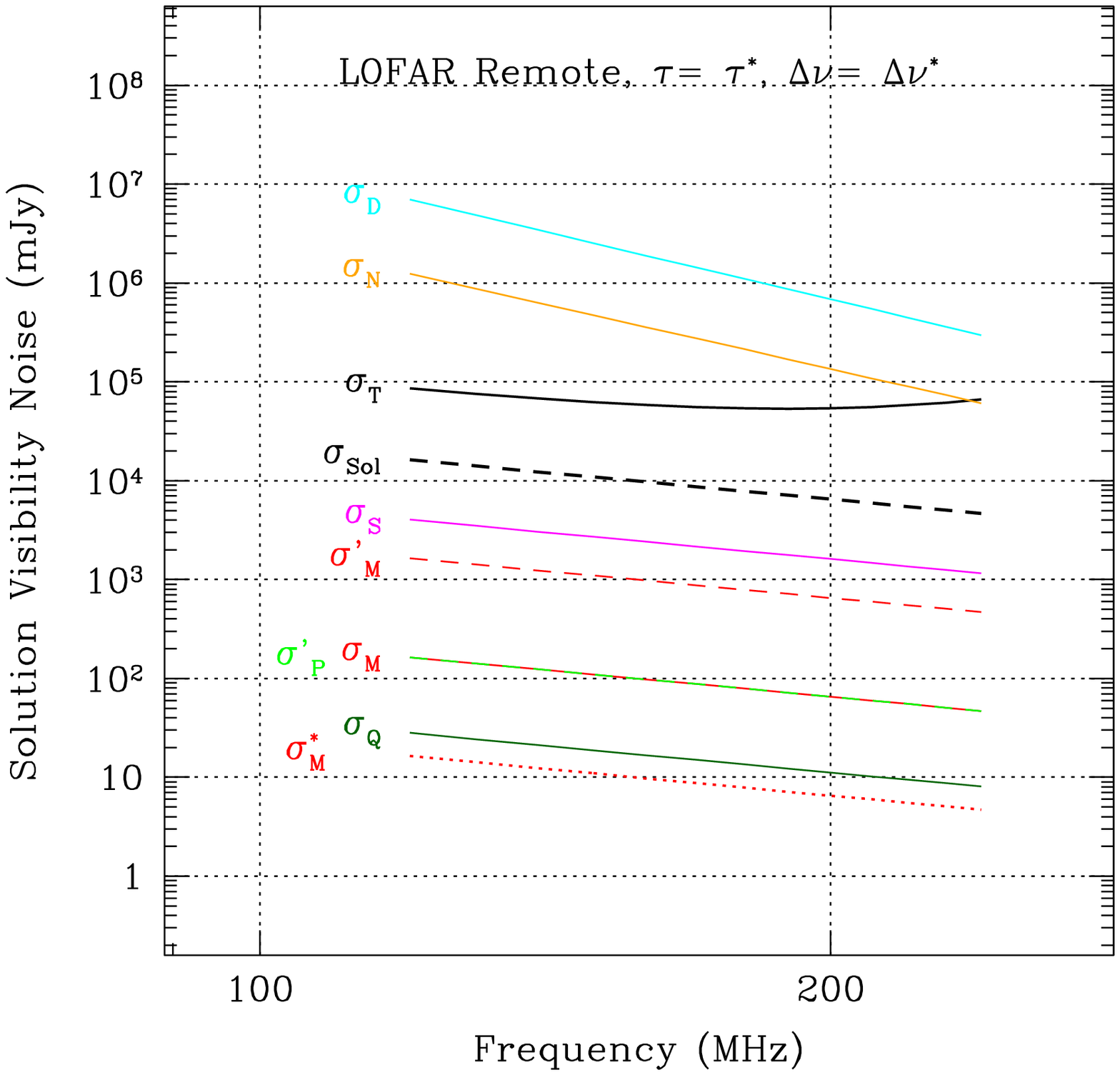},\includegraphics[bb=40 180 535 660]{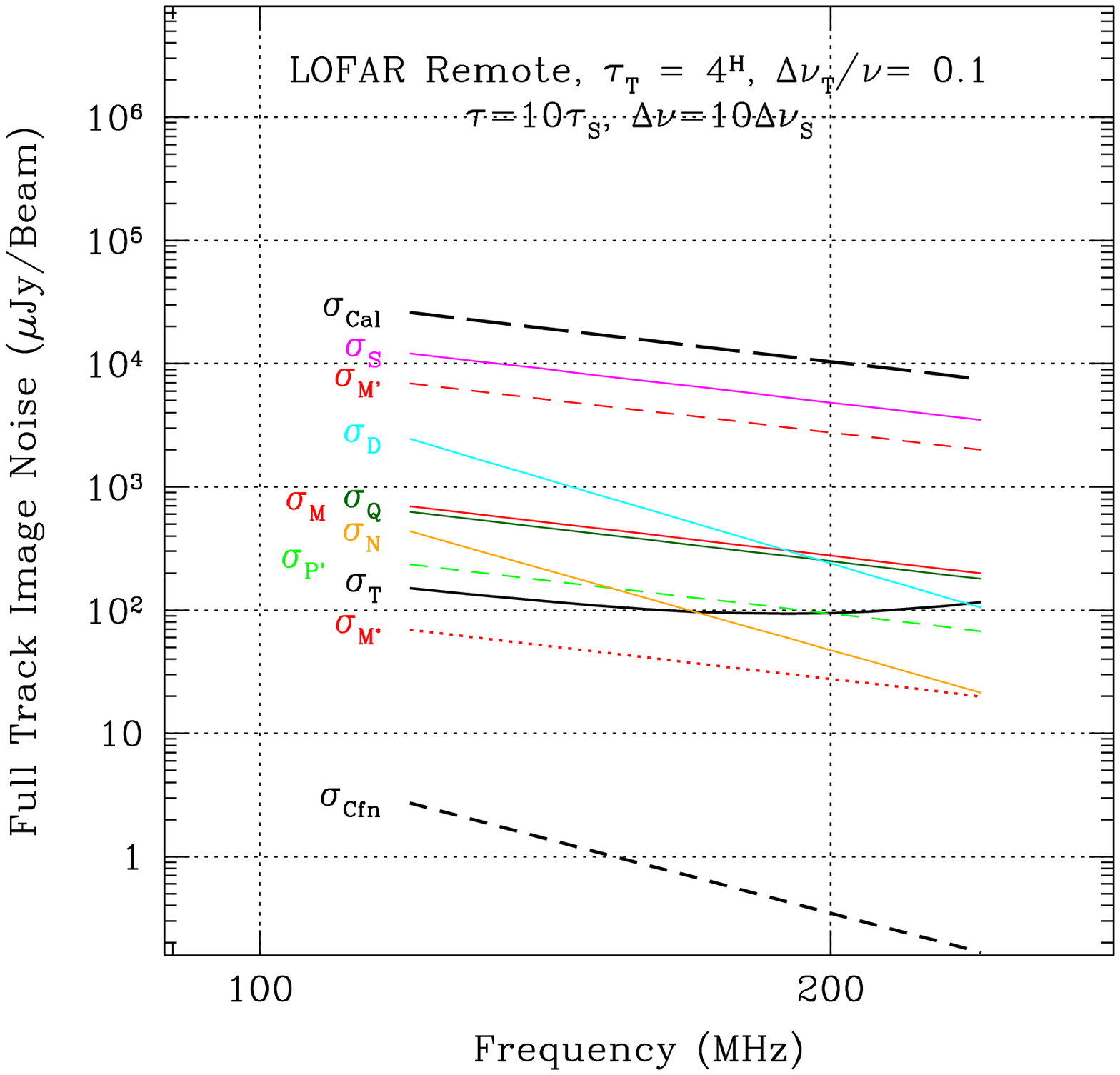},\includegraphics[bb=40 180 535 660]{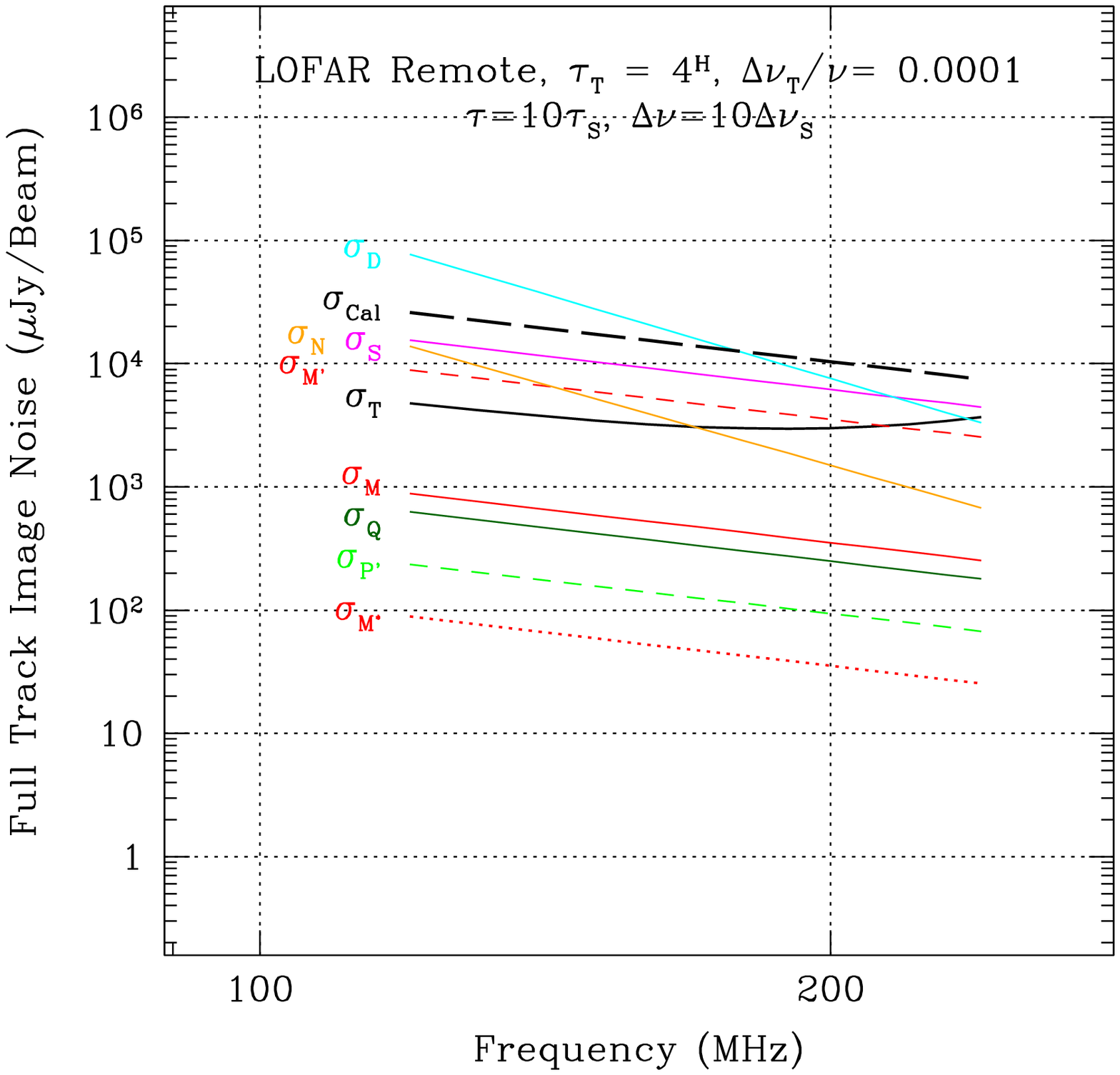}}
\caption{Noise budget for the LOFAR High Band Remote telescope as in
  Fig.~\ref{fig:vlad}.}
\label{fig:lofarm}
\end{figure*}

The noise budget for the combined Core plus Remote configuration is
shown in Fig.~\ref{fig:lofarm}. The nominal solution timescale for
this much larger value of $B_{Max}$ does not provide adequate
signal-to-noise for self-calibration. A substantial degree (factor 10)
of time and frequency averaging is necessary to bring $\sigma_T$ down
below $\sigma_{Sol}$ which will have an adverse effect on source
modelling within the main beam. Such averaging will provide attenuation
of the far sidelobe contributions by an order of magnitude. Even so, a
further reduction of $\sigma_D$ or $\sigma_N$ by at least another
order of magnitude is needed to allow for the possibility of self-cal
convergence. As discussed above, a minimum requirement will be
modelling and excision of the $S_{1.4GHz} > 30$~Jy source category. The
full-track continuum noise budget shows that source confusion noise
has now become negligible relative to other terms. Unfortunately,
there are many error contributions that exceed the thermal noise in a
single track and self-calibration is clearly necessary to approach
that limit. We will assume that the far sidelobe contributions have
been reduced from the outset by about a factor on ten through modelling
of the $S_{1.4GHz} > 30$~Jy population. The next limitation becomes
that of unmodelled sources in the near-in sidelobes, followed by
modelling errors associated with main beam sources. Reaching the
thermal noise in a four hour track will require a source modelling
precision that exceeds 0.3\%. This will be particularly challenging in
view of the high degree of visibility smearing (factor 10) required to
provide solution signal-to-noise. The modelling of the time variable
response to sources within the near-in sidelobes will also need to
achieve a 1\% acccuracy, in order to avoid this potential
limitation. It is likely to be very challenging to reach the thermal
noise in such four hour tracks.  Even deeper continuum integrations
would need to improve on all of the previously mentioned error
contributions. 

The full-track spectral line case is significantly less challenging
than the continuum for the Remote configuration, although again
self-calibration is required across the entire band. Once the far sidelobe
effects have been reduced to the point where self-cal becomes
possible, the primary limitation come from adequate modelling of
sources within the near-in sidelobes. This must be reduced by about a
factor of two to reach $\sigma_T$. Deeper spectral line integrations
will need to improve further on the $\sigma_S$ component and may then
also need to improve on the far sidelobe contributions $\sigma_N$.

\subsection{Murchison Wide-field Array (MWA)}

The MWA \citepads{5164979} is currently expected to consist of $N =
128$ station tiles, each a square of $4.4\times4.4$~m dimensions that
cover the frequency band 80 -- 300 MHz. The
configuration is planned to be centrally concentrated with $B_{Max} =
3$~km and $B_{Med} \approx 300$~m. The system temperature is expected to vary as,
\begin{equation}
T_{Sys} \approx T_{Rcv} + 60\lambda_m^{2.55} \quad {\rm K},
\label{eqn:mwatsys}  
\end{equation}
in terms of a receiver temperature, $T_{Rcv}$ and the wavelength,
$\lambda_m$ in meters. While the receiver temperature specification is
not currently published, we will assume $T_{Rcv} = 150$~K as a
representative value. The effective collecting area per station is
expected to vary as,
\begin{eqnarray}
\label{eqn:mwaae}
 A_{eff} & = & 16\bigg({\lambda^2 \over 3}\bigg) \qquad (\lambda <
 \lambda_T) \\ 
\nonumber & = & 16\bigg({\lambda_T^2 \over 3}\bigg) \qquad (\lambda >
\lambda_T)  \quad {\rm m^2}
\end{eqnarray}
for 16 dual polarisation antenna elements in each tile that are
sparsely distributed for $\lambda < \lambda_T = 2.2$ m. We will assume
an electronic pointing error associated with gain fluctuations of the
individual antenna elements of about $\epsilon_P^\prime = 0.01$ and
timescale of $\tau_P^\prime = 1$ minute. The rectangular tile geometry
will give rise to large departures from axisymmetry for the main beam
shape.  We will also assume a residual beam squash error of
$\epsilon_Q = 0.1$ that applies to the full track observations of
$\tau_T = 4$ hour duration. Untapered aperture weighting is assumed
that might yield a near-in sidelobe contribution $\epsilon_S =
0.1$. The far sidelobe response will be characterised by $\eta_F
\approx 0.5$.

\begin{figure*}
\resizebox{\hsize}{!}{\includegraphics[bb=40 180 535 660]{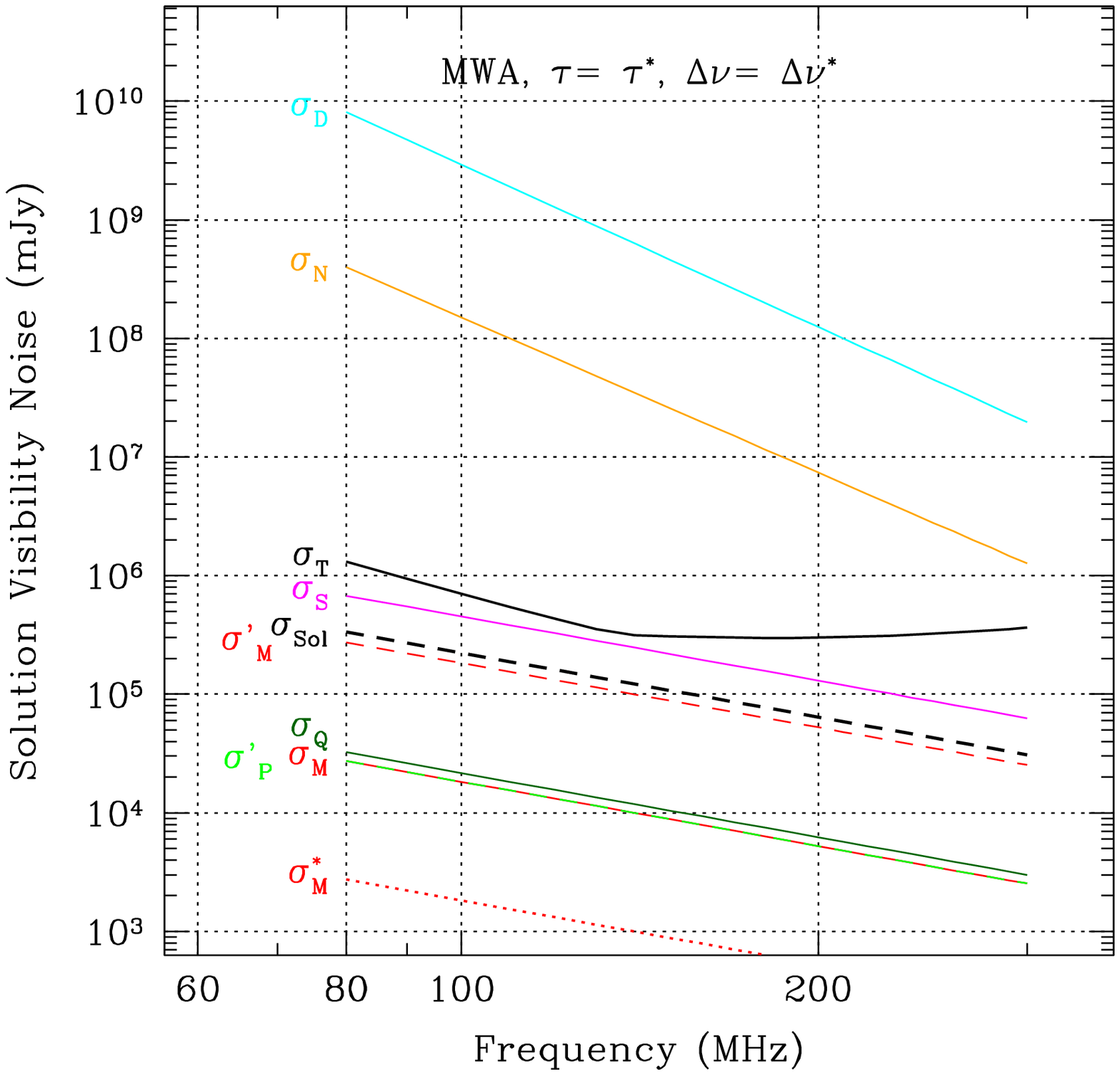},\includegraphics[bb=40 180 535 660]{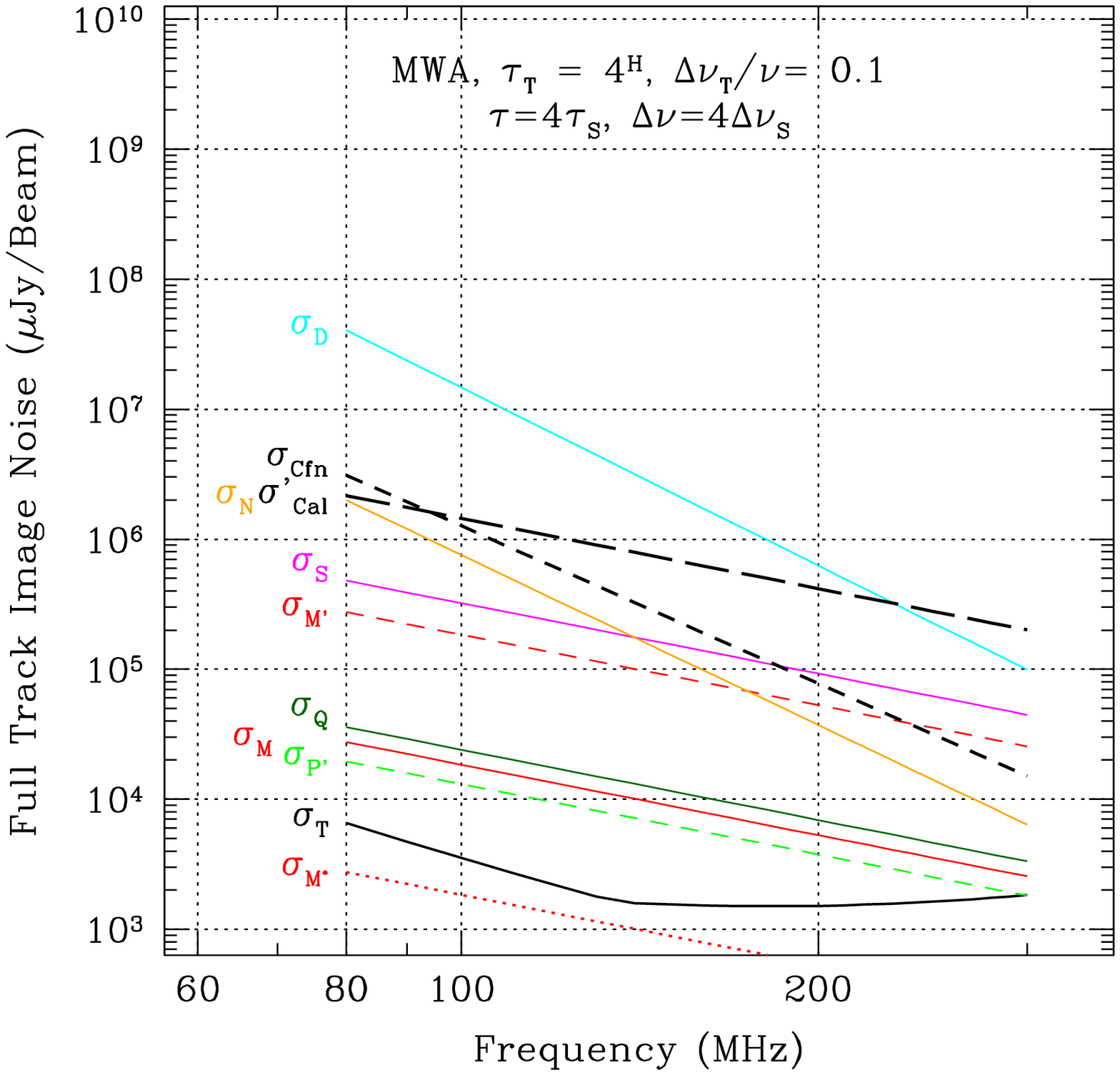},\includegraphics[bb=40 180 535 660]{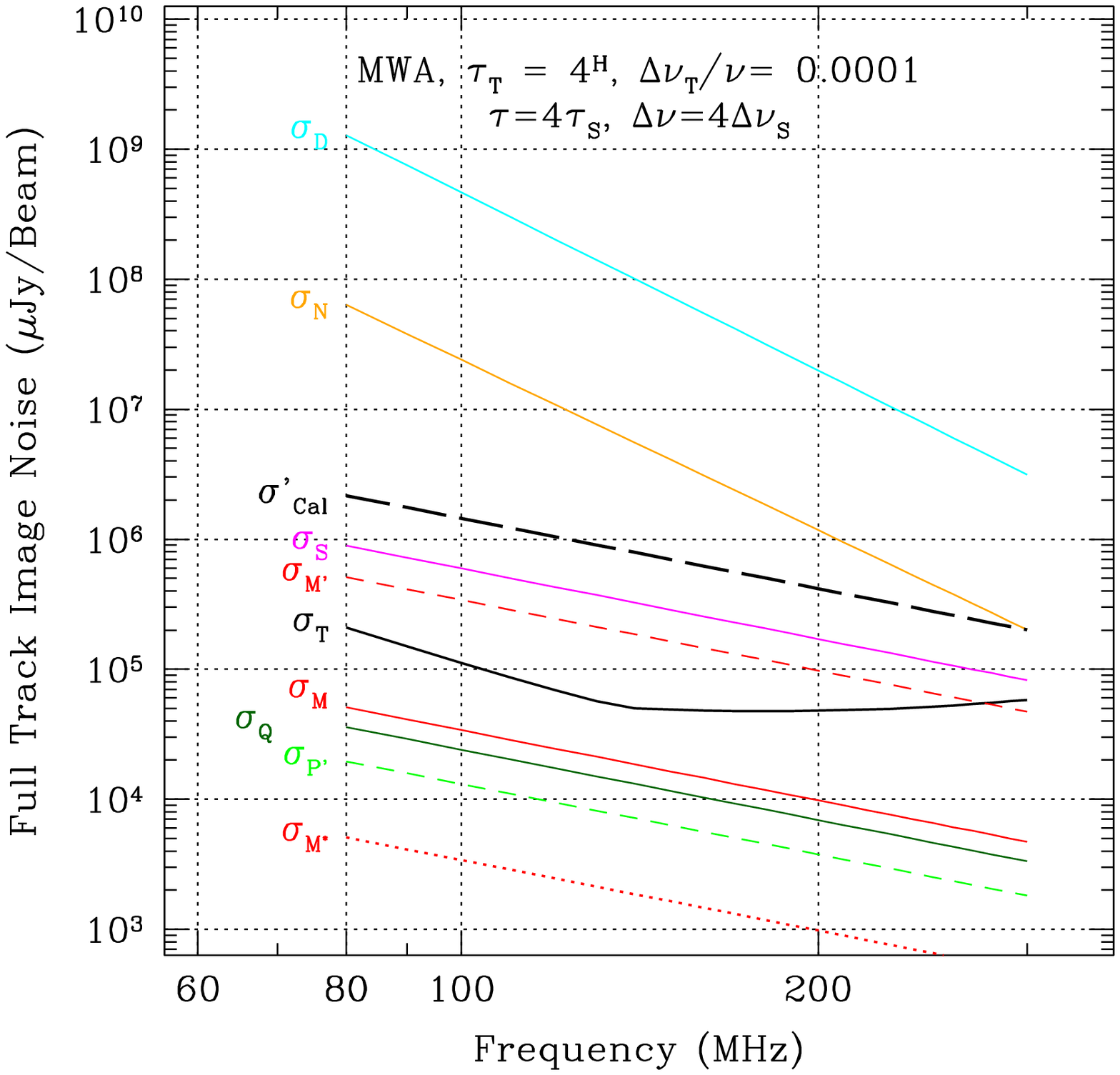}}
\caption{Noise budget for the MWA telescope as in
  Fig.~\ref{fig:vlad}.}
\label{fig:mwa}
\end{figure*}

The noise budget plot of Fig.~\ref{fig:mwa} demonstrates the regime
that this facility will operate in. In view of the extremely large
field of view of each tile, that varies between 200 and 3000 deg$^2$
across the band, this is a case where the ``quiet'' sky approximation
is clearly not relevant and equation \ref{eqn:srmsp} rather than
\ref{eqn:srms} is employed in calculating noise terms.  Far sidelobe
pickup of the sky will dominate over all other contributions by two to
three orders of magnitude on a natural solution timescale. Even full
track continuum and spectral line observations are completely
dominated by this all-sky response to a similar degree. Comparison
with Fig.~\ref{fig:sign} suggests that even the $S_{1.4GHz} < 0.1$~Jy
population is contributing to visibility modulations in this
regime. Modelling of the time variable response to each of 10,000+
sources on the sky does not seem likely to be practical. Even the 8000
instantaneous visibilities are less in number than the number of
relevant sources on the sky, without considering any instrumental
variables. Traditional self-calibration is not a viable option in this
circumstance, although it appears to be essential to provide some
improvement to the noise budget for both continuum and spectral line
observations. A telescope operating in this regime
will need to rely on the quality of external calibration that can be
acheived and seems unlikely to achieve thermal noise limited performance. 

\subsection{SKA1-Low}

The Square Kilometre Array Phase 1 low frequency aperture array is
currently expected to consist of 50 stations of $d = 180$~m diameter
operating between 70 and 450 MHz. The array configuration is planned
to consist of 25 stations within a diameter of 1~km, a further 10 in
the annulus of 5~km diameter and a final 15 distributed out to perhaps
100~km. The system temperature is expected to vary as,
\begin{equation}
T_{Sys} \approx 150 + 60\lambda_m^{2.55} \quad {\rm K},
\label{eqn:ska1ltsys}  
\end{equation}
in terms of the wavelength, $\lambda_m$ in meters. The effective
collecting area per station is expected to vary as,
\begin{eqnarray}
\label{eqn:ska1lae}
 A_{eff} & = & 11200\bigg({\lambda^2 \over 3}\bigg) \qquad (\lambda <
 \lambda_T) \\ 
\nonumber & = & 11200\bigg({\lambda_T^2 \over 3}\bigg) \qquad (\lambda >
\lambda_T)  \quad {\rm m^2}
\end{eqnarray}
given the 11200 dual polarisation antenna elements in each station
that are sparsely distributed for $\lambda < \lambda_T = 2.6$ m. Many
of the same considerations discussed previously for the LOFAR analysis
apply here as well. We will assume an electronic pointing error
associated with gain fluctuations of the individual antenna elements
of about $\epsilon_P^\prime = 0.01$ and timescale of $\tau_P^\prime =
1$ minute. We will also assume a residual beam squash error of
$\epsilon_Q = 0.01$ that applies to the full track observations of
$\tau_T = 4$ hour duration. As a starting assumption we will assume
untapered aperture weighting that yields a near-in sidelobe
contribution $\epsilon_S = 0.1$. The far sidelobe response will be
characterised by $\eta_F \approx 0.5$. Just as done for the case of
LOFAR we will consider both a core configuration with $N = 35$,
$B_{Max} = 5$~km and $B_{Med} = 500$~m, as well as an extended
configuration with $N = 50$, $B_{Max} = 100$~km and $B_{Med} = 2500$~m.

\begin{figure*}
\resizebox{\hsize}{!}{\includegraphics[bb=40 180 535 660]{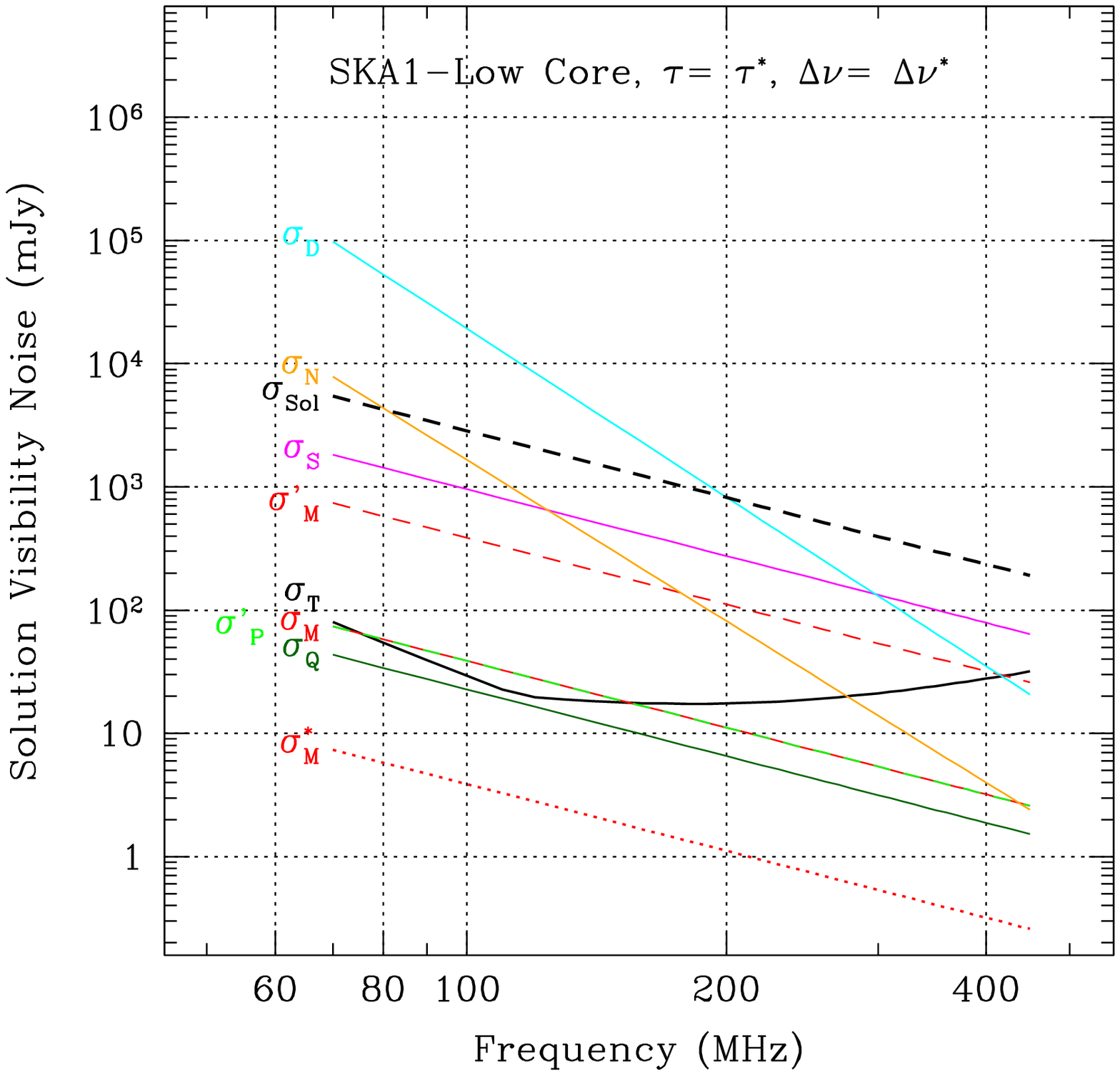},\includegraphics[bb=40 180 535 660]{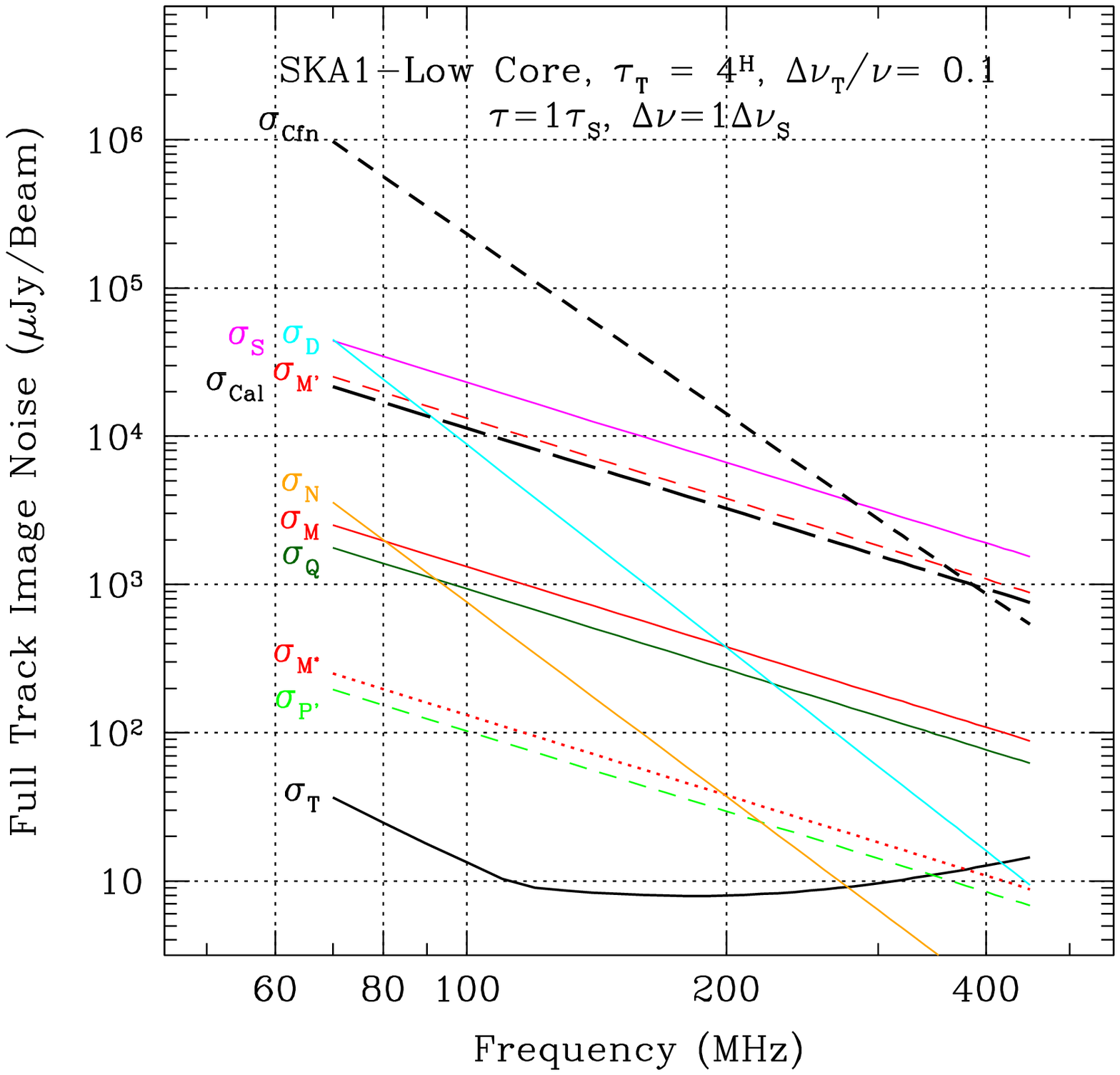},\includegraphics[bb=40 180 535 660]{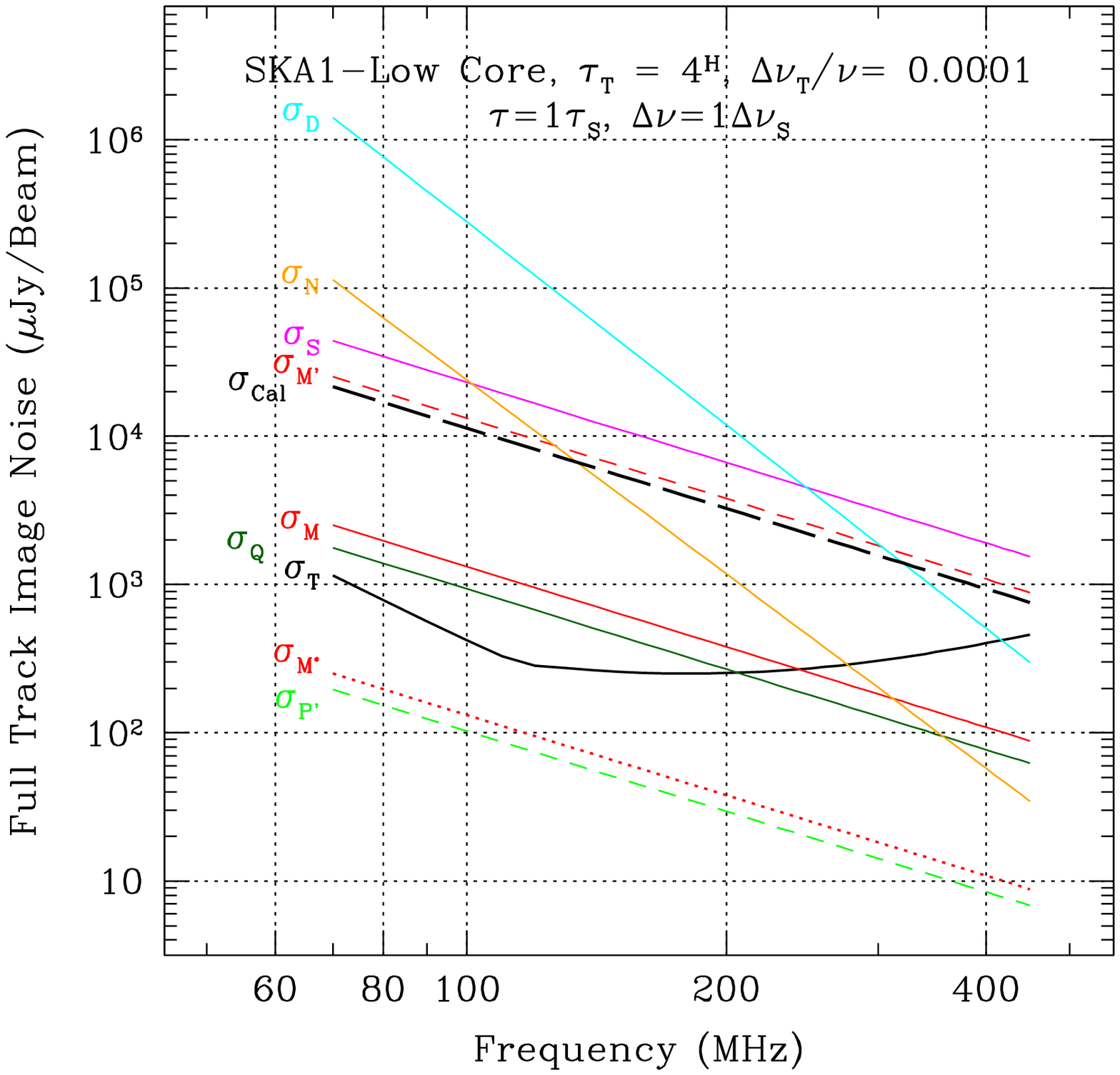}}
\caption{Noise budget for the SKA1-Low Core telescope as in
  Fig.~\ref{fig:vlad}.}
\label{fig:ska1lc}
\end{figure*}

The noise budget for the SKA1-Low core is shown in
Fig.~\ref{fig:ska1lc}. Although the thermal noise is well below the
self-cal solution limit on the natural timescale, there are a number
of other contributions that exceed $\sigma_T$ by large factors. Far
sidelobe pickup is chief among these, followed by near-in sidelobe
contributions and source modelling errors. The brightest 10-20 dB of
the nighttime sky will likely require modelling to allow useful
self-calibration to be undertaken. Full track continuum performance is
completely source confusion dominated, while the spectral line case is
analogous to that seen on the solution timescale. Self-cal will be
required to approach the thermal noise, but far sidelobe
modulations dominate by two to three orders of magnitude over the
thermal noise, followed by near-in sidelobe sources and source
modelling errors. A source modelling precision of about 0.3\% is required, once
the $\sigma_N$ and $\sigma_S$ contributions have been reduced to the
point that they no longer dominate. Adequate modelling of the nighttime
sky will be quite challenging, since its contribution will need to be
reduced by about 20 dB below about 120~MHz. As noted previously, this
will entail modelling down to the $S_{1.4GHz} < 0.1$~Jy level, where
more than $10^4$ sources are above the horizon at any moment. 

\begin{figure*}
\resizebox{\hsize}{!}{\includegraphics[bb=40 180 535 660]{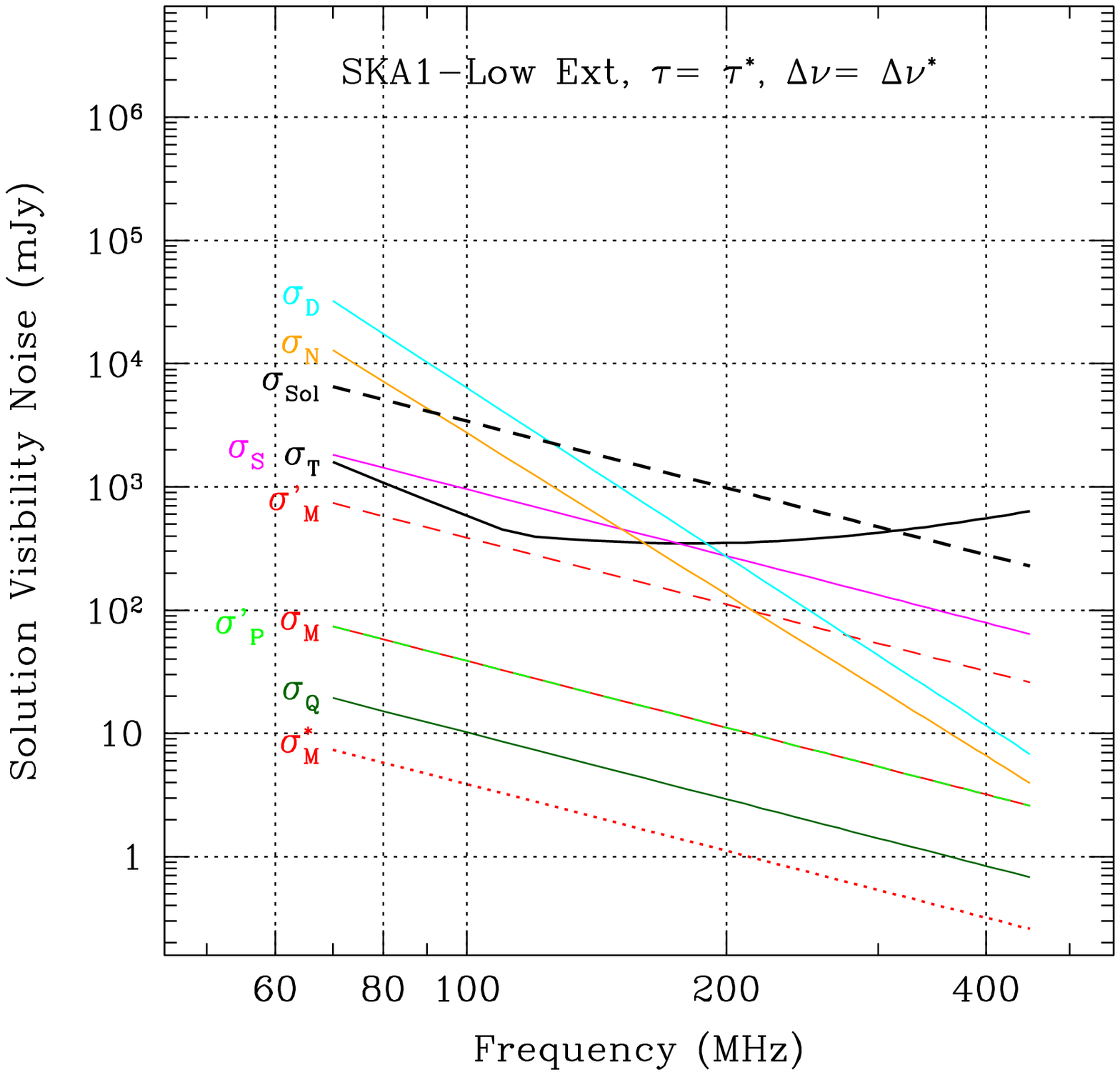},\includegraphics[bb=40 180 535 660]{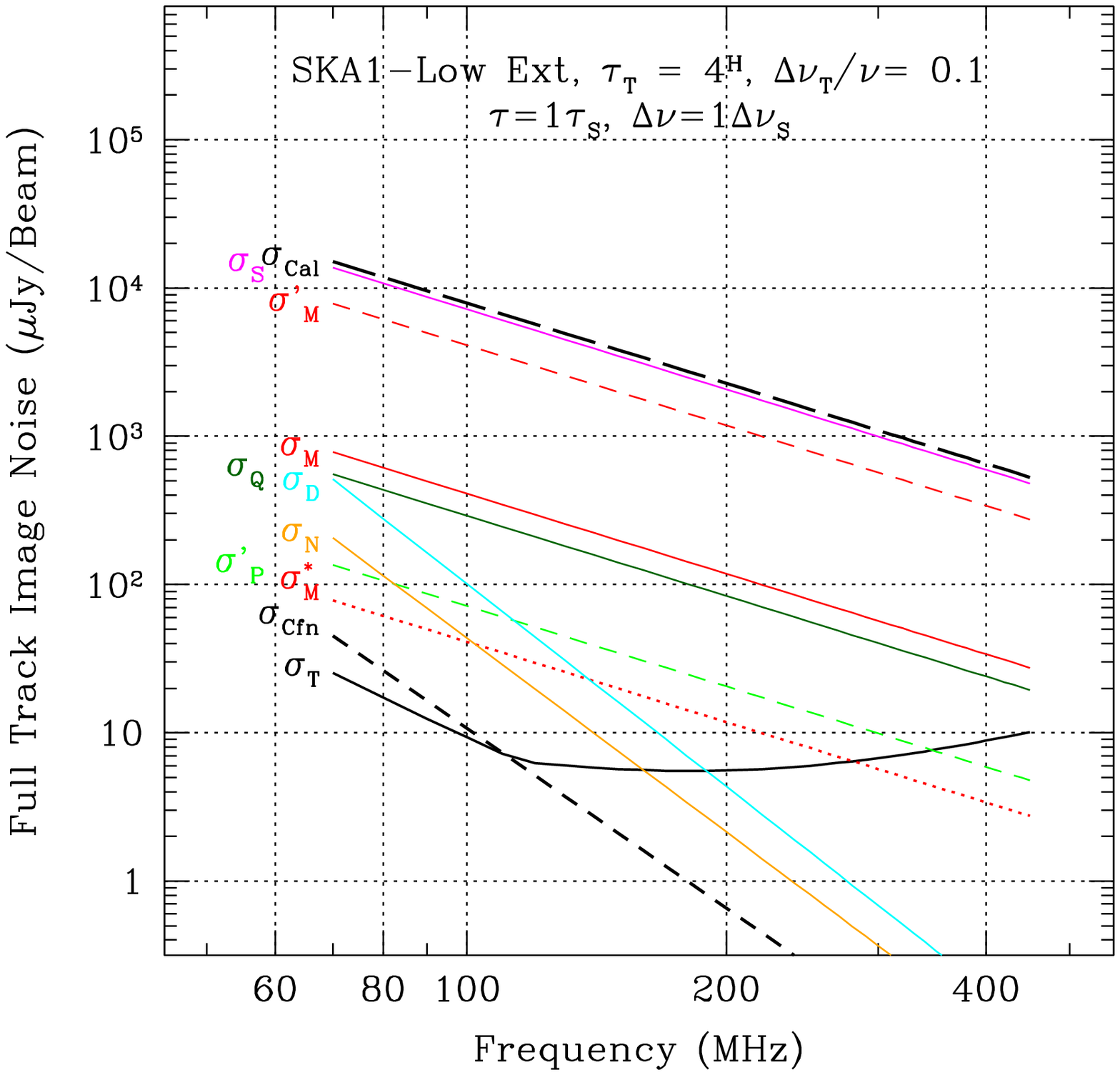},\includegraphics[bb=40 180 535 660]{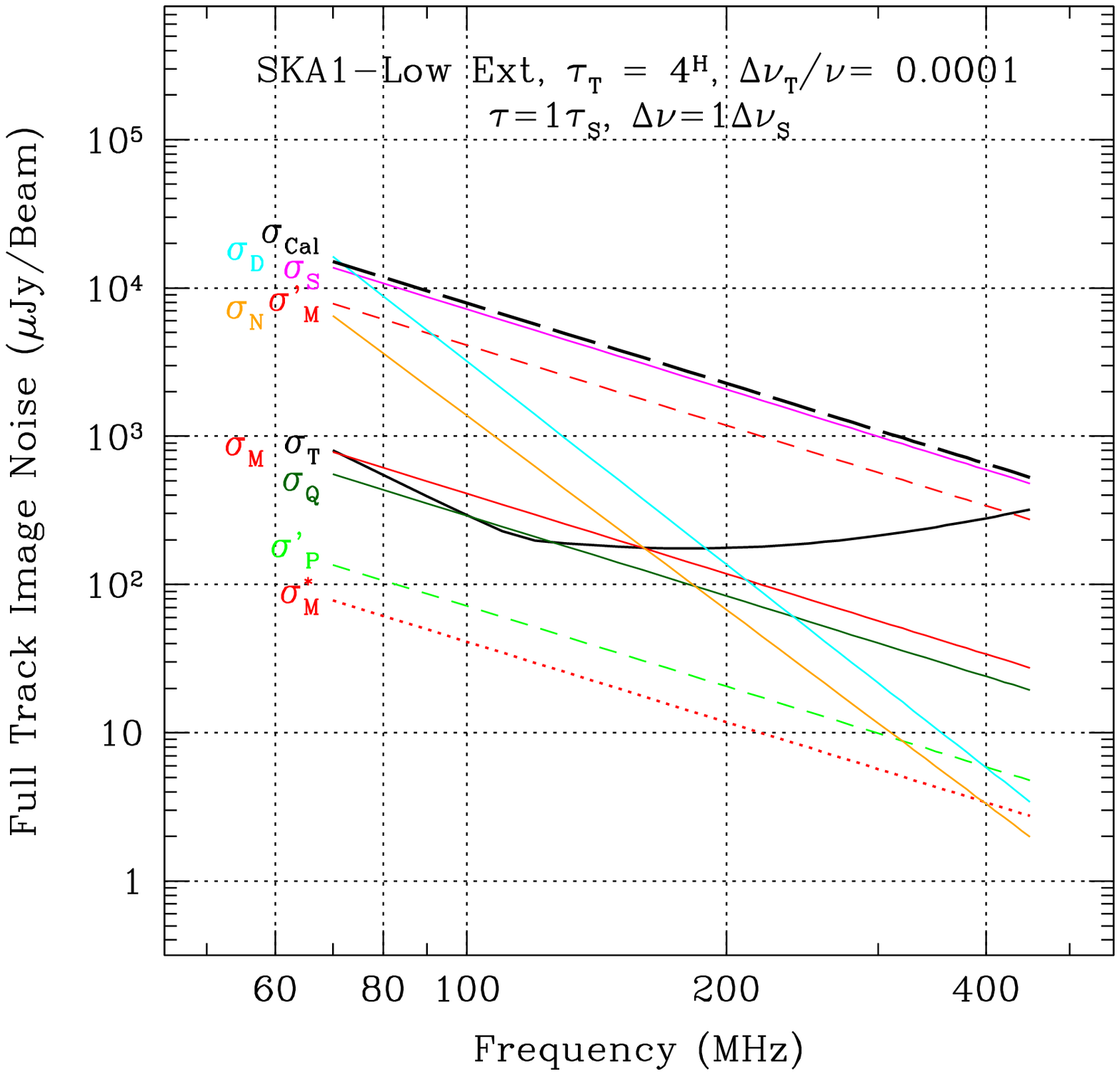}}
\caption{Noise budget for the SKA1-Low extended telescope as in
  Fig.~\ref{fig:vlad}.}
\label{fig:ska1lm}
\end{figure*}

The noise performance of the extended SKA1-Low array is shown in
Fig.~\ref{fig:ska1lm}. While some far sidelobe modelling or data
averaging is needed to enable self-calibration, this is less extreme
than for the more compact core. The full track continuum performance
faces a wide range of practical challenges. All of the effects under
consideration exceed the thermal noise below 200~MHz; the most severe
being unmodelled sources in the near-in sidelobe which must be reduced
by some 20 dB. The full track spectral line case is somewhat less
challenging, although some 10 dB are required here as
well. Self-calibration is required for both continuum and spectral
line applications.

\section{Discussion\label{section:disc}}

\subsection{Dishes\label{section:dish}}

Consideration of the various telescope systems in
Section~\ref{sec:exam} allows some general conclusions to be
drawn. The ``traditional'' 25~m diameter class dishes of existing
synthesis radio telescopes operate in a regime where the thermal noise
of the receiver system dominates the noise budget on a self-cal
solution interval, typically by a large margin. Furthermore, at least
for relatively compact configurations, there is adequate
signal-to-noise provided by sources occurring randomly within the
field to allow self-calibration to be undertaken at close to the
``natural'' solution interval that keeps time and bandwidth smearing
effects at a small fraction of the synthesied beamwidth over the
entire primary beam field of view. This constitutes the most
advantageous of circumstances, since simple imaging strategies can be
employed that use only a single version of the data, in contrast to
more complex strategies that might require simultaneous processing of
multiple phase centres. Even so, full track continuum observations
must overcome several potential obstacles, at the 10 dB level, before
thermal noise limited performance is achieved below 2~GHz. This is
consistent with practical experience with the JVLA, the WSRT and the
ATCA where considerable care must be excercised with precise modelling
of sources that occur within the field and in the case of the JVLA,
explicit modelling of the non-axisymmetric beam shape and its rotation
on the sky during tracking, to keep these effects from becoming
dominant. Above about 2~GHz and in more extended array configurations
it becomes necessary to introduce more data averaging prior to
self-calibration than is optimum for simple imaging
strategies. However, only a limited number of complications need be
addressed to achieve thermal
noise limited performance in full-track continuum observations in
these circumstances. The full-track spectral line performance of such
arrays is less subject to complications and only requires
self-calibration to reach the thermal noise floor below about 2 GHz. Only
in the most compact configurations and at frequencies approaching
1~GHz does the Sun begin to present some problems.

This is in stark contrast to the situation encountered with synthesis
arrays that employ significantly smaller dishes. The most striking
comparison is with the ATA, where the combination of small dish
diameter and compact array configuration yield a very different
regime. For a nominal self-cal solution interval, the thermal noise is
dominated by the far sidelobe response to the entire sky. The same is
true even after integrating over an entire 12 hour track, both for
spectral line and continuum applications. Continuum observations are
further challenged by a wide range of other noise contributions which
dominate over the thermal noise. Its not clear whether high dynamic
range imaging in this regime will be possible.

The case of 12 to 15~m class dishes in more extended configurations is
improved over that with only 6~m dishes. Even so, more than nominal
data averaging is required to provide the signal-to-noise needed for
self-calibration with this dish size, which will add complexity to
wide-field imaging. Higher source modelling precision than what has
been needed for the 25~m VLA dishes, by about a factor of two will be
required, together with explicit modelling of any main beam asymmetries
in cases where the parallactic angle is not fixed on the sky during an
observation. In this respect, the ASKAP telescope benefits from both
the greater circular symmetry of polarisation beams possible with a
phased array feed as well as the polarisation rotation axis that keeps
the beam orientation fixed on the sky during tracking. Single pixel
fed systems with an $(alt,az)$ mount will face significantly greater
challenges to reach the thermal noise floor. Deep field continuum
observations will need to achieve extremely high source modelling
precision that extends out into the near-in sidelobe region.

Exactly the same considerations apply to the SKA, if 15~m class dishes
are deployed as envisaged within the current baseline
design. Achieving thermal noise limited continuum performance will be
more challenging than with the 25~m dishes of the JVLA and sub-$\mu$Jy
imaging would need to overcome a whole range of complex modelling
issues. One of the greatest threats to thermal noise-limited
performance is the departure from circular symmetry of the
polarisation beams and the rotation of this pattern on the sky. The
use of phased array feeds provides one means of reducing the risk
associated with beam asymmetries, while the adoption of an equatorial
or $(alt,az,pol)$ mount system provides another for eliminating their
time variability.

Adoption of a larger dish diameter would provide a significant risk
reduction and this option should be considered very seriously during
cost/performance optimsation of the final SKA dish design.

\subsection{Aperture arrays\label{section:aas}}

The aperture arrays considered in Section~\ref{sec:exam} display many
of the same phenomena. The small station sizes of LOFAR (30.8~m) and
particularly the MWA (4.4~m) result in visibilities that are
completely dominated by noise-like modulations from sources appearing
in the far sidelobe response. Very extensive modelling of the
visibility response to these distant sources becomes necessary before
self-calibration within the main beam can be contemplated. A major
additional complication is introduced by the geometric foreshortening
of each station aperture while tracking a source. While an absence of
moving parts is often cited as a major advantage of these systems, the
continuously changing response of the sky-telescope system is the
price that is paid. Efforts are underway to improve predictive
modelling of both the far sidelobe response patterns
\citepads{5355494} as well as compensation for the time-variable main
beam shape \citepads{2009IEEEP..97.1472R}. Current LOFAR performance
employing baselines up to 30~km and a relative bandwidth,
$\Delta\nu/\nu \sim 0.35$, in a well-characterised reference field
have reached noise levels of about 100--180 $\mu$Jy/beam, or within a
factor of about 1.4 of the thermal noise
\citepads{2013arXiv1301.1630Y} for a 10 hour effective duration
observation.  While the calibration challenge is daunting for LOFAR,
where some 20 dB of reduction is needed to approach the thermal noise
on a solution interval, it is more extreme for the MWA where 30-40 dB
are needed. Although self-calibration seems to be required to reduce
the magnitude of gain calibration errors it does not appear to be a
viable option for the MWA. The source population responsible for
residual visibility errors is so large, consisting of more than $10^4$
relevant sources above the horizon at any moment, that it may not
prove practical to model the time-dependent system response to these
sources at all. This may be an intrinsically under-determined problem
with the finite information available within the solution interval
over which the instrumental response is time-invariant. While it has
been suggested \citepads{2009MNRAS.400.1814M} that alternate
calibration and imaging strategies might be utilised to circumvent
these challenges, the fundamental question of sufficient independent
data constraints on a solution timescale has not been demonstrated in
either theory or practise.

The SKA1-Low concept calls for much larger aperture arrays, of 180~m
diameter, to combat the problem of far sidelobe response. While this
approach helps, it is still the case that many noise contributions
exceed the thermal noise, both on a self-cal solution interval as well
as in full-track spectral line and continuum images. Some 30 dB of
attenuation is necessary for a wide range of phenomena to approach the
thermal noise in a single 4 hour track. This will be extremely
challenging to overcome, since even the 10 dB level issues faced by
the VLA 25~m dishes at GHz frequencies have required many years of
algorithm development to partially address. While large station
diameters seem to offer the best prospects, it will still require
very significant progress in both algorithm development and efficient
computational implementations over what is available currently to
approach thermal noise limited performance for $\nu < 200$ MHz. 

\begin{acknowledgements}
We acknowledge useful discussions with, and comments on earlier drafts
of this manuscript from, Tim Cornwell, Ron Ekers, Stuart Hay, Mike
Kesteven, Rick Perley, Maxim Voronkov, Stefan Wijnholds and an anonymous referee.
\end{acknowledgements}

\bibliographystyle{aa}
\bibliography{man}

\end{document}